\documentclass[aps,pra,epsf,superscriptaddress,amsmath,amssymb,amsfonts,twocolumn,showpacs,nofootinbib]{revtex4-1}

\usepackage{graphicx}
\usepackage{epsfig}
\usepackage{dcolumn}
\usepackage{bm}
\usepackage{braket}
\usepackage{amsmath}
\usepackage{mathtools}
\usepackage{graphicx,color,xcolor}
\usepackage{hyperref}
\newcommand{\abs}[1]{\left| #1 \right|} 
\usepackage{hyperref}
\usepackage{multirow}

\usepackage[normalem]{ulem} 

\begin{document}
\title{Correlated dynamics of collective droplet excitations\\ in a one-dimensional harmonic trap}

\author{I. A. Englezos}
\affiliation{Center for Optical Quantum Technologies, Department of Physics, University of Hamburg, 
Luruper Chaussee 149, 22761 Hamburg Germany}
\author{S. I. Mistakidis}
\affiliation{ITAMP, Center for Astrophysics $|$ Harvard $\&$ Smithsonian, Cambridge, MA 02138 USA} 
\affiliation{Department of Physics, Harvard University, Cambridge, Massachusetts 02138, USA}
\author{P. Schmelcher}
\affiliation{Center for Optical Quantum Technologies, Department of Physics, University of Hamburg, 
Luruper Chaussee 149, 22761 Hamburg Germany} \affiliation{The Hamburg Centre for Ultrafast Imaging,
University of Hamburg, Luruper Chaussee 149, 22761 Hamburg, Germany}

\date{\today}

\begin{abstract}

We address the existence and dynamics of one-dimensional harmonically confined quantum droplets, appearing in two-component mixtures by deploying a nonperturbative approach.
We find that, in symmetric homonuclear settings, beyond Lee-Huang-Yang correlations result in flat-top droplet configurations for either decreasing intercomponent attraction or larger atom number.
Asymmetric mixtures feature spatial mixing among the involved components with the more strongly interacting or heavier one exhibiting flat-top structures.
Applying quenches on the harmonic trap we trigger the lowest-lying collective droplet excitations.
The interaction-dependent breathing frequency, being slightly reduced in the presence of correlations, shows a decreasing trend for stronger attractions. 
Semi-analytical predictions are also obtained within the Lee-Huang-Yang framework.
For relatively large quench amplitudes the droplet progressively delocalizes and higher-lying motional excitations develop in its core.
Simultaneously, enhanced intercomponent entanglement and long-range two-body intracomponent correlations arise.
In sharp contrast, the dipole motion remains robust irrespectively of the system parameters.
Species selective quenches lead to a correlation-induced dephasing of the droplet or to irregular dipole patterns due to intercomponent collisions.

\end{abstract}

\maketitle

\section{Introduction}

Ultracold atoms constitute versatile platforms for probing correlated quantum many-body (MB) phases of matter~\cite{BlochNature2012} such as self-bound quantum droplets~\cite{Petrov2015,KadauDropExp,PfauReview,MalomedLuoReview,MalomedReview}.
The latter manifest the existence of quantum correlations in macroscopic mixtures, represented to first order by the Lee-Huang-Yang (LHY) correction term~\cite{LeeHuangYang1957}, stabilizing the system against collapse due to  mean-field (MF) effects~\cite{Petrov2015,PfauReview,MalomedLuoReview,MalomedReview}.
Importantly, these self-bound states have been realized in both homonuclear~\cite{CabreraTarruellDropExp,CheineyTarruellDropExp,SemeghiniFattoriDropExp} and heteronuclear~\cite{FortHeteroExp} short-range interacting bosonic mixtures in three-dimensions as well as in single component~\cite{KadauDropExp,Bottcher2019SupersolidDrop,chomaz2022dipolar} and binary dipolar gases~\cite{Bisset2021,Smith2021} and mixtures thereof~\cite{MalomedLuoReview,PfauReview}.
Focusing on short-range interacting droplets, several experiments have addressed their dynamical formation~\cite{SemeghiniFattoriDropExp,ModugnoFerioliDynamicalFormation}, the droplet to a gaseous BEC crossover~\cite{CheineyTarruellDropExp,CappellaroCrossoverDynamics,Cui2021PeriodicConfinement} and their collisional properties~\cite{FattoriCollisions} with their self-evaporation and three-body recombination being central issues~\cite{CabreraTarruellDropExp,CheineyTarruellDropExp,SemeghiniFattoriDropExp,FortHeteroExp}.

On the theoretical side, droplets have been found to emerge also in Bose-Fermi mixtures with~\cite{CuiSpinOrbitBoseFermi} and without spin-orbit coupling~\cite{GajdaBoseFermi,Wang_2020BoseFermi}, as well as in Bose-Bose mixtures in the presence of three-body interactions~\cite{Nishida3body,Morera3body1D}.
Moreover, collective excitations~\cite{Sturmer2021} and the properties of topological excitations~\cite{KartashovVortex2018,Malomed2DropVortex,ReimannRotatingAndVortex,ZedaLin2Dvortex,Kavoulakis2DDropRotation} e.g. vortices~\cite{KartashovVortex2018} embedded in a droplet background have been investigated to a certain extent.
Additionally, bosonic droplets can be accommodated in optical lattices in both one-~\cite{MoreraAstrakharchikLattice,MoreraAstrakharchikDimerizedDroplets} and two-dimensions~\cite{2Dlattice,KasamatsuLatticeMott} as well as in semi-discrete settings~\cite{MalomedDualCoreTrap,MalomedSemidiscrete}.
Their excitation spectrum was studied especially in one-dimension (1D)~\cite{AstrakharchikMalomed1DDynamics,Collective1D,AbdullaevSuperGaussian,tengstrand2022droplet,MithunMI} and in the crossover from three-dimensions to 1D~\cite{IdziaszekDimensionalCrossover}. 
Droplets spatial configurations acquire a flat-top (FT) profile for larger atom number~\cite{Petrov2015}, while associated thermal instabilities leading to their self-evaporation have also been reported~\cite{HuiThermal,AstrakharchikThermal,BoudjemaaHigherOrderThermal,BoudjemaaThermal,mithun2021statistical}.

The lifetime of droplets is expected to be prolonged in 1D~\cite{Astrakharchik_2006,BourdelCrossover} and in the case of heteronuclear mixtures, due to their lower density~\cite{FortHeteroExp}.
In spite of this advantage, heteronuclear mixtures, where correlation effects should be enhanced due to the mixed character of the ensuing droplet, have not been extensively studied thus far~\cite{FortModugnoSelfEvaporation,AncilottoLocalDensity,Mistakidis2021}.
This is in part due to the complicated form of the corresponding extended Gross-Pitaevskii equation (eGPE)~\cite{FortModugnoSelfEvaporation,FattoriCollisions}.
It is a partial aim of our study to explore the role of correlations in harmonically trapped heteronuclear mixtures and the associated droplet configurations.
In 1D, droplets have been primarily described within the eGPE framework~\cite{AstrakharchikMalomed1DDynamics,Collective1D,AbdullaevSuperGaussian,MithunMI,PetrovLowD},with notable exceptions where non-perturbative approaches were utilized to unveil beyond-LHY physics in free space~\cite{ParisiMonteCarlo2019,ParisiGiorginiMonteCarlo,Mistakidis2021} and in lattice settings~\cite{MoreraAstrakharchikLattice,MoreraAstrakharchikDimerizedDroplets}.
Interestingly, however, the formation of 1D harmonically trapped droplets remains largely unexplored~\cite{Mistakidis2021,DEBNATH2022MGPtrap,PathakHarmonic2022}.
Within the eGPE framework~\cite{Mistakidis2021,DEBNATH2022MGPtrap}, FT droplet structures were shown to be suppressed, while it was argued that they can exist using temporally varying interactions~\cite{PathakHarmonic2022}.
 
In this sense, it is highly desirable to inspect the interplay of correlations for both stationary droplet entities as well as in their dynamics when trap effects, being commonly un-avoidable in experiments~\cite{CabreraTarruellDropExp,CheineyTarruellDropExp,SemeghiniFattoriDropExp,FortHeteroExp}, are not omitted.
The conditions under which FT configurations occur in an external harmonic trap and importantly whether the presence of beyond-LHY correlations favors their formation remain unclear~\cite{Mistakidis2021}.
Moreover, droplet collective excitations, being crucial for understanding e.g. their ability to support nonlinear excitations~\cite{Kavoulakis2DDropRotation}, can be triggered owing to the tunability of the external confinement, without exciting higher order correlations as is the case e.g. for interaction quenches~\cite{Mistakidis2021}.
While the behavior of the droplet breathing frequency in 1D has been studied to some extent~\cite{AstrakharchikMalomed1DDynamics,Mistakidis2021}, the dynamical response of the system and the build-up of intercomponent correlations still remain elusive especially when species selective quenches are applied.
In the present work, we study the ground state and collective dynamics of 1D harmonically trapped  droplet structures in both homonuclear and heteronuclear mixtures which experience repulsive (attractive) intra- (inter-)component interactions.
For addressing correlation effects on the formation of harmonically trapped droplets beyond the LHY approximation we rely on the nonperturbative multi-layer multi-configuration time-dependent Hartree method for atomic mixtures (ML-MCTDHX)~\cite{Kronke_2013,Cao2013,cao2017unified}.
Additionally, in order to explicate the role of correlations at different levels we compare our results with the common MF treatment~\cite{pethick2008bose} as well as the predictions of the eGPE~\cite{PetrovLowD}.

We showcase that FT signatures stemming from beyond-LHY correlations are present in symmetric homonuclear mixtures for either decreasing attraction or an increasing atom number, in contrast to the predictions of the eGPE~\cite{DEBNATH2022MGPtrap,Mistakidis2021}.
Otherwise, a larger attraction leads to an alteration of the droplet configuration from a FT to a Gaussian-shaped one as in free space~\cite{ParisiGiorginiMonteCarlo,Mistakidis2021}.
Another central result of our findings is that similar structures occur also for interaction (mass) imbalanced bosonic mixtures, whose 1D eGPE is not known, where the more strongly repulsive (heavier) component features FT signatures for reduced intercomponent attraction and the setting is spatially mixed.
In all cases, the droplets show an anti-bunching (bunching) behavior at the same (different) locations.

The breathing dynamics of the droplet is initiated through a quenching the trap frequency.  
Interestingly, using relatively large quench amplitudes higher-lying motional excitations build-upon the droplet core and simultaneously density portions are expelled, a process that is reminiscent of the self-evaporation mechanism in 3D~\cite{SemeghiniFattoriDropExp,ModugnoFerioliDynamicalFormation}. 
This self-evaporation mechanism, however, has not been previously reported in 1D settings~\cite{Collective1D,Mistakidis2021}, exposing the crucial effect of the external confinement on the dynamical response of quantum droplets.
We show herein that this overall unstable dynamics is accompanied by enhanced intercomponent entanglement and the development of long-range two-body intracomponent spatial correlations.
Analytical predictions for the breathing frequency are provided at relevant limits by extending the variational approximation within the LHY theory introduced in Ref.~\cite{AstrakharchikMalomed1DDynamics}, while numerical estimations show that it tends to the ideal gas case for strong attractions and it is reduced towards the FT regime in accordance with Refs.~\cite{ParisiGiorginiMonteCarlo,ParisiMonteCarlo2019,Mistakidis2021}.
Another important result is the development of a pronounced correlation driven dephasing in the dynamics of heteronuclear mixtures, where the relevant eGPE equation is not available.
This dephasing is owed to the tendency of the individual components to oscillate in-phase (with a phase difference) for strong (weak) attractions.
In contrast to free space, here the confinement allows to also excite the droplet dipole motion, via a sudden displacement of the trap position, which is found to be insensitive to interactions.
Moreover, utilizing relevant species selective quenches for weak attractions gives rise to irregular dipole patterns as a result of intercomponent collisions and consequent energy transfer, while at strong attractions the droplets prefer to maximize their overlap.

This work is organized as follows.
In Section~\ref{sec:setup} we introduce the bosonic mixture under investigation and briefly discuss the established eGPE theory and the nonperturbative ML-MCTDHX approach used for the description of quantum droplets.
Section~\ref{sec:Ground State Droplets} elaborates on the static properties of harmonically confined droplets, with an emphasis on their correlation aspects.
The non-equilibrium droplet dynamics upon considering quenches of the trap is subsequently examined focusing on their collective excitations and in particular their breathing mode [Sec.~\ref{sec:Collective excitations of Droplets}] and dipole motion [Sec.~\ref{Dipole dynamics}].
We conclude offering also perspectives for future work in Sec.~\ref{sec:SummaryAndOutlook}.
Appendix~\ref{app:var} provides the ingredients of the variational approximation and the time-dependent Gaussian ansatz employed for a complementary interpretation of the droplet properties.

\section{Attractively interacting mixture and theory models}\label{sec:setup} 

\subsection{Many-body two-component bosonic system}

We employ a particle-balanced bosonic mixture with $N_A=N_B=N$ atoms confined in a weak 1D harmonic trap.
To address confined droplet structures in both homonuclear and heteronuclear mixtures we shall consider mass ratios $m_A/m_B=1$ and $m_B/m_A \approx 2.1$ respectively.
Such two-components systems can be experimentally emulated using two hyperfine states of $^{39}$K ~\cite{SemeghiniFattoriDropExp,CabreraTarruellDropExp,CheineyTarruellDropExp} or in the heteronuclear case a mixture composed of $^{42}$K and $^{87}$Rb ~\cite{FortHeteroExp}.
Our setting focuses in the ultracold temperature limit where s-wave scattering dominates~\cite{olshanii1998atomic}.
In this sense, inter-particle interactions correspond to contact potentials characterized by effective intra- ($g_{A}$, $g_{B}$) and intercomponent ($g_{AB}$) coupling strengths.
They can be tuned either through the three-dimensional s-wave scattering lengths using Feshbach resonances~\cite{chin2010feshbach,kohler2006production} or via the transversal confinement with confinement induced resonances~\cite{olshanii1998atomic}.
The MB Hamiltonian reads 
\begin{align} \label{MB_Hamilt}
\begin{split}
    H =& \sum_{\sigma=A,B} \sum_{i=1}^{N_{\sigma}} \left(-\frac{\hslash^2}{2m_\sigma}   \left(\frac{\partial^2}{\partial {x_i^\sigma}^2}\right)   + \frac{1}{2}m\omega^2(x_i^\sigma)^2\right) \\
    &+ \sum_{\sigma=A,B}g_{\sigma} \sum_{i<j}^{N_{\sigma}} \delta(x_i^{\sigma} -x_j^{\sigma}) + g_{AB} \sum_{i=1}^{N_{A}} \sum_{j=1}^{N_B} \delta(x_i^A -x_j^B).    
\end{split}
\end{align}
The frequency of the longitudinal ($\omega_{{\rm x}}$) over the transverse ($\omega_{{\rm \perp}}$) trapping frequencies is fixed to $\omega=\omega_{{\rm x}}/\omega_{{\rm \perp}}=0.01$.
Similar values are commonly employed experimentally to realize 1D settings~\cite{Ketterle2001LowDexp}.
In the following, we rescale the Hamiltonian in terms of $\hslash\omega_\perp $.
This means that the length, time and interaction strengths are expressed with respect to $\sqrt{\hslash/(m\omega_{\rm \perp}) }$, $1/\omega_{\rm \perp}$ and $\sqrt{\hslash^3\omega_{\rm \perp} /m}$ respectively.

\subsection{Droplet region and the extended Gross-Pitaevskii equation} 

Within the local density approximation and under the impact of the first order quantum correction term (LHY contribution), Bogoliubov theory leads to the so-called eGPE framework~\cite{Petrov2015,PetrovLowD}.
It has been demonstrated that the eGPE, in the absence of confinement ($\omega=0$) and in the thermodynamic limit, describes the formation of quantum droplets~\cite{PetrovLowD}.
Considering the average repulsion $g=\sqrt{g_{A}g_{B}}$, the droplet interval is quantified by the measure $\delta g=g+g_{AB}$ with $0 < \delta g\ll g$~\cite{PetrovLowD}, which implies interspecies attraction $g_{\rm AB}<0$.
For a symmetric mixture, i.e. $m_A=m_B \equiv m$, $N_A=N_B\equiv N$ and  $g_{A}=g_{B}\equiv g$, the genuine two-component system is described by a reduced  single-component eGPE equation which in the presence of a sufficiently weak ($\omega \ll 1$) harmonic trap reads  
\begin{align}
\label{MGP}
\begin{split}
    i\hslash &\frac{\partial \Psi(x,t)}{\partial t}= - \frac{\hslash^2}{2m}\frac{\partial^2 \Psi(x,t)}{\partial x^2} + \delta g |\Psi(x,t)|^2\Psi(x,t) \\
    &- \frac{\sqrt{2m}}{\pi\hslash}g^{\frac{3}{2}}|\Psi(x,t)|\Psi(x,t) + \frac{1}{2}m\omega^2x^2\Psi(x,t).
\end{split}
\end{align}

Within the interaction regime $0<\delta g \leq g$ a structure reminiscent of a liquid puddle appears being characterized by a FT density profile, which saturates at $n_0=8mg^3/(9\pi^2\hslash^2\delta g^2)$~\cite{PetrovLowD}.
However, an increasing attraction such that $0<\delta g\ll g$ results in a transition behavior towards more localized solutions having a Gaussian-shape.
In both cases, these self-bound localized structures emerge due to the competition between the overall quadratic MF repulsion and the linear LHY attraction and thus constitute a beyond MF effect. 
Moreover, in the case that the interspecies MF attraction balances exactly the respective intraspecies repulsion, namely $\delta g=0$, the eGPE equation depends purely on the quantum fluctuation LHY term and the so-called LHY fluid arises~\cite{Skov2021LHYfluidexp}.
Finally, turning to $\delta g<0$ the eGPE approach admits various soliton-type solutions, including  "bubble" or "W-shaped" configurations under suitable conditions, see for details Refs.~\cite{BARASHENKOV1993Bubbles,Optik2017Belic,MithunMI}.

The inclusion of a harmonic trap ($\omega \neq 0$) leads to localized Gaussian-shaped (soliton type) configurations in the MF realm, independently of the value of the interspecies interaction $g_{AB}$~\cite{pethick2008bose}.
The corresponding density distributions exhibit a larger width for decreasing $\abs{g_{AB}}$~\cite{pethick2008bose,Stringari2016BEC}.
Hence, as we shall demonstrate below, the more pronounced BMF effect that is expected in the presence of confinement is the existence of a FT density profile restricted around the trap center, where trap effects are diminished.

Concluding, the eGPE~\eqref{MGP} was derived in the absence of confinement ($\omega=0$) and it is valid for macroscopic systems close to the MF balance point $\delta g \approx 0$~\cite{PetrovLowD}.
However, it has been demonstrated that its predictions can be in good qualitative agreement with non-perturbative methods even for finite values of $\delta g$ and $N$, i.e. for mesoscopic systems~\cite{ParisiMonteCarlo2019,ParisiGiorginiMonteCarlo,Mistakidis2021}.
The inclusion of a harmonic trap is expected to affect the Bogoliubov modes and therefore the form of the LHY term~\cite{BourdelCrossover}.
Nevertheless, throughout this work, we employ the eGPE framework since it provides an adequate phenomenological description of quantum droplets and in order to exemplify when effects beyond the standard LHY theory become important.

\subsection{Many-body wave function approach}\label{Variational Method}

To expose the impact of beyond LHY correlation effects on the ground state and quench dynamics of quantum droplets we shall utilize the {\it ab-initio} ML-MCTDHX method~\cite{Kronke_2013,Cao2013,cao2017unified}.
It enables us to numerically solve the time-dependent MB Schr\"odinger equation of the mixture and specifically its multi-layer structure of the total MB wave function is tailored to account for both intra- and intercomponent correlations. Comprehensive discussions on the ingredients, successful applicability and benchmarking of this approach to various multicomponent settings such as impurity setups, cavities and spinor systems and reductions to other approaches can be found in the recent reviews~\cite{mistakidis2022cold,lode2020colloquium}.

Particularly, to address the intercomponent correlations of the bosonic mixture, the MB wave function is written as a truncated Schmidt decomposition~\cite{horodecki2009quantum} based on $D$ different species functions, $\ket{\Psi_k^\sigma(t)}$, for each component $\sigma=A,B$. 
Accordingly 
\begin{equation}
\ket{\Psi(t)}= \sum_{k=1}^{D} \sqrt{\lambda_k(t)} \ket{\Psi_{k}^{A}(t)} \ket{\Psi_{k}^{B}(t)}.
\label{SpeciesFunctionLevel}
\end{equation}
The time-dependent Schmidt weights $\sqrt{\lambda_k(t)}$ characterize the degree of intercomponent correlations (or entanglement) of the system, since if only $\lambda_1(t)=1$ is non-zero  and $\lambda_{k>1}(t)=0 $, then the MB ansatz is a product (non-entangled) state.
Conversely, the wave function is in a superposition and the system is referred to as entangled~\cite{horodecki2009quantum,mistakidis2022cold}.

Subsequently, intracomponent correlations are incorporated, by expanding each species function as a linear superposition of time-dependent number states $\ket{\mathbf{n_k}^\sigma}$ with time-dependent expansion coefficients $A_{n_k}^{\sigma}(t)$. 
The number states $\ket{\mathbf{n}_k^\sigma}$ refer to the full set of permanent states defined by $d_\sigma$ time-dependent variationally optimized single-particle functions (SPF's) or orbitals $\ket{\Phi_i^\sigma}$ with occupation numbers $\mathbf{n}=(n_1,...,n_{d_\sigma})$.
Moreover, the $d_{\sigma}$ time-dependent SPFs evolve in the single-particle Hilbert space spanned by the time-independent basis $\{ \ket{r_j^k} \}_{j=1}^{ \mathcal{M}}$.
The latter, in our case, is taken to be an $\mathcal{M}$ dimensional discrete variable representation with $\mathcal{M}=1000$ grid points.
The equations of motion for the coefficients of the ML-MCTDHX wave function ansatz describing the MB  Hamiltonian of Eq.~\eqref{MB_Hamilt} are found, for instance, using the Dirac-Frenkel variational principle~\cite{frenkel1934wave,cao2017unified}, $\braket{\delta\Psi|( i\hbar \partial_t-\Hat{H})|\Psi}=0$.

Summarizing, a major asset of the above outlined approach is the expansion of the MB wave function in terms of time-dependent and variationally optimized basis sets. 
In this sense, the relevant Hilbert space is efficiently spanned employing a computationally feasible basis size as compared to other approaches such as exact diagonalization using a time-independent basis. Naturally, as in every \textit{ab-initio} method, the superposition of multiple orbitals restricts its applicability to mesoscopic systems when the degree of correlations is enhanced. 
For instance, in the current attractively interacting mixture where intercomponent correlations become appreciable the truncated Hilbert space is characterized by the orbital configuration space $C=(D;d_A;d_B)=(10,4,4)$ for $N_A=N_B=N=20$ and $C=(10,6,6)$ for $N_A=N_B=N=5$. 
As a consequence, the number of the respective equations of motion that are numerically solved is tractable. 
Notice that due to the variational character of the method, its convergence has been carefully checked, namely the used observables remain unchanged within a desired accuracy upon increasing the basis size. 

The ML-MCTDHX wave function ansatz naturally reduces to the usual MF one~\cite{pethick2008bose}, where all correlation are absent, when only a single Schmidt coefficient and SPF for each species are used ($D=d_A=d_B=1$).
In this case, the corresponding wave function takes a simple product form, and considering a variational principle yields the widely used coupled set of Gross-Pitaevskii equations for the bosonic mixture~\cite{pethick2008bose,Stringari2016BEC}.
Namely 
\begin{equation}
\label{MFAnsatz}
\begin{split}
        &i\hslash \frac{\partial \Phi^\sigma (x,t)}{\partial t}= - \frac{\hslash^2}{2m}\frac{\partial^2 \Phi^\sigma (x,t)}{\partial x^2} +\frac{m \omega^2x^2}{2}\Phi^\sigma (x,t)  \\
        &+g_\sigma |\Phi^\sigma (x,t)|^2\Phi^\sigma (x,t)+g_{\rm \sigma\sigma'} |\Phi^{\sigma'}|^2\Phi^\sigma(x,t). \\
\end{split}
\end{equation}
By comparing the MF predictions to the MB ones, we are able to infer the impact of interparticle correlations on the formation and dynamics of confined quantum droplets.

\begin{figure}[ht]
\includegraphics[width=0.48\textwidth]{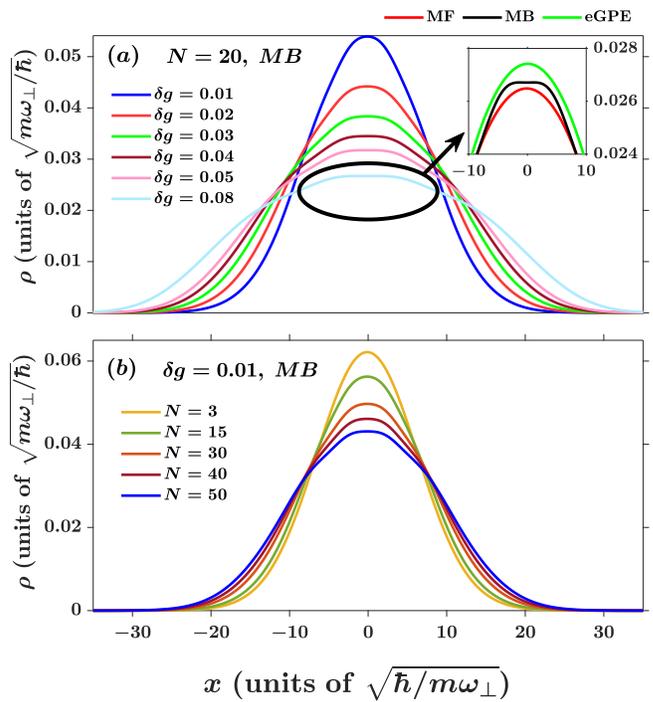}
\caption{Ground state droplet densities of a symmetric mixture in a harmonic trap as obtained in the MB approach.
The cases of (a) different $\delta g$ and fixed $N_A=N_B=N=20$ and (b) varying atom number $N$ for constant interaction $\delta g=0.01$ are presented.
A transition from a Gaussian-type distribution to a FT one for increasing either $\delta g$ or $N$ occurs.
In (b) a FT profile appears for $N=50$.
Inset of (a): Density around the trap center within the MF, MB and the eGPE approach for $\delta g=0.08$. Notice that only the MB calculation captures the FT profile.
}
\label{fig:GSBalanced}
\end{figure} 

\section{Ground State Droplets}\label{sec:Ground State Droplets} 

We begin by studying the formation of 1D harmonically trapped quantum droplets appearing in two-component short-range interacting bosonic mixtures. 
Our analysis is mainly based on the above-discussed MB ML-MCTDHX approach which allows us to systematically account for beyond MF corrections.
To expose the latter we also occasionally compare with the predictions of the eGPE treatment and the common MF method.

\subsection{Symmetric bosonic mixtures}\label{SymmetricGS}

Initially, we consider a fully symmetric homonuclear mixture characterized by $N_A=N_B\equiv N$, $m_A=m_B \equiv m$, $g_{A}=g_{B} \equiv g=0.1$ while the interspecies attraction $g_{\rm AB}<0$ is tuned.
Recall that in this case the two components behave equivalently~\cite{Petrov2015,AstrakharchikMalomed1DDynamics,mistakidis2022cold}, since they also experience the same external trap and thus the mixture reduces to a single-component system.
The respective one-body density~\footnote{In the case of the symmetric mixture the observables associated with the two components are the same. For instance, the densities $\rho_A(x)=\rho_B(x)\equiv \rho (x)$ as well as the reduced density matrices [Eq.~(\ref{2bodyDens})] $\rho^{(2)}_{AA}(x_1,x_2)=\rho^{(2)}_{BB}(x_1,x_2)\equiv \rho^{(2)}(x_1,x_2)$.} 
$\rho (x)$, which is throughout normalized to unity, is depicted in Fig.~\ref{fig:GSBalanced}(a).
Apparently, there is a transition from the Gaussian density profile [see, for instance, $\delta g=0.01$]
to a more delocalized FT one [e.g. for $\delta g=0.08$]
around the trap center for increasing $\delta g=g+g_{AB}$.
The FT structure appears only for the MB case and it is an effect of residual beyond LHY correlations since it does not appear in the eGPE case, see the inset of Fig.~\ref{fig:GSBalanced}(a).
However, it should be mentioned that the above-described delocalization trend of the density for larger $\delta g$ is also captured within the eGPE (not shown) and it can be explained by the scaling of the healing length $\xi \propto \delta g/g^{3/2}$~\cite{Collective1D}. 
Accordingly, also the MF fails to capture this FT structure showing a relatively more delocalized density distribution.

Naturally, in our trapped system the FT features are not as prominent as in free space. 
Quantitatively, if the system resides in its ground state with density close to its saturation value in free space, i.e.  $|\Psi(x,t)|^2\approx n_0=8g^3/(9\pi^2\delta g^2)$, then it is described to a good approximation by the static eGPE in the Thomas-Fermi limit, namely $\delta g n_0 - \frac{\sqrt{2}}{\pi\hslash}g^{\frac{3}{2}}\sqrt{n_0} + \frac{1}{2}\omega^2x^2=0$. 
The effect of the harmonic trap then becomes comparable to that of the MF and LYH terms for $x\approx\sqrt{8g^3/(9\pi^2\delta g)}\omega^{-1}\lessapprox 10$, for the parameter values considered here, which gives an estimation for the size of the observed FT profiles. 
This suggests that the FT profile is not expected to grow significantly in size even for very large particle numbers. 
Of course, a systematic study of the first- (LHY) and potentially higher-order perturbative corrections in the presence of the trap would be required to fully characterize the FT configurations, a study that lies beyond the scope of this work.

\begin{figure}[ht]
\centering
\includegraphics[width=0.48\textwidth]{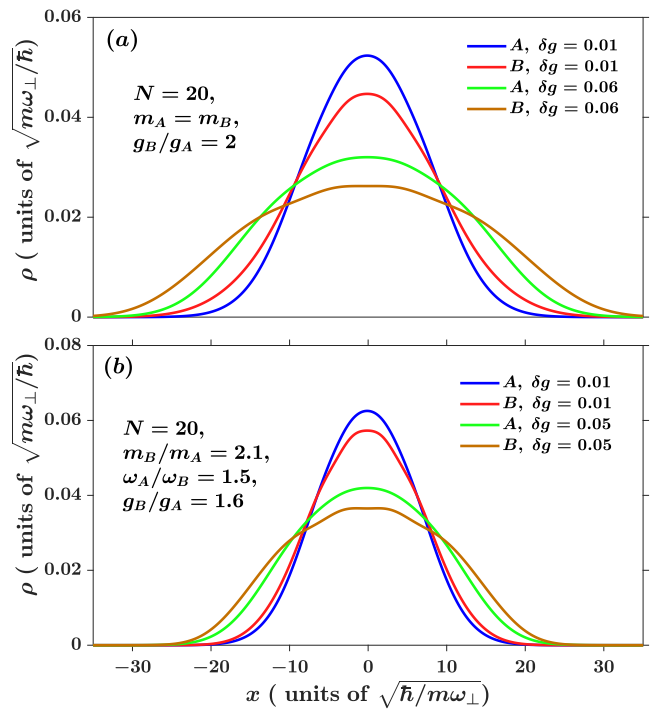}
\caption{Density profiles of harmonically confined droplets within the MB approach for varying $\delta g$ in (a) an interaction-imbalanced ($g_B=2g_A=0.1$) and (b) a mass-imbalanced ($m_B=2.1m_A$) bosonic mixture.
A tendency towards a FT droplet building upon the B component being either (a) more strongly interacting or (b) heavier takes place for larger $\delta g$.
Intercomponent spatial mixing is also enhanced for increasing $\delta g$. 
In both cases $N_A=N_B=N=20$, while in the mass-imbalance setting $g_B=1.6g_A=0.08$.}
\label{fig:GSImb}
\end{figure}

A similar structural deformation from a Gaussian to a FT configuration can be realized for a fixed interspecies attraction e.g. $\delta g=0.01$ and by varying the atom number as illustrated in Fig.~\ref{fig:GSBalanced}(b).
This behavior can be understood in terms of the eGPE predictions in free space.
Particularly, it has been shown~\cite{PetrovLowD} that in the absence of a trap, the droplet exhibits a FT profile when its particle density saturates towards $n_0=8g^3/(9\pi^2\delta g^2)\approx0.9$ for $g=0.1$ and $\delta g=0.01$.
In our case where $\omega \neq 0$ we can assume that the mixture resides within the spatial region $|x|<2 a_{{\rm osc}} = 20$.
As such, its peak density is of the order of $N/4 a_{{\rm osc}}$ with $a_{\rm {osc}}=1/\omega$.
Then the above-mentioned free space saturation density is reached as long as $N_s/4 a_{\rm osc}\approx n_0$, which corresponds to a critical atom number $N_s\approx 36$ for saturation in the trap.
This prediction is found to be consistent with our MB calculations [Fig.~\ref{fig:GSBalanced}(b)], where the FT profile emerges only for $N>40$.
Note, however, that this argument does not apply for larger values of $\delta g$.
As an example, for $g=0.1$ and $\delta g=0.08$, the saturation density in free-space is $n_0\approx 0.014$ and the corresponding critical particle number for saturation in the trap is $N_s \approx  0.5$.
The latter implies that a FT should occur for any atom number, which is of course not confirmed within our simulations (not shown).
Let us mention that in the absence of an external trap it was argued, by employing a Quantum Monte-Carlo approach, that for larger values of $\delta g$ droplet formation is inhibited due to the generation of dimers that appear due to beyond-LHY correlations~\cite{ParisiMonteCarlo2019,Hui2020LowD}.
This can be understood from the fact that the droplet saturation density $n_0=8g^3/(9\pi^2\delta g^2)$ lies below the dimer threshold $2n_0/g\leq 1$ for sufficiently large $\delta g \approx g$.
Evidently, in our setup, the harmonic trap prevents the mixture from reaching such low densities and, for instance, in the case of $N=20$, $g=0.1$ and $\delta g=0.08$ a liquid-like state with saturation density $n_{tr}(\delta g =0.08) =N \rho_{{\rm max}}(\delta g =0.08) = 0.5342$, where $\rho_{{\rm max}}$ denotes the droplet peak density, is established as observed in Fig.~\ref{fig:GSBalanced}(a). 

\begin{figure*}[ht]
\includegraphics[width=0.8\textwidth]{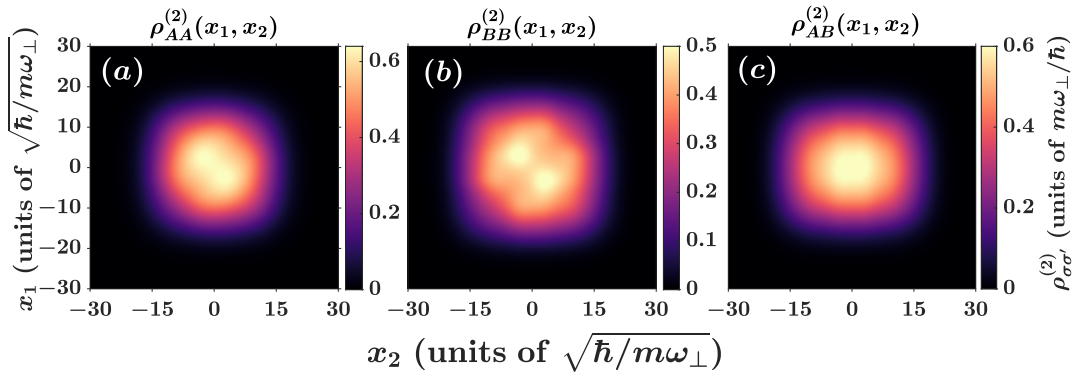}
\caption{Two-body reduced intracomponent density for species (a) $A$ and (b) $B$ as well as (c) the respective intercomponent density in the ground state of the mass and interaction-imbalanced mixture with $m_B=2.1m_A$, $g_B=1.6g_A=0.08$, $\delta g=0.05$ and $N_A=N_B=N=20$.
An intraspecies anti-bunching behavior at the center of the droplet occurs [see the diagonal of $\rho^{(2)}_{AA}(x_1,x_2)$, $\rho^{(2)}_{BB}(x_1,x_2)$] which is more prominent for the more strongly interacting $B$ component.
Intercomponent mixing is identified in $\rho^{(2)}_{AB}(x_1,x_2)$.}
\label{fig:2BDGS}
\end{figure*}

Overall, we can conclude that in our setting the interaction region where FT structures appear (for fixed particle number) is shifted  to larger values of $\delta g$ when compared to their free space counterparts.
Moreover, the Lieb-Liniger parameter~\cite{giamarchi2003quantum,mistakidis2022cold} $\gamma=mg/(\hslash^2n_{\rm max})$ takes the value $\gamma = 0.19$ [$\gamma = 0.047$] for the parameters where FT structures occur in the case of $N=20$ and $\delta g=0.08$ [$N=50$ and $\delta g=0.01$].
Notice that the eGPE framework is expected to be valid for $\gamma \ll 1$~\cite{Collective1D} and indeed it provides a somewhat adequate description for the $N=50$ case, while it fails in the case of $N=20$ as discussed above. 
Therefore, the increased localization and hence peak density caused by the presence of the trap, shifts the validity region of the eGPE further towards macroscopic weakly interacting systems i.e. $\delta g\ll g$ and $ N \gg 1$.

\subsection{Asymmetric two-component settings}\label{AsymetricGS}

We address now droplet configurations arising in asymmetric bosonic mixtures. 
First for a homonuclear mixture, corresponding for instance to two hyperfine states of $^{39}$K, e.g. $\ket{1,-1}$ and $\ket{1,0}$ as in the experiments of Ref.~\cite{CabreraTarruellDropExp,CheineyTarruellDropExp,SemeghiniFattoriDropExp}; here $g_B=2g_A=0.1$ whilst $m_A=m_B \equiv m$ and $N_A=N_B=N=20$.
Tuning the interspecies coupling, only the ground state density of the more strongly interacting component B exhibits a transition from a Gaussian profile ($\delta g = 0.01$) to one featuring a FT signature ($\delta g = 0.06$) [Fig.~\ref{fig:GSImb}(a)].
Instead, the more weakly interacting component A features a significantly more localized Gaussian profile for all attractions.
This is attributed to the implicit violation of the particle number condition $N_A/N_B =\sqrt{g_B/g_A}$~\cite{Petrov2015,PetrovLowD}.
We aim to address this intriguing question in more detail in a future work. 
Lastly, since $g_A\neq g_B$, the SU(2) symmetry of the mixture is broken and thus the components are not equivalent~\cite{SemeghiniFattoriDropExp}.
As a result intercomponent spatial mixing is induced independently of $\delta g$ and becomes more prominent for larger $\delta g$. 

Similarly, the components are distinguishable and spatially mixed for heteronuclear (i.e. mass-imbalanced) settings [Fig.~\ref{fig:GSImb}(b)].
To support this argument we employ a mixture of $N=20$ in a species selective harmonic trap with $\omega_A=1.5\omega_B=0.015$, a mass ratio $m_B/m_A=2.1$ and intraspecies interactions $g_A=0.05$ and $g_B=0.08$ inspired by the experiment of Ref.~\cite{FortHeteroExp} where the isotopes $^{41}$K and $^{87}$Rb have been exploited.
Apparently, an intercomponent spatial mixing trend between the components occurs.
This is a result of the mass-imbalance which counteracts the interaction-imbalance as well as the weaker confinement of the heavier species.
The two latter contribute towards a larger width  of the heavier component density distribution.
We note that this effect will have a significant impact on the dipole dynamics of mass-imbalanced mixtures, as we shall argue below in Sec.~\ref{Dipole dynamics}.

\subsection{Two-body droplet configurations}\label{2b_ground}

\begin{center}
\begin{figure*}[ht]
\includegraphics[width=0.9\textwidth]{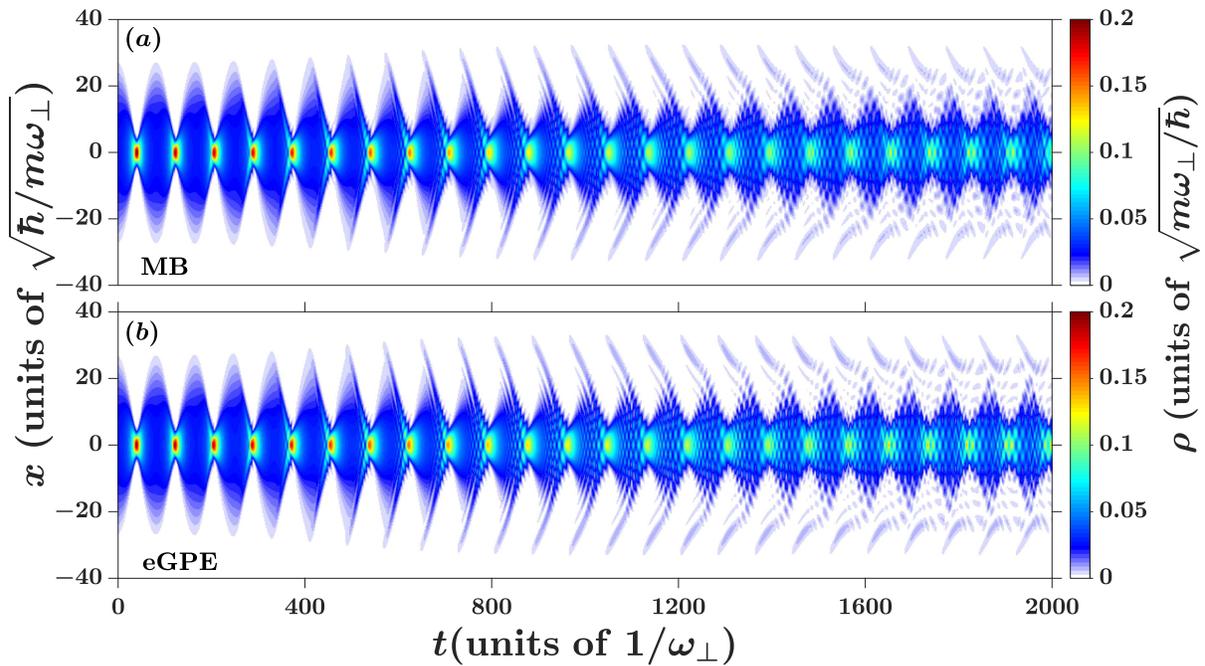}
\caption{Density evolution of a weakly attractive, symmetric mixture upon considering a sudden increase of the trap frequency, namely  $\omega^{\rm f}=4\omega^{\rm i}=0.04$ within (a) the ML-MCTDHX and (b) the perturbative eGPE approach.
The droplet undergoes a breathing motion, while higher-lying motional excitations arise in the droplet for long evolution times ($t>1000$).
Apparently, the eGPE prediction slightly overestimates the delocalization of the droplet density profile and the peak density of the localized excitations around the trap center at long evolution times.
The bosonic mixture contains $N_A=N_B=N=20$ atoms with $g_{A}=g_{B}=0.1$ and $\delta g=0.08$.}
\label{fig:BreathingBalancedDensity}
\end{figure*}
\end{center}

To further probe the superposition nature of the droplet MB state we examine the respective two-body reduced densities

\begin{equation}\label{2bodyDens}
\begin{split}
    \rho^{(2)}_{\sigma\sigma'}(x_1,x_2)= & \langle \Psi|\hat{\Psi}_\sigma ^\dagger (x_1) \hat{\Psi}_{\sigma'} ^\dagger (x_2)\\
    & \times \hat{\Psi}_\sigma (x_1) \hat{\Psi}_{\sigma'} (x_2) |\Psi \rangle,
\end{split}
\end{equation}
where $\hat{\Psi}_\sigma (x_i)$ denotes the bosonic field operator annihilating a $\sigma$-species particle at position $x_i$. $\rho^{(2)}_{\sigma\sigma'}(x_1,x_2)$ refers to the probability of simultaneously measuring one boson of species $\sigma$ at position $x_1$ and a boson of species $\sigma'$ located at $x_2$~\cite{Sakmann2008Coher,Naraschewski1999Coher}. 

Focusing on the case of a mass-imbalanced mixture, we observe that each species features a tendency towards an anti-correlated behavior, see Fig.~\ref{fig:2BDGS}(a), (b).
This is evident by the suppressed amplitude of the diagonal of the intraspecies two-body densities which implies a reduced probability of two $\sigma$-species atoms to reside at the same position.
However, it is more likely two atoms of the same component to be symmetrically placed close to $x=0$, as it can be seen from the two-body density humps appearing in $\rho^{(2)}_{\sigma\sigma}(0.5<x_1<6,-6<x_2<-0.5)$.
Such two-body patterns occur also in the the absence of a trap~\cite{Mistakidis2021}.
Naturally, the anti-correlation is more pronounced for the heavier component which is the more strongly interacting one.
In contrast, intercomponent two-body correlations take place among the species especially at the trap center where the FT profile forms [Fig.~\ref{fig:2BDGS}(c)].
The above discussed two-body correlation patterns are found to be robust for the different systems considered herein (not shown) and become enhanced for stronger repulsions or increasing particle numbers.
It should also be noted that we do not observe a sharp increase of interspecies correlations for increasing $\delta g$.
Such an enhancement would be associated with the generation of dimers predicted in free-space due to beyond-LHY correlations~\cite{ParisiMonteCarlo2019,Hui2020LowD}.
However, as we argued in Sec.~\ref{SymmetricGS} the presence of the harmonic confinement suppresses dimer formation.

\section{Collective excitations of Droplets}\label{sec:Collective excitations of Droplets} 

Having determined the ground state of harmonically confined droplets, let us now investigate the impact of correlations during their non-equilibrium dynamics.
The presence of the external harmonic trap enables us to seed dynamical scenarios that have not been addressed previously.
Specifically, in order to trigger the time-evolution of quantum droplets, a quench of either the frequency [Sec.~\ref{Homonuclear mixture Breathing dynamics},~\ref{Heteronuclear mixture Breathing dynamics}] or the position [Sec.~\ref{Dipole dynamics}] of the external trap is applied.
These protocols naturally excite the breathing and dipole motion of the quantum droplet respectively.
Moreover, by employing species selective quenches we are able to break the SU(2) symmetry of the symmetric mixture and consequently monitor the emergent droplet interspecies collisions.
Notice that performing these trap quenches, while keeping the interactions intact, provides the possibility to study the build-up of correlations both at the FT and the Gaussian-type phase independently.
Concluding, the dynamics of the above-mentioned low-lying collective modes is monitored in heteronuclear mixtures [Sec.~\ref{Heteronuclear mixture Breathing dynamics},~\ref{Dipole dynamics}], which are found to feature enhancement of correlations and intercomponent mixing.

\subsection{Homonuclear mixtures: Breathing dynamics}\label{Homonuclear mixture Breathing dynamics}

To excite the lowest-lying breathing mode of the droplet, building upon the homonuclear mixture, a quench of the trap frequency is performed from an initial $\omega^{\rm i}$ to a final value $\omega^{\rm f}$.
The bosonic setting consists of $N_A=N_B=N=20$ atoms with $g_A=g_B=g=0.1$ and different intercomponent attractions $g_{AB}<0$.
Since we aim to also reveal the interplay between the quench amplitude and the droplet excitation dynamics two different postquench frequencies are employed, namely  $\omega^{\rm f}=4\omega^{\rm i}=0.04$ and $\omega^{\rm f}=2\omega^{\rm i}=0.02$.

\begin{center}
\begin{figure}[ht]
\includegraphics[width=0.48\textwidth]{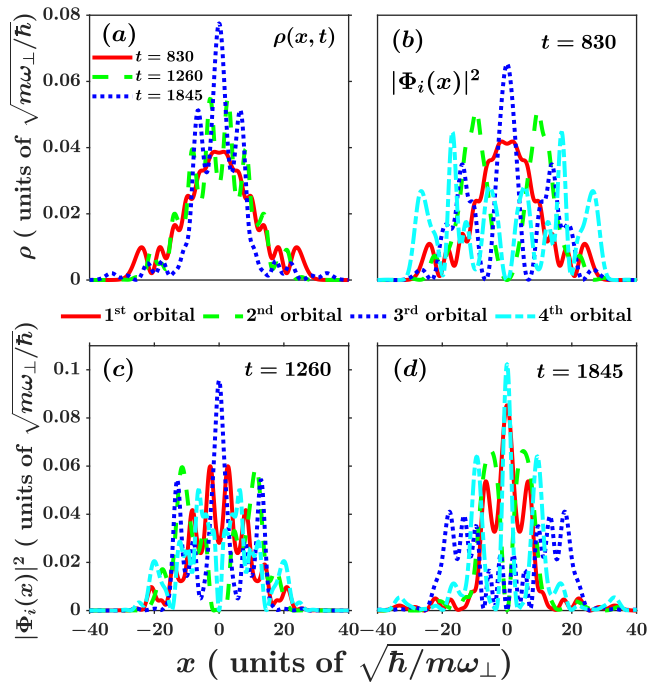}
\caption{Density profiles of (a) the symmetric homonuclear droplet and (b), (c), (d) the respective orbitals $\abs{\Phi_i(x)}^2$ with $i=1,\dots,4$ (see legend) at different evolution times. 
Namely, at (b) $t=830$, (c) $t=1260$ and (d) $t=1845$. 
Evidently, higher-lying orbitals support the spatial delocalization and excitation patterns of the density, while the lowest orbital has the dominant contribution to the density. 
The breathing dynamics of the symmetric mixture, characterized by $g_{A}=g_{B}=0.1$, $N_A=N_B=N=20$, is triggered by a frequency quench where $\omega^{\rm f}=4\omega^{\rm i}=0.04$.}
\label{fig:Snapshots}
\end{figure}
\end{center}

To track the dynamics of the ensuing breathing motion of the droplet cloud we initially invoke its one-body density.
The time-evolution of this observable for a weakly attractive mixture with $\delta g=0.08$ and a postquench frequency $\omega^{\rm f}=4\omega^{\rm i}=0.04$ is presented in Fig.~\ref{fig:BreathingBalancedDensity}(a), (b) within the MB and the eGPE approach respectively.
Overall, a periodic expansion and contraction of the droplet is observed within both methods for timescales $t<500$ followed by a progressive delocalization of the density profile.
Later on, for $t>800$, prominent spatial undulations arise in the droplet density especially around the trap center manifesting its excited nature, see also selective density snapshots in Fig.~\ref{fig:Snapshots}(a).
They initially appear as relatively small density humps around $t=830$ [solid red line in Fig.~\ref{fig:Snapshots}(a)] and eventually dominate the droplet profile rendering the FT background no longer visible, see e.g. the dashed green line in Fig.~\ref{fig:Snapshots}(a).
It is also worth mentioning that even during the contraction process there are delocalized density tails, see for instance $\abs{x} \approx 20$ of the dotted blue line at $t=1845$ in Fig.~\ref{fig:Snapshots}(a).
These motional excitations can be understood in terms of the MB wave function superposition and in particular stem from the non-negligible occupation of higher-lying natural orbitals being the eigenstates of the one-body reduced density matrix.
To support this argument the densities of the first four orbitals, $\abs{\Phi_i(x)}^2$ where $i=1,\dots,4$, are also provided for the same time instants in Fig.~\ref{fig:Snapshots}(b)-(d).
As expected, higher-order orbitals exhibit a hierarchy in terms of their nodal structure accompanied by an enhanced spatial delocalization.
Hence, the existence of higher-lying orbitals is responsible for both the delocalization and the excitation of the droplet.

Notably, the eGPE provides a somewhat accurate description of the observed MB dynamics, as it captures both the delocalization trend and the spatial undulations appearing in the time-evolved one-body density, see Fig.~\ref{fig:BreathingBalancedDensity}. However, it should be emphasized that the delocalized density tails are slightly more pronounced within the eGPE treatment indicating a tendency towards a slightly less stable droplet state 
as compared to the MB approach. 
A closer inspection of Fig~\ref{fig:BreathingBalancedDensity}(a), (b) for $t>1600$ and $|x|>20$ or of the corresponding variances (not shown) indicates an increased delocalization in the eGPE compared to the MB approach of the order of $\sim4\%$ of the initial dropltet width.
Also, the spatial undulations predicted by the eGPE are characterized by more prominent density peaks and thus a higher-degree of localization, see Fig~\ref{fig:BreathingBalancedDensity}(a), (b). 
These differences may be interpreted at the microscopic level in terms of the superposition of the orbitals shown in Fig.~\ref{fig:Snapshots}(b)-(d).
The fact that beyond-LHY correlations are not substantial during the evolution is partly attributed to the quench protocol related to the single-particle potential.
The above-described dynamical phenomena appear in experimentally accessible evolution times, e.g. for customarily used 1D trap frequencies $\omega_{{\rm x}}=2\pi \times 1.5$Hz and $\omega_{\perp}=2\pi \times 300$Hz~\cite{Ketterle2001LowDexp,katsimiga2020observation,bersano2018three} the excitations become evident after $t\approx 265$ms and the delocalization around $t\approx 530$ms.
While exact estimations of the lifetimes of 1D quantum droplets remain still elusive, experimental observations in higher dimensions typically report lifetimes of the order of $10$ms for homonuclear mixtures~\cite{SemeghiniFattoriDropExp} and significantly higher ones for heteronuclear systems, i.e. $\sim 120$ms~\cite{FortHeteroExp}.
Importantly, three-body losses which lead to droplet decay, are commonly reduced by one order of magnitude in 1D settings~\cite{Tolra20043BR} and they have been recently argued to become negligible in the droplet regime~\cite{AstrakharchikThermal,BourdelCrossover}, resulting in even longer droplet lifetimes.
These substantially improved lifetimes constitute a major benefit for studying 1D systems and especially heteronuclear droplets and suggest that the long-time dynamics presented herein, might be experimentally accessible.

The aforementioned delocalization behavior suggests a dissociation tendency of the droplet, resembling the well-known self evaporation mechanism observed in three-dimensional experiments~\cite{Petrov2015,CabreraTarruellDropExp,CheineyTarruellDropExp}. 
This dynamical response has not yet been reported in 1D free space~\cite{Collective1D,Mistakidis2021} and it is thus inherently related to the presence of the trap.
On the other hand, the robustness of the excitation patterns in the central droplet region for long evolution times supports the assumption that the FT configuration can maintain nonlinear structures, e.g. solitons, a study that is interesting to be pursued in the future.

\begin{center}
\begin{figure}[ht]
\includegraphics[width=0.48\textwidth]{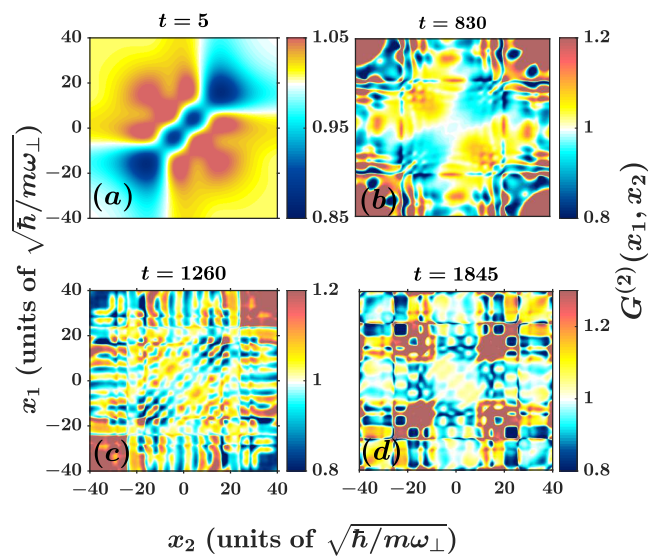}
\caption{
(a)-(d) Profiles of the two-body coherence $G^{2}(x_1,x_2)$ in the breathing dynamics of the symmetric mixture with $g_{A}=g_{B}=0.1$, $N_A=N_B=N=20$ at distinct time-instants (see legend) after a quench to $\omega^{\rm f}=4\omega^{\rm i}=0.04$.
Suppression of two-body correlations in the bulk ($G^{2}(x_1,x_2=x_1)$) occurs for long evolution times while a correlated behavior is evident for bosons located at different edges of the droplet ($G^{2}(x_1,x_2=-x_1)$).
}
\label{fig:CoherenceSnaps}
\end{figure}
\end{center}

\subsection{Correlation and entanglement dynamics}

A feature that is inherently related with the droplet formation is their correlated character~\cite{ParisiMonteCarlo2019,Mistakidis2021,MoreraAstrakharchikLattice}, while the fate of dynamical droplet correlations remains largely unexplored. 
Specifically, in our setting we aim to reveal the correlation patterns that correspond to the above-mentioned droplet excitation processes.
Below, we infer the build-up of intraspecies two-body correlations during the breathing evolution of the droplet in a spatially resolved manner by investigating the so-called two-body coherence function~\cite{Sakmann2008Coher,Naraschewski1999Coher}.
\begin{align}
\label{coherences}
G_{\rm \sigma\sigma}^{(2)}(x_1,x_2,t)=&\frac{\rho_{\rm \sigma\sigma}^{(2)}(x_1,x_2,t)}{\rho_\sigma(x_1,t)\rho_\sigma(x_2,t)}. 
\end{align}
The two-body reduced density matrix $\rho_{\rm \sigma\sigma}^{(2)}(x_1,x_2,t)$ is defined in Eq.~(\ref{2bodyDens}).
Naturally, in the symmetric mixture case $G_{AA}^{(2)}(x_1,x_2)=G_{BB}^{(2)}(x_1,x_2)=G^{(2)}(x_1,x_2)$.
Apparently, a two-body correlated (anti-correlated) behavior occurs for $G^{(2)}(x_1,x_2,t)>1$ ($G^{(2)}(x_1,x_2,t)<1$), whilst the case of $G^{(2)}(x_1,x_2,t)=1$ is said to be two-body uncorrelated. 

Snapshots of $G^{(2)}(x_1,x_2,t)$ are illustrated in Fig.~\ref{fig:CoherenceSnaps}(a)-(d).
At the initial stages of the dynamics [Fig.~\ref{fig:CoherenceSnaps}(a)] the droplet maintains the two-body anti-correlated behavior of its ground state [see Sec.~\ref{sec:Ground State Droplets}] for two bosons at the same location [i.e. $G^{(2)}(x_1,x_2=x_1,t)<1$] while two atoms placed symmetrically with respect to the FT exhibit a correlated tendency, namely $G^{(2)}(x_1,x_2=-x_1,t)>1$.
For longer evolution times where the density delocalization is observed [Fig.~\ref{fig:BreathingBalancedDensity}(a)], a suppression of two-body correlations takes place at the central bulk region since $G^{(2)}(x_1,x_2=x_1,t)\approx1$, see Fig.~\ref{fig:CoherenceSnaps}(b)-(d).
However, the delocalized density tails (see e.g. $\rho (|x|>20,t>1000)$ in Fig.~\ref{fig:BreathingBalancedDensity}(a)) develop a noticeable two-body correlated behavior among each other as it can be deduced, for instance, from the anti-diagonal of $G^{(2)}(x_1,x_2=-x_1,t)>1$ depicted in Fig.~\ref{fig:CoherenceSnaps}(d) (e.g. at $x_{1}=-x_{2}\approx 20$).

\begin{center}
\begin{figure}[ht]
\includegraphics[width=0.48\textwidth]{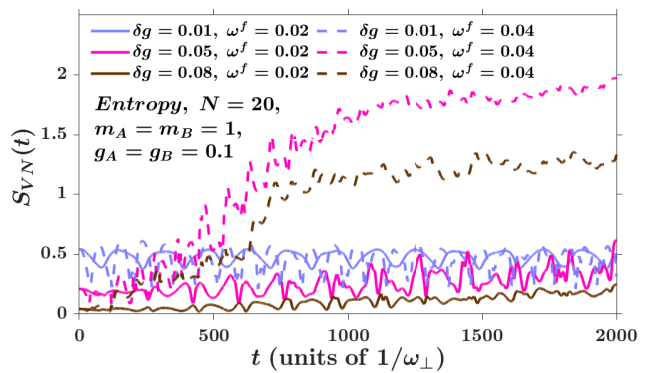}
\caption{Time-evolution of the Von-Neumann entropy in the course of the droplet breathing motion induced by various quenches of the trap frequency at distinct interspecies couplings (see legend).
The symmetric bosonic mixture with $N_A=N_B=N=20$ experiences $g_{A}=g_{B}=0.1$ (see legend).
The finite entropy evinces the presence of intercomponent entanglement, while its sharp increase for sufficiently large $\delta g$ and $\omega^{\rm f}$ signals the prominent excitation dynamics of the droplet.}
\label{fig:Entropy}
\end{figure}
\end{center}

Regarding the impact of intercomponent correlations (entanglement) on the breathing dynamics of quantum droplets we analyze the corresponding Von-Neumann entropy~\cite{Horodecki2009}
\begin{equation}
\label{VNentropy}
S_{\rm VN}(t)=-\sum_{k=1}^{D} \lambda_k(t)\ln[\lambda_k(t)].
\end{equation}
In this expression, $\lambda_k(t)$ refer to the Schmidt coefficients of the MB wave function ansatz~\eqref{SpeciesFunctionLevel} which are essentially the eigenvalues of the species reduced density matrix~\cite{horodecki2009quantum}.
This entropic measure captures the degree of intercomponent entanglement, namely a non-entangled state corresponds to $\lambda_1(t)=1$ and $\lambda_{k\neq1}(t)=0$ leading to $S_{\rm VN}(t)=0$.
We note that the entanglement is crucial for quantum droplets since the widely used MGP approach primarily takes into account the effects stemming from intraspecies quantum fluctuations.
Instead the interspecies coupling processes should be generally introduced perturbatively in terms of $\delta g/g$ as it has been argued in Refs.~\cite{Petrov2015,ParisiMonteCarlo2019,MithunMI,Hui20203D,Hui2020LowD}.

The harmonically confined droplets discussed herein, offer a promising setting for exploring beyond-LHY effects in the course of the time-evolution.
Indeed, the dynamics is induced by applying a quench on the trap frequency while keeping fixed the interaction parameters ($g_A$, $g_B$ and $g_{AB}$).
This allows us to study the built-up of intercomponent correlations both in the Gaussian and the FT regimes which occur at strong and weak attractions respectively.

For postquench amplitudes $\omega^{\rm f}=2\omega^{\rm i}=0.02$ we observe that $S_{\rm VN}(t)$ fluctuates around a mean value determined by the strength of $g_{AB}$, see Fig.~\ref{fig:Entropy}.
Generically, the degree of entanglement is larger for stronger attractions.
This behavior is related to the fact that the breathing motion of the droplet is nearly stable, exhibiting regular periodic expansion and contraction for this quench amplitude independently of $\delta g$ and excitations do not appear, e.g. on the one-body density level (not shown).
Turning to the case of larger quench amplitudes, i.e. $\omega^{\rm f}=4\omega^{\rm i}=0.04$, the entanglement dynamics of the droplet is more involved.
Namely, for Gaussian-shaped droplets ($\delta g=0.01$) the entropy oscillates around $S_{\rm VN}(t)\approx0.5$.
In this regime again the droplet can not sustain excitations due to its narrow width.
Remarkably, within the FT regime ($\delta g=0.08$) $S_{\rm VN}(t)$ features initially a moderately increasing tendency and thereafter exhibits a strong increase when localized excitations arise in the droplet core as presented in Fig.~\ref{fig:BreathingBalancedDensity}(a) e.g. for $|x|<20$ and $t>700$.
Subsequently, the delocalization of the droplet edges as seen in Fig.~\ref{fig:BreathingBalancedDensity}(a) e.g. for $|x|>20$ and $t>1200$ is related to a saturation trend of $S_{\rm VN}(t)$ e.g. towards $S_{\rm VN}(t) \approx 1.25$ for $\delta g=0.08$.
Therefore, the appearance of higher-lying motional excitations at the droplet center is accompanied by a noticeable increase of interspecies correlations which then approach a plateau behavior as long as the droplet delocalizes.

\begin{center}
\begin{figure}[ht]
\includegraphics[width=0.48\textwidth]{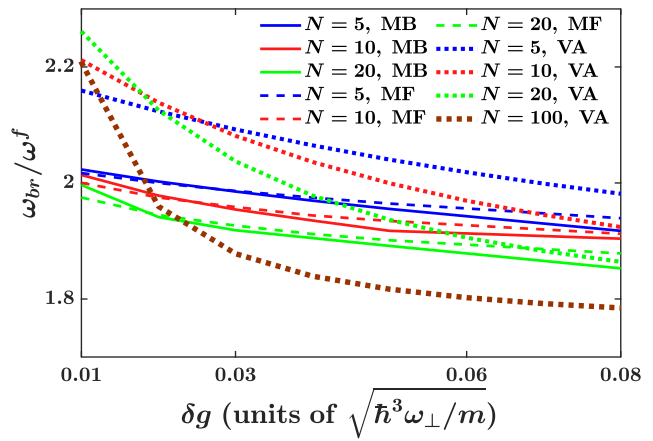}
\caption{Breathing mode frequency ($\omega_{\rm br}/\omega^{\rm f}$) of the symmetric mixture with $g_{A}=g_{B}=0.1$ in units of the postquench trap frequency for varying interspecies attraction and atom number (see legend).
The predictions of $\omega_{\rm br}/\omega^{\rm f}$ are given within the MB approach (solid lines), the MF (dashed lines) and variational (dotted lines) approximation.
A monotonic decreasing behavior of $\omega_{\rm br}/\omega^{\rm f}$ is observed for increasing either $\delta g$ or $N_A=N_B=N$.
The breathing frequency is (slightly) smaller in the MB approach as compared to the MF and the variational approximation.}
\label{fig:BreathingFrequency}
\end{figure}
\end{center} 

\subsection{Droplet breathing mode frequency}

Next, we employ the position variance of the bosonic cloud
\begin{equation}
\braket{X_\sigma^2(t)}= \braket{\Psi(t)|\hat{x}_\sigma^2|\Psi(t)}- \braket{\Psi(t)|\hat{x}_\sigma|\Psi(t)}^2,
\label{Variance}
\end{equation}
\noindent
where $\hat{x}_\sigma$ denotes the $\sigma$-species position operator and $\braket{\Psi(t)|\hat{x}_\sigma|\Psi(t)}=0$ due to the preserved ``parity" symmetry.
This experimentally accessible measure via time-of-flight imaging~\cite{ronzheimer2013expansion,fukuhara2013quantum}, captures the breathing motion of the cloud and its spectrum contains the respective breathing mode frequency, $\omega_{\rm br}$.
The latter is provided in Fig.~\ref{fig:BreathingFrequency}, upon considering a quench with $\omega^{\rm f}=4\omega^{\rm i}=0.04$, as a function of $\delta g$ for various particle numbers composing the droplet and within different approaches.
These approaches refer to the full MB treatment including the correlations of the mixture, the MF approximation where correlations are neglected and a variational approximation (VA) within the LHY theory.
The latter was also exploited in Ref.~\cite{AstrakharchikMalomed1DDynamics} to estimate $\omega_{\rm br}$ of 1D droplets in the absence of an external trap.
It is based on a time-dependent Gaussian ansatz [see details in Appendix~\ref{app:var}] providing an approximate analytical solution of the eGPE~\eqref{MGP} by utilizing a Gaussian wave function with variationally optimal width.
This explicit time-dependence of this ansatz can provide semi-analytical insights into the dynamical droplet properties, e.g. the breathing frequency (see the discussion below). 

Focusing on the outcome of the MB and the MF approaches it is found that overall $\omega_{\rm br}$ is close to the prediction in the ideal gas limit, i.e. $\omega_{\rm br}\approx 2\omega^{\rm f}$, for strong interspecies attraction (or otherwise small $\delta g$), while it shows a weakly decreasing tendency towards the decoupled scenario corresponding to increasing $\delta g$.
This behavior of $\omega_{\rm br}$ in terms of $\delta g$ holds also independently of $N$, but $\omega_{\rm br}$ also decreases for a larger atom number since for an increasing $N$ the effective MF interactions ($\propto \delta gN$) are enhanced. Thus, deviations from the non-interacting limit become more prominent. 
This reduced trend of $\omega_{{\rm br}}$ for smaller $g_{AB}$ has also been reported in free space using a Quantum Monte-Carlo method~\cite{ParisiGiorginiMonteCarlo}. 
Moreover, by closely inspecting $\omega_{\rm br}$ it can be deduced that within the MF approximation it is smaller for stronger attractions and larger in the reverse case as compared to the MB result.
Interestingly, the crossing point with respect to $\delta g$ between the two predictions shifts towards the MF balance point ($\delta g=0$) for larger $N$.
This is due to the fact that the contribution of the LHY term to the breathing frequency is positive for $\delta g  \approx 0$ and negative when $\delta g \neq 0$, as we shall explicate below using the VA method, see in particular Eq.~\eqref{breathing_VA1}.
Additionally, the deviations of the MB and the MF results become more pronounced in the few-body limit, e.g. $N=5$, due to the involvement of higher-order correlations.

Turning to the VA method we observe that it fails to capture $\omega_{\rm br}$, especially for few-body systems (e.g. $N=5$).
Still, it improves significantly for increasing $N$ and it approaches more closely the MB prediction as compared to the MF approximation at the FT regime, e.g. around $\delta g=0.08$ for $N=20$.
We remark that this agreement between the VA and the MB cases solely occurs for the breathing frequency, whilst the density profile is not appropriately captured by the VA (not shown).
However, the VA approximation completely overestimates $\omega_{\rm br}$ for stronger attractions such as $\delta g=0.01$.
In particular, for large $N \gg 1$ and stronger attractions such that $\delta g\approx0$ (LHY fluid) it can be proven, following the minimization of the underlying effective potential [see Eq.~\eqref{EffPotVA}], that the breathing frequency scales as

\begin{equation}
\omega_{\rm br}^{\rm VA}\propto N^{2/3}.
\end{equation}
Hence, within the VA a diverging breathing frequency at the limit of the LHY fluid for increasing particle numbers is encountered.

On the other hand, for finite values of $\delta g$ and $\omega$ the scaling of the optimal width of the Gaussian wave function is dominated by the average MF repulsion, with the LHY term providing only a higher-order correction leading to a stronger localization, namely $W\approx (\frac{\delta g N}{m\omega^2\sqrt{2\pi}})^{1/3} - \frac{8^{1/4}}{3\pi \hslash \omega}\sqrt{(g^3/\delta g)}$.
In this case within VA the breathing frequency reads
\begin{equation}
\omega_{{\rm br}}^{{\rm VA}}\approx\sqrt{3}\omega-\mathcal{O}\big(g^{3/4}(\delta gN)^{-1/6}\big),\label{breathing_VA1} \end{equation}
where the correction term originates from the LHY contribution. 
It is also worth mentioning that, in the thermodynamic limit $N \to \infty$, where the above-discussed results become exact, the VA prediction reduces to the one of the usual Gross-Pitaevskii equation~\cite{Salasnich2000}.
Indeed, for $N \to \infty$ we obtain a breathing frequency $\omega_{\rm br}^{\rm VA}\approx\sqrt{3}\omega$, which is nearly reached e.g. for $N=100$ as shown by the dotted brown line in Fig.~\ref{fig:BreathingFrequency}.

\begin{center}
\begin{figure}[ht]
\includegraphics[width=0.48\textwidth]{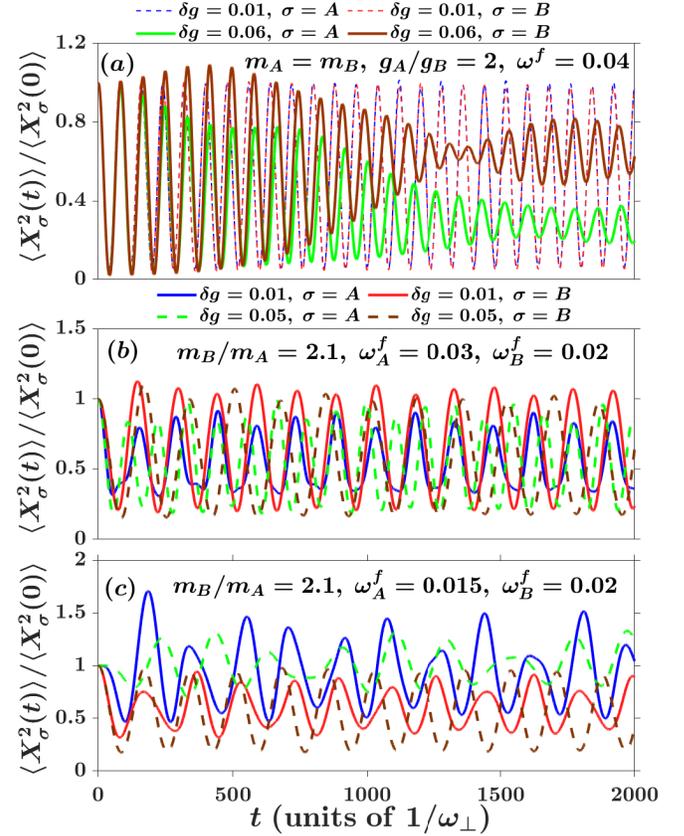}
\caption{
Time-evolution of the position variance of droplets within the MB method. 
(a) The case of a trap quenched ($\omega^{\rm f}=4\omega^{\rm i}=0.04$) homonuclear mixture, characterized by $g_{B}=2g_{A}=0.1$, $N=20$, for varying interspecies attractions (see legend) is depicted. 
The collapse and revival pattern observed in the weakly attractive mixture constitutes an imprint of the droplet excitations during its breathing motion. (b), (c) Same as (a) but for a heteronuclear mixture with  $m_B=2.1m_A$, $g_B=1.6g_A=0.08$, and $N=20$, following (b) a sudden increase of the trap frequency according to $\omega^f=2\omega^i$ and (c) a species selective quench where $\omega_{\rm B}^{\rm f}=2\omega_{\rm B}^{\rm i}=0.02$ and  $\omega_{\rm A}^{\rm f}=\omega_{\rm A}^{\rm i}=0.015$. 
The dynamics change from being out-of-phase to phase locked among the two components for increasing attraction.
}
\label{fig:BreathingVariance}
\end{figure}
\end{center}

\subsection{Dynamics of interaction-imbalanced mixtures}

\begin{center}
\begin{figure*}[ht]
\includegraphics[width=0.9\textwidth]{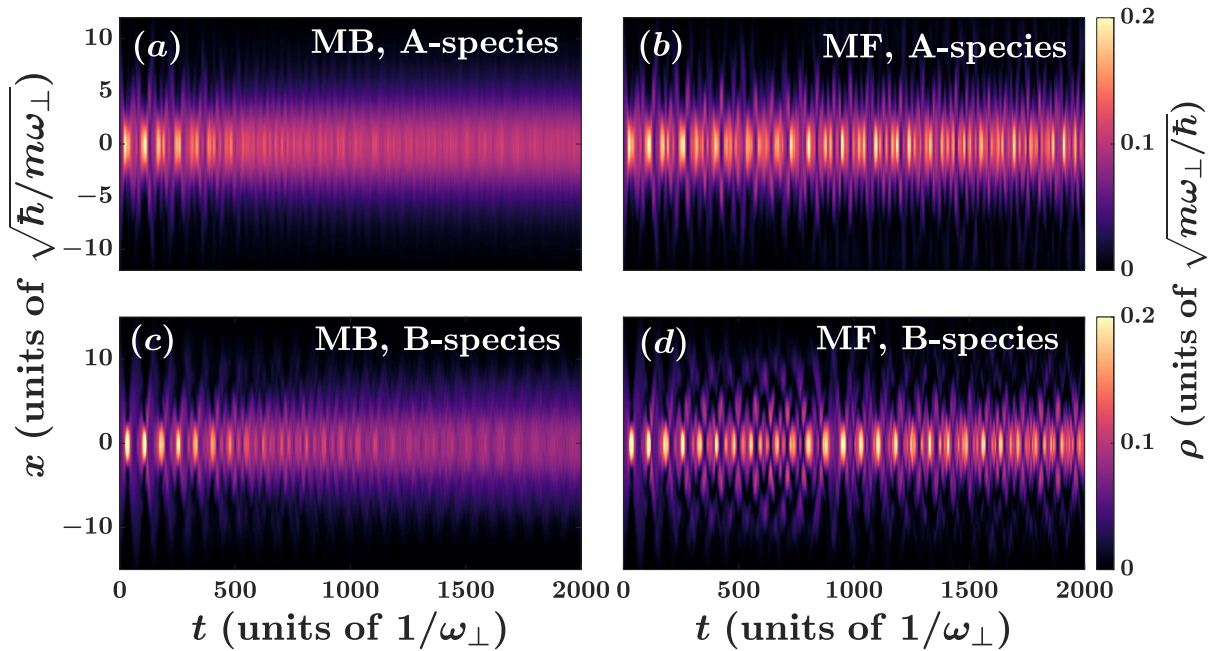}
\caption{Density evolution of the $\sigma$-component (see legends) of a heteronuclear mass-imbalanced mixture with $m_B=2.1m_A$, $g_B=1.6g_A=0.08$, $N_A=N_B=N=20$ and strong intercomponent attraction $\delta g=0.01$. 
The dynamics induced by quenching the trap frequency, $\omega_\sigma^{\rm f}=4\omega_\sigma^{\rm i}$, is monitored within (a), (c) the MB and (b), (d) MF approaches. Evidently, the build-up of correlations leads to a dephasing of the breathing mode amplitude at long evolution times.}
\label{fig:heteroDensities}
\end{figure*}
\end{center}

In an attempt to generalize the persistence of the droplet excitation processes in the course of its breathing motion we also investigate interaction-imbalanced homonuclear mixtures with $g_{B}=2g_{A}=0.1$, $g_{\rm AB}<0$, $m_A=m_B\equiv m$ and $N_A=N_B=N=20$, for its ground state characteristics.
Generically imbalanced mixtures are, among others, particularly prone to experience higher-order correlation phenomena, due to the different degrees of miscibility that may emerge.
These systems are far less explored and they are described by a set of coupled eGP equations as it has been reported e.g. in Refs.~\cite{MithunMI,Mistakidis2021} instead of the simple reduced single-component eGPE~\eqref{MGP}.
In order to seed the breathing motion of the mixture we perform a quench of the trap frequency towards $\omega^{\rm f}=4\omega^{\rm i}=0.04$ and track the dynamics through the $\sigma$-species variance $\braket{X_\sigma^2(t)}$ [Eq.~\eqref{Variance}] shown in Fig.~\ref{fig:BreathingVariance}(a).
For stronger attractions ($\delta g=0.01$), both $\braket{X_A^2(t)}$ and $\braket{X_B^2(t)}$ undergo almost constant amplitude oscillations being in-phase among each other, see in particular the dashed lines in Fig.~\ref{fig:BreathingVariance}(a).
The minor deviations between the oscillation amplitude of $\braket{X_A^2(t)}$ and $\braket{X_B^2(t)}$ in the long time dynamics ($t>1300$) evinces the suppressed degree of intercomponent mixing in this interaction regime which is caused by the strong attractive coupling.
Notably, the persistence of the amplitude of the position variances implies that excitations do not form in this two-component droplet scenario.

However, for weaker interspecies attractions ($\delta g=0.06$) the ensuing motion of the two components is drastically altered.
Specifically, the components expand and contract in a periodic manner initially ($0<t<150$) with almost the same amplitude.
The latter progressively differentiates between $\braket{X_A^2(t)}$ and $\braket{X_B^2(t)}$, while afterwards a pronounced damping (i.e. reduction of the oscillation amplitude) is evident followed by a revival pattern at least for the more strongly repulsively interacting $B$ component which exhibits a FT profile, see the solid lines in Fig.~\ref{fig:BreathingVariance}(a).
This distinct behavior of $\braket{X_A^2(t)}$ and $\braket{X_B^2(t)}$ along with their damping is an imprint of the excitations building upon each component in the course of its breathing motion. 
In particular, the density of the $B$ component experiences similar structural deformations in this case ($\delta g=0.06$) with the interaction balanced mixture depicted in Fig.~\ref{fig:BreathingBalancedDensity}(a).
Instead, the $A$ component having a Gaussian-shaped ground state density profile features a significantly lower degree of spatial delocalization.

\subsection{Heteronuclear mixtures: Breathing dynamics}\label{Heteronuclear mixture Breathing dynamics} 

Next, we examine the main features of the breathing dynamics of mass-imbalanced heteronuclear bosonic mixtures, e.g. consisting of $^{41}$K and $^{87}$Rb isotopes that have been experimentally realized~\cite{FortHeteroExp}.
In this sense, we consider a mass-imbalanced mixture $m_B=2.1m_A$, with fixed intraspecies repulsions $g_B=1.6g_A=0.08$, particle numbers $N_A=N_B=N=20$, and varying interspecies attraction $g_{AB}<0$, see also Sec.~\ref{AsymetricGS} for the ground state properties of this system.
In order to study the response of this setting, a sudden change of the original trap frequencies $\omega_{\rm A}^{\rm i}=1.5\omega_{\rm B}^{\rm i}=0.015$ is applied and the breathing motion of each cloud is initiated.
Below, our investigations are restricted within the ML-MCTDHX framework and the common MF treatment.
We remark that the respective eGPE for mass-imbalanced mixtures in three-dimensions was used in Refs.~\cite{FortHeteroExp,AncilottoLocalDensity} but with the LHY contribution possessing a somewhat complicated form.
To the best of our knowledge, the exact form of the one-dimensional, mass-imbalanced eGPE has not yet been reported.

We first excite the breathing mode of the system, by suddenly doubling the harmonic trap frequencies for each species, i.e. $\omega_\sigma^{\rm f}=2\omega_\sigma^{\rm i}$.
Monitoring the time-evolution of the $\sigma$-species position variance [Eq.~\eqref{Variance}] it becomes apparent that each component expands and contracts [Fig.~\ref{fig:BreathingVariance}(b)] but importantly the response of $\braket{X_{\sigma}^2(t)}$ depends strongly on $g_{AB}$.
Specifically, for weak attractions e.g. $\delta g=0.05$ the widths $\braket{X_{\sigma}^2(t)}$ feature a phase difference and distinct frequencies ($\omega_{\rm br}^{\rm A}\approx 1.3\omega_{\rm br}^{\rm B}$) during the dynamics.
Notice that this breathing frequency ratio is slightly smaller than that of the respective traps, namely $\omega^{\rm A}\approx 1.5\omega^{\rm B}$.
This reduced frequency ratio evinces that the heavier component $B$ possesses a higher breathing frequency compared to the mass-balanced case, see also the stable breathing mode of the strongly attractive interaction-imbalanced mixture depicted in Fig.~\ref{fig:BreathingVariance}(a).
Furthermore, the oscillation amplitude of the heavier species is nearly constant, in contrast to the one of the lighter component which becomes significantly suppressed when it oscillates out-of-phase with the heavier species.
On the other hand, for stronger attractions $\delta g=0.01$, the heavier component "effectively" traps the lighter one and they oscillate in-phase with $\omega_{\rm br}\approx2\omega^{\rm f}_{B}$.

To further exploit the asymmetries of our system, we apply a species selective quench on the harmonic trap of the heavier species, i.e. $\omega_{\rm B}^{\rm f}=2\omega_{\rm B}^{\rm i}$ whilst $\omega_{\rm A}^{\rm f}=\omega_{\rm A}^{\rm i}$ [Fig.~\ref{fig:BreathingVariance}(c)].
A close inspection of $\braket{X_{\sigma}^2(t)}$ for stronger attractions such as $\delta g = 0.01$, reveals that the heavy component imparts at the initial stages of the dynamics ($t<100$) part of its energy to the lighter species which is subsequently set to breathing motion.
The two components oscillate almost in-phase but with distinct amplitudes among each other and also varying in the course of the evolution.
We remark that this response is in sharp contrast to the high degree of delocalization that species selective quenches excite on mass-balanced mixtures (not shown).
Turning to weaker attractions ($\delta g = 0.05$), we observe a somewhat delayed energy transfer towards the lighter species.
The latter consequently undergoes breathing dynamics characterized by two dominant breathing frequencies while exhibiting a phase difference with the heavier species as shown by the dashed lines in Fig.~\ref{fig:BreathingVariance}(c).
$\braket{X_{A}^2(t)}$ exhibits distorted oscillations leading to a slightly larger frequency $\omega_A^{\rm br}$ at long evolution times ($t>1500$), than at the initial stages of the dynamics ($t<500$).
As a consequence, at long evolution times ($t>1500$) there is a breathing frequency ratio $\omega_B^{\rm br}/\omega_A^{\rm br}\approx 1.23$ between the two components. 
This ratio is slightly smaller than that of their respective postquench traps, i.e. $\omega_B^{\rm f}/\omega_A^{\rm f}\approx 1.33$, similar to the breathing evolution of the mass-imbalanced setting discussed above [Fig.~\ref{fig:BreathingVariance}(b)]. 
It is worth to be mentioned that small deviations between the MB and the MF predictions are again evident in the respective one-body density evolution, especially for smaller particle numbers, e.g. $N=5$, which manifests that few-body effects come into play (not shown).

\begin{center}
\begin{figure}[ht]
\includegraphics[width=0.48\textwidth]{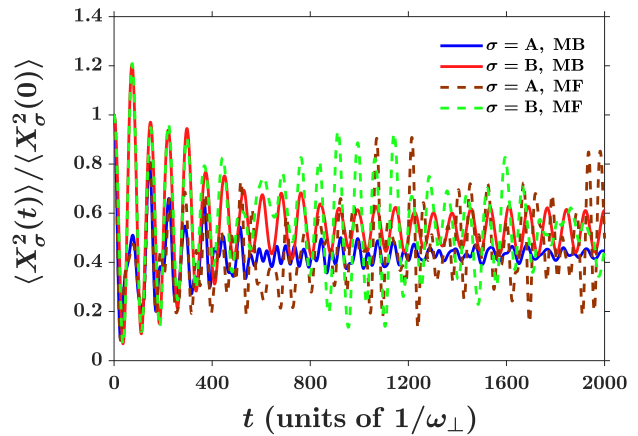}
\caption{Time-evolution of the $\sigma$-species position variance of a mass-imbalanced heteronuclear mixture characterized by $m_B=2.1m_A$, $g_B=1.6g_A=0.08$, $N_A=N_B=N=20$ and strong intercomponent attraction $\delta g=0.01$.
The dynamics induced by a quench where $\omega_\sigma^{\rm f}=4\omega_\sigma^{\rm i}$ is monitored within the MB (solid lines) and MF (dashed lines) approach.
A correlation induced dephasing behavior is observed in the MB case in sharp contrast to the MF dynamics where an irregular breathing motion persists.}
\label{fig:heteroVariance}
\end{figure}
\end{center}

Performing a more intense quench characterized by $\omega_\sigma^{\rm f}=4\omega_\sigma^{\rm i}$, we observe a prominent beyond MF effect for the strongly attractive ($\delta g=0.01$) mass-imbalanced droplet configurations [Fig.~\ref{fig:heteroDensities}].
At short timescales ($0<t<400$) they exhibit a breathing motion [Fig.~\ref{fig:heteroDensities}(a), (c)] whose amplitude afterwards decays [Fig.~\ref{fig:heteroVariance}] as a result of a dephasing mechanism due to the build-up of both intra- and intercomponent correlations, with the former being enhanced for the heavier species. 
This dephasing is established faster in the lighter component [Fig.~\ref{fig:heteroVariance}].
Evidently, this response is not captured by the MF approximation, where each component performs a breathing motion of non-negligible amplitude [Fig.~\ref{fig:heteroVariance}] throughout the time-evolution accompanied by density delocalization during the expansion of the clouds, see Fig.~\ref{fig:heteroDensities}(b), (d).
We also note in passing that motional excitations do not emerge in the course of the evolution which suggests that the observed response corresponds to a collective mechanism of the mixture.

\begin{center}
\begin{figure*}[ht]
\includegraphics[width=0.9\textwidth]{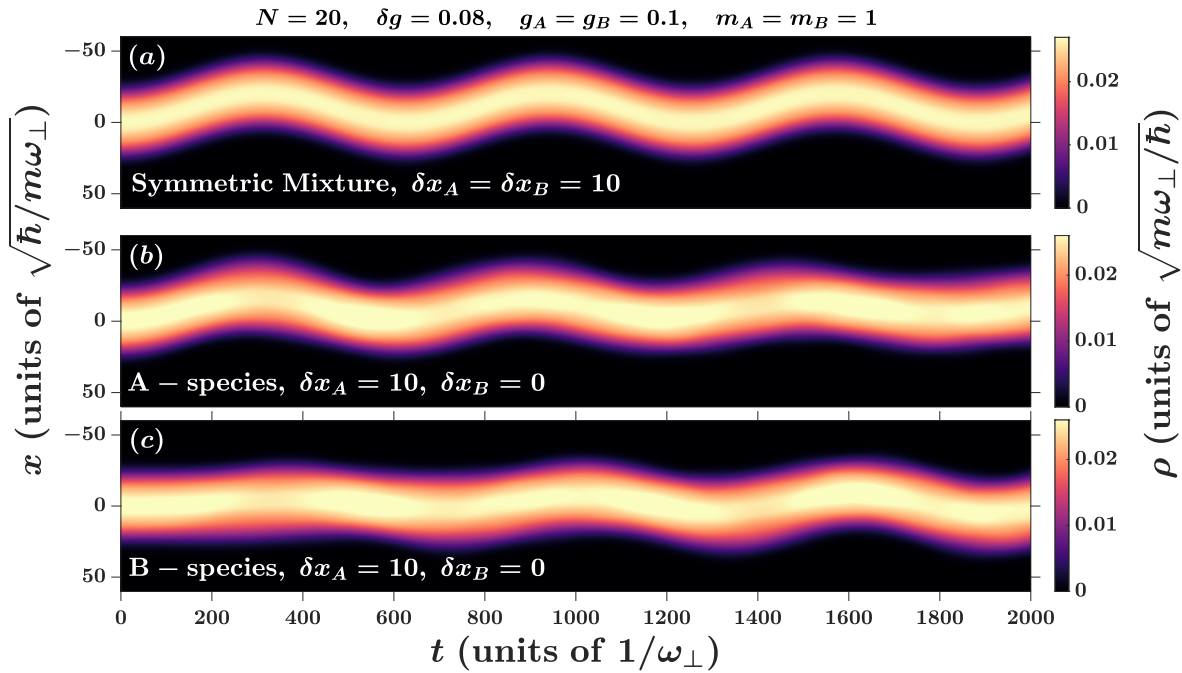}
\caption{One-body density evolution of a droplet building upon a weakly attractive symmetric mixture with $N_A=N_B=N=20$, $g_{A}=g_{B}=0.1$ and $\delta g=0.08$ after an abrupt displacement of the trap center by $\delta x_{\sigma}$. 
(a) The displacement is performed in both components, where $\delta x_A=\delta x_B=10$, resulting in a stable dipole motion having a frequency equal to the one of the trap. 
(b), (c) A species selective quench with $\delta x_A=10$ and $\delta x_B=0$ leads to an energy transfer from species $A$ to $B$ and consequent out-of-phase dipole oscillations of the individual components. 
All results were obtained within the MB approach.}
\label{fig:disquenchDensities}
\end{figure*}
\end{center}

The dephasing effect is attributed to the competition between the tendency of the two components to oscillate in-phase in the strongly attractive case 
and the pronounced difference of their postquench confinement frequency.
In this case, the heavier species $B$ cannot instantly trap the rapidly oscillating lighter species $A$, as in the case of the less intense quench presented in Fig~\ref{fig:BreathingVariance}(b).
Instead, significant intercomponent collisions occur initially ($t<500$) rendering the breathing modes of both components highly distorted, as can also be seen from the position variance of the mixture [Fig.~\ref{fig:heteroVariance}].
Subsequently, the heavier species $B$ imposes its oscillation frequency on the lighter one, while the respective amplitudes of both species reduce significantly.
However, species $A$ maintains also a significant admixture of its initial frequency along with the one of species $B$ resulting in a prominent dephasing behavior of $\braket{X_{A}^2(t)}$, see e.g. Fig.~\ref{fig:heteroVariance} in the interval $t=[500-900]$.
The above description also applies in the MF case [Fig.~\ref{fig:heteroDensities}(c), (d)], see the agreement with the MB results in the variance for $t<200$ in Fig.~\ref{fig:heteroVariance}.
However, in the absence of correlations, the heavier component $B$ is not able to impose its oscillation frequency to the lighter one [Fig.~\ref{fig:heteroVariance}].
Rather, intercomponent collisions dominate throughout the time-evolution, leading to irregular breathing dynamics and a high degree of spatial delocalization, see for instance $t\approx500$ and $t\approx1600$ in Fig.~\ref{fig:heteroDensities}(c), (d).
This discrepancy, among the MF and MB predictions, could be interpreted as a manifestation of the self-bound nature of quantum droplets.
Indeed, we overall observe an enhanced stability in the dynamics of heteronuclear mixtures, which is in accordance with the recent experimental observations in three-dimensions~\cite{FortHeteroExp}.

\section{Dipole droplet dynamics}\label{Dipole dynamics}

\begin{center}
\begin{figure*}[ht]
\includegraphics[width=0.9\textwidth]{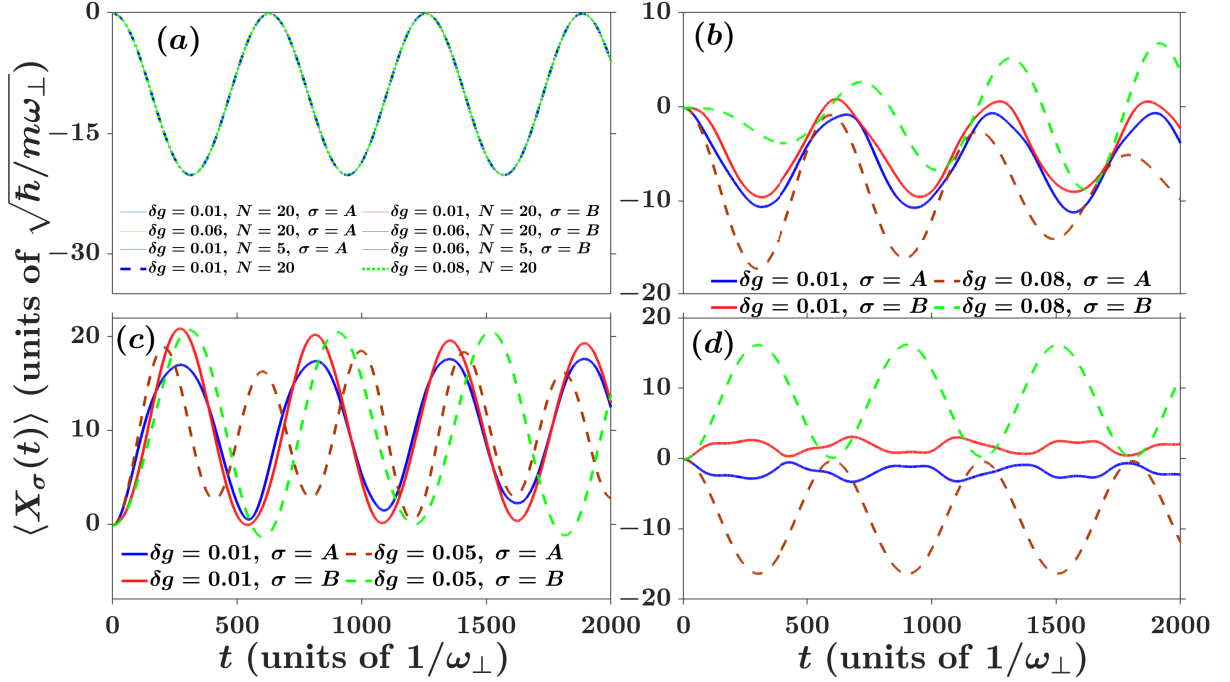}
\caption{Dynamics of the spatially averaged position of the $\sigma$-species center-of-mass $\braket{X_{\sigma}(t)}$ following a sudden displacement of the trap center by an amount $\delta x_{\sigma}$ within the MB approach.
(a) The case of homonuclear interaction-imbalanced $g_B=2g_A=0.1$ (solid lines) and balanced with $g_B=g_A=0.1$ (dashed lines) bosonic mixtures subjected to quenches with $\delta x_A=\delta x_B=10$ for different interactions and atom numbers (see legends) is depicted.
The emergent dipole motion is independent of the system parameters. 
Species selective quenches with (b) $\delta x_A=10$, $\delta x_B=0$ and (d) $\delta x_A= - \delta x_B=10$ in a symmetric mixture characterized by $g_{A}=g_{B}=0.1$, $N_A=N_B=N=20$. 
(c) Dipole motion induced by $\delta x_A=\delta x_B=-10$ in a mass-imbalanced heteronuclear mixture where $m_B=2.1m_A$, $g_B=1.6g_A=0.08$ and $N_A=N_B=N=20$.
Apparently, following either species selective quenches or using mass-imbalanced mixtures gives rise to a more complex dipole dynamics, with a strong dependence on $\delta g$.}
\label{fig:disquenchMeanPos}
\end{figure*}
\end{center}

The presence of the external trap allows us to excite the dipole motion of the droplet by quenching the position of the trap center according to the protocol $V(x)\rightarrow V(x+\delta x_\sigma)$.
The ensuing dynamics can be monitored through the $\sigma$-species one-body densities and the average position of the center-of-mass  $\braket{X_{\sigma}(t)}=\braket{\Psi(t)|\hat{x}_\sigma|\Psi(t)}$ whose spectrum provides the dipole mode frequency.
The properties of this collective droplet mode have not been previously addressed in detail, since these self-bound structures have been predominantly studied in flat geometries, where the translational invariance of the system leads to a vanishing dipole mode~\cite{Collective1D}.

We first consider a FT droplet building upon a symmetric mixture characterized by $N_A=N_B=N=20$, $g_A=g_B\equiv g=0.1$, $\delta g=0.08$ and $m_A=m_B$, while being subjected to a sudden displacement of the trap position, i.e. $\delta x_A=\delta x_B=10$, of both species.
The emergent dipole motion of this self-bound configuration corresponding to a collective oscillation of the droplet cloud around $x=-10$ with a frequency equal to the trap one is showcased in Fig.~\ref{fig:disquenchDensities}(a).
We have verified that it is perfectly stable for long evolution times and independent of the interparticle interactions [see the dashed lines in Fig.~\ref{fig:disquenchMeanPos}(a)]. 
The interaction-independent character of the droplet dipole mode persists for different atom numbers, as well as for interaction-imbalanced mixtures irrespectively of the FT or Gaussian-shape of the homonuclear droplet.
This behavior can be readily inferred from the insensitivity of the $\braket{X_{\sigma}(t)}$ for various system parameters depicted in Fig.~\ref{fig:disquenchMeanPos}(a).
The only impact of the value of the involved interaction strengths on the dipole motion is on the constant spatial width of the oscillating droplet, which is determined by the respective ground state configuration [Fig.~\ref{fig:GSBalanced} and Fig.~\ref{fig:GSImb}].
As we shall explicate below this stable dipole response is a characteristic of homonuclear droplets after symmetrically quenching both components.
The above-described insensitivity of the dipole mode takes equally place also within the different approaches, i.e. the MF and the eGPE (not shown), by means that its frequency and oscillation amplitude remain the same but the width of the droplet changes among the distinct frameworks, a result that can be traced back to the impact of correlations on the droplet initial width.

To exploit the inherent two-component nature of the symmetric droplet we perform a species selective quench on the trap position.
Namely, we shift the trap of species $A$ by $\delta x_A=10$ while keeping the trap of species $B$ intact ($\delta x_B=0$).
Within the FT regime ($\delta g =0.08$), the quenched species $A$ is set to dipole motion while slowly transferring energy to component $B$,  due to the finite interspecies coupling~\cite{mistakidis2022cold,mistakidis2021radiofrequency,mistakidis2020pump}, and thus inducing to it a small amplitude dipole motion, see Fig.~\ref{fig:disquenchDensities}(b), (c) and Fig.~\ref{fig:disquenchMeanPos}(b).
Subsequently, component $B$ is further perturbed due to its collisions with component $A$ 
see e.g. Fig.~\ref{fig:disquenchMeanPos}(b) at $t\approx 500$.
As such, both components exhibit an oscillatory motion around the center of their respective trap, i.e. $x_A=-10$ and $x_B=0$.
Their oscillations are characterized by a temporally varying amplitude and a phase difference stemming from their periodic collisions and the initial slow energy transfer respectively.
For instance, the oscillation amplitude of species $A$ reduces from $\Delta X_A^{\rm max}(t\approx 297)\approx 8.6$ to $\Delta X_A^{\rm max}(t=1789)\approx 4.9$, while the respective amplitude of species $B$ increases from $\Delta X_B^{\rm max}(t=400)\approx 3.9$ to $\Delta X_B^{\rm max}(t=1916)\approx 6.7$ as it can seen in Fig.~\ref{fig:disquenchMeanPos}(b).
This change of amplitudes further indicates the non-negligible energy transfer from the quenched component $A$ to the externally unperturbed component $B$.
An interesting perspective from the above-described process would be to study under which conditions a periodic energy exchange among the components takes place.
In contrast, for stronger attractions e.g. $\delta g =0.01$ the quenched component $A$ induces a dipole motion to component $B$ almost instantly as shown in Fig.~\ref{fig:disquenchMeanPos}(b).
Afterwards, both components perform a nearly in-phase dipole motion centered towards the midpoint of their respective trap origins (i.e. $x=-5$) and having an amplitude which is half of the symmetric dipole mode [Fig.~\ref{fig:disquenchMeanPos}(a)].
The small asymmetries appearing in the amplitude of each component motion are attributed to their mutual interactions.

Next, we explore the impact of intercomponent mass-imbalance on the droplet dipole mode which is induced by a common displacement of the $\sigma$-species trap center, i.e. $\delta x_A=\delta x_B =-10$.
Indeed, the dipole dynamics of the heteronuclear mixture ($m_B=2.1m_A$, $g_B=1.6g_A=0.08$ and $N_A=N_B=N=20$) is interaction-dependent, see Fig.~\ref{fig:disquenchMeanPos}(c) in sharp contrast to homonuclear settings [Fig.~\ref{fig:disquenchMeanPos}(a)]. 
In particular, similarly to the breathing mode response analyzed in Sec.~\ref{Heteronuclear mixture Breathing dynamics} for mass-imbalanced systems, a stronger interspecies attraction (e.g. here $\delta g =0.01$) enforces the two components to oscillate with the same dipole frequency as shown in Fig.~\ref{fig:disquenchMeanPos}(c).
Conversely, weaker attractions such as $\delta g =0.05$ lead to distinct dipole motions among the two components characterized by frequencies slightly higher than the ones of their respective traps.
Moreover, the presence of the heavier $B$ species causes the lighter one ($A$) to exhibit an enhanced or reduced oscillation amplitude when it evolves in-phase or with $\pi$-phase difference with respect to $B$ [Fig.~\ref{fig:disquenchMeanPos}(c)].
Additionally, the dipole mode frequency of the heavier component, B, reduces with decreasing attraction, namely $\omega_{\rm B}^{\rm dip}(\delta g =0.05)\approx 0.9\omega_{\rm B}^{\rm dip}(\delta g =0.01)$~\footnote{Note that a similar comparison for the lighter component $A$ is not possible since it adopts the dipole frequency of the heavier species in the more strongly attractive case ($\delta g=0.01$), as we discussed above [Fig.~\ref{fig:disquenchMeanPos}(c)].}.
This is to be opposed with the dipole motion of homonuclear mixtures [Fig.~\ref{fig:disquenchMeanPos}(a)], which is independent of the interspecies attraction.

Concluding, a counter displacement of the component trap centers i.e. $\delta x_A=-\delta x_B=10$ is applied on the symmetric mixture aiming to induce intercomponent collisions, see Fig.~\ref{fig:disquenchMeanPos}(d).
Close to the decoupling limit e.g. $\delta g =0.08$, the droplets undergo stable in time and nearly independent dipole oscillations, with opposite phase, whose amplitudes are reduced by approximately $15\%$ as compared to the common dipole motion of the symmetric case depicted in Fig.~\ref{fig:disquenchDensities}(a) and Fig.~\ref{fig:disquenchMeanPos}(a).
Importantly, within this weakly attractive regime the droplets feature elastic intercomponent collisions around $x=0$ in a periodic manner as can be deduced by their constant oscillation amplitude. 
It is also worth mentioning that in spite of the smooth center-of-mass droplet oscillations, the respective densities do not remain unaffected in the course of the evolution.
Instead, they develop enhanced density peaks at their collision events which become suppressed at maximum separation.
Moreover, the droplets maintain their FT profile at maximum separation while exhibiting a flattened "wavefront" close to their oscillation centers (not shown).
Turning to strong attractions, e.g. $\delta g =0.01$, the droplets remain tightly self-bound at the origin.
However, certain density portions are emitted from the droplet edges and re-attach to it periodically.
This results in the fluctuations captured by the mean position of the cloud [Fig.~\ref{fig:disquenchMeanPos}(d)]. 

\section{Summary and Outlook}\label{sec:SummaryAndOutlook}

We have studied the ground state properties and quench dynamics of one-dimensional harmonically confined droplet configurations appearing both in homonuclear (interaction balanced or imbalanced) and heteronuclear (mass-imbalanced) bosonic mixtures. 
To appreciate the role of beyond-LHY correlations different theoretical approaches are employed and compared, namely the {\it ab-initio} ML-MCTDHX approach the so-called eGPE as well as the standard MF approximation.

Regarding the ground state of the trapped interaction balanced mixtures we identify a transition from Gaussian shaped to FT droplets due to beyond-LHY correlations for either weak attractions or larger atom numbers. 
This result is in contrast to the predictions of both the MF and the eGPE frameworks, with the former recovering FT droplet structures when tending to the homogeneous limit as discussed in Ref.~\cite{DEBNATH2022MGPtrap}. 
For interaction or mass-imbalanced mixtures it is shown that the involved components become spatially mixed.  
In the case of interaction-imbalanced systems the strongly repulsive component experiences FT signatures for reduced intercomponent attraction, while the remaining one has a Gaussian profile.
A similar behavior occurs for mass-imbalanced mixtures where the heavier component exhibits FT structures for weaker attractions. 
Furthermore, a robust anti-bunching behavior is identified at the same location of the droplet, while two bosons are correlated when placed at opposite sides of the droplet.

Following, a quench of the trap frequency seeds the droplet breathing dynamics.
For homonuclear mixtures we explicate that in addition to the expected contraction and expansion each cloud also experiences a complex excitation process in the long-time evolution.
Specifically, a progressive spatial delocalization of the droplet takes place accompanied by the build-up of motional excitations around its core along with the simultaneous expulsion of density portions.
These excitation mechanisms originate from the participation of higher-lying orbitals of the MB wave function and are associated with the development of enhanced intercomponent entanglement and long-range two-body correlations.
The breathing frequency of the droplet is close to the ideal gas prediction for stronger attractions, while it shows a reduction for weaker ones where the FT is attained. 
Our results are corroborated through analytical estimations of the breathing frequency in the large atom limit by invoking a variational time-dependent Gaussian ansatz. 
In heteronuclear mixtures the components undergo an in-phase breathing at strong attractions, while experiencing a phase difference towards the decoupling limit.
Interestingly, a pronounced dephasing develops for sufficiently strong quench amplitudes emanating from the competition between the tendency to phase-lock and the difference of the species trap frequencies.
Instead, in the absence of correlations solely intercomponent collisions dominate the dynamics.

The droplet dipole motion is triggered by a sudden displacement of the trap's position which turns out to be remarkably stable and insensitive to parametric variations of the homonuclear mixture.
However, the individual components of heteronuclear mixtures feature phase-locked dipole motions for strong interspecies attraction, otherwise they exhibit a phase difference.
Furthermore, irregular dipole patterns occur due to component collisions and intercomponent energy transfer, when considering species selective quenches of the trap position.
Concluding, after a counter displacement of each component trap center the droplets experience elastic collisions for weak attractions while they remain to a large extent bound for stronger ones.

Our findings pave the way for various interesting future research directions.
A fruitful prospect is to utilize beyond-LHY correlations to reveal droplet mixed states in species selective traps and study their dynamical response following time-dependent rampings of the intercomponent attraction across the identified phases.
Also, an explicit derivation of the LHY or higher-order contributions in the presence of external confinement would be at least theoretically desirable. 
Furthermore, the investigation of self-bound state formation in the crossover from highly particle imbalanced to balanced settings is another interesting perspective. 
Certainly, the interplay of beyond-LHY correlations and entanglement for the droplet formation in three-component mixtures~\cite{ma2021borromean,keiler2021polarons} is another promising route to follow.

\section*{Acknowledgements} 

This work (P.S. and I.A.E.) has been funded by the Deutsche Forschungsgemeinschaft (DFG, German Research Foundation) - SFB 925 - project 170620586. S. I. M. acknowledges support from the NSF through a grant for ITAMP at Harvard University.
This research was supported in part by the National Science Foundation under Grant No. NSF PHY-1748958.

\appendix

\section{Variational ansatz for confined droplets}\label{app:var}

To provide further insights into the stationary and dynamical properties of quantum droplets in the main text we have employed, besides the MB ML-MCTDHX and the eGPE approaches, also a so-called variational approximation (VA)~\cite{Salasnich2000}.
It relies on the eGPE framework and utilizes a time-dependent Gaussian ansatz.
This method was recently employed to solve the reduced single-component eGPE in free-space~\cite{AstrakharchikMalomed1DDynamics}, while relevant generalizations have also been reported for dipolar settings~\cite{chomaz2022dipolar}.
Particularly, it was argued that it fails to capture the ground state one-body density of the system, especially close to the FT regime. 
However, it can provide adequate estimates regarding the frequency of the droplet collective excitations.
Here, we seek to apply this method to confined droplets, where it is anticipated that the density of the system can be better approximated by a Gaussian profile, due to the presence of the harmonic trap.  

To construct this VA scheme we initially define the Lagrangian density of the eGPE~\eqref{MGP}  
\begin{equation}
\label{VALagDen}
\begin{split}
    \mathcal{L}=&\frac{i\hslash}{2} (\Psi\Psi^*_t- \Psi^*\Psi_t)+ \frac{\hslash^2}{2m}|\Psi_x|^2+ \frac{ \delta g}{2} |\Psi|^4 \\
    &- \frac{2\sqrt{2m}}{3\pi\hslash}g^{\frac{3}{2}}|\Psi|^3 + \frac{1}{2}m\omega^2x^2|\Psi|^2.
\end{split}
\end{equation}
The subscripts $x$, $t$ refer to the space and time derivatives respectively.
Then, a Gaussian ansatz characterized by time-dependent amplitude ($A(t)$), width ($W(t)$) and phases ($\phi(t)$, $b(t)$) is introduced 
\begin{equation}
\label{VAAnsatz}
    \Psi(x,t)=A(t)\exp\left[i\phi(t)+ib(t)x^2-\frac{x^2}{2W(t)^2}\right].
\end{equation}
It is normalized to the particle number, namely  $N=A(t)^2W(t)\sqrt{\pi}$. 

Substituting this wave function ansatz into the Lagrangian density of Eq.~(\ref{VALagDen}) and integrating over the entire space [-$\infty$, $\infty$], we arrive at the effective Lagrangian per particle
\begin{equation}
\label{VALag}
\begin{split}
    \frac{L_{\rm VA}}{N}=&\hslash\Dot{\phi}+ (\hslash\Dot{b}+\frac{2\hslash^2}{m}(b^2+\frac{1}{4W^4})+\frac{m\omega^2}{2} )\frac{W^2}{2}\\
    &+ \frac{ N\delta g}{2\sqrt{2\pi}W} -\frac{\sqrt{2mg^3}}{3^{\frac{3}{2}}\pi^{\frac{5}{4}}\hslash} \sqrt{ \frac{N}{W}}.
\end{split}
\end{equation}
The corresponding Euler-Lagrange equations of motion in terms of $\phi$, $W$ and $b$ reduce to a classical equation of motion for the width, namely  $m\ddot{W}=-\frac{dU_{\rm eff}}{dW}$.
Hence, in the framework of the VA method the stationary optimal Gaussian solution corresponds to the minimum of the effective potential 
\begin{equation}
    \label{EffPotVA}
\begin{split}
    U_{\rm eff}(W)=&\frac{m\omega^2}{2}W^2+\frac{\hslash^2}{2mW^2}+\frac{N\delta g}{\sqrt{2\pi}W}\\
    &-2\frac{\sqrt{2mg^3}}{\pi^{\frac{5}{4}}\hslash} \sqrt{ \frac{N}{W}}. 
\end{split}
\end{equation} 
Having determined the optimal width ($W_{\rm min}$) of our Gaussian ground state ansatz, the energy per particle of the system is given by $E=U_{\rm eff}(W_{\rm min}/2)$ and the frequency of the breathing mode is the lowest eigenvalue of the Hessian matrix
\begin{equation}
    \label{VAfreq}
    [\omega^{\rm VA}_{\rm br}]^2=\frac{1}{m}\frac{d^2U_{\rm eff}}{dW^2}\Bigg|_{W_{min}}.
\end{equation}
As discussed in the main text, this approximation can adequately describe the breathing frequency of quantum droplets especially in the vicinity of the FT regime. 
However, it fails to capture the underlying density profiles in most of the cases. 
Interestingly, the obtained analytical predictions, in the large particle number limit, provide invaluable insights on the scaling of the droplet breathing frequency in terms of the system parameters and importantly on the LHY contribution.

\bibliographystyle{apsrev4-1}
\bibliography{ref_drops}	

\begin{thebibliography}{89}%
\makeatletter
\providecommand \@ifxundefined [1]{%
 \@ifx{#1\undefined}
}%
\providecommand \@ifnum [1]{%
 \ifnum #1\expandafter \@firstoftwo
 \else \expandafter \@secondoftwo
 \fi
}%
\providecommand \@ifx [1]{%
 \ifx #1\expandafter \@firstoftwo
 \else \expandafter \@secondoftwo
 \fi
}%
\providecommand \natexlab [1]{#1}%
\providecommand \enquote  [1]{``#1''}%
\providecommand \bibnamefont  [1]{#1}%
\providecommand \bibfnamefont [1]{#1}%
\providecommand \citenamefont [1]{#1}%
\providecommand \href@noop [0]{\@secondoftwo}%
\providecommand \href [0]{\begingroup \@sanitize@url \@href}%
\providecommand \@href[1]{\@@startlink{#1}\@@href}%
\providecommand \@@href[1]{\endgroup#1\@@endlink}%
\providecommand \@sanitize@url [0]{\catcode `\\12\catcode `\$12\catcode
  `\&12\catcode `\#12\catcode `\^12\catcode `\_12\catcode `\%12\relax}%
\providecommand \@@startlink[1]{}%
\providecommand \@@endlink[0]{}%
\providecommand \url  [0]{\begingroup\@sanitize@url \@url }%
\providecommand \@url [1]{\endgroup\@href {#1}{\urlprefix }}%
\providecommand \urlprefix  [0]{URL }%
\providecommand \Eprint [0]{\href }%
\providecommand \doibase [0]{http://dx.doi.org/}%
\providecommand \selectlanguage [0]{\@gobble}%
\providecommand \bibinfo  [0]{\@secondoftwo}%
\providecommand \bibfield  [0]{\@secondoftwo}%
\providecommand \translation [1]{[#1]}%
\providecommand \BibitemOpen [0]{}%
\providecommand \bibitemStop [0]{}%
\providecommand \bibitemNoStop [0]{.\EOS\space}%
\providecommand \EOS [0]{\spacefactor3000\relax}%
\providecommand \BibitemShut  [1]{\csname bibitem#1\endcsname}%
\let\auto@bib@innerbib\@empty
\bibitem [{\citenamefont {Bloch}\ \emph {et~al.}(2012)\citenamefont {Bloch},
  \citenamefont {Dalibard},\ and\ \citenamefont
  {Nascimbène}}]{BlochNature2012}%
  \BibitemOpen
  \bibfield  {author} {\bibinfo {author} {\bibfnamefont {I.}~\bibnamefont
  {Bloch}}, \bibinfo {author} {\bibfnamefont {J.}~\bibnamefont {Dalibard}}, \
  and\ \bibinfo {author} {\bibfnamefont {S.}~\bibnamefont {Nascimbène}},\
  }\href {\doibase 10.1038/nphys2259} {\bibfield  {journal} {\bibinfo
  {journal} {Nature Phys.}\ }\textbf {\bibinfo {volume} {8}},\ \bibinfo {pages}
  {267} (\bibinfo {year} {2012})}\BibitemShut {NoStop}%
\bibitem [{\citenamefont {Petrov}(2015)}]{Petrov2015}%
  \BibitemOpen
  \bibfield  {author} {\bibinfo {author} {\bibfnamefont {D.~S.}\ \bibnamefont
  {Petrov}},\ }\href {\doibase 10.1103/PhysRevLett.115.155302} {\bibfield
  {journal} {\bibinfo  {journal} {Phys. Rev. Lett.}\ }\textbf {\bibinfo
  {volume} {115}},\ \bibinfo {pages} {155302} (\bibinfo {year}
  {2015})}\BibitemShut {NoStop}%
\bibitem [{\citenamefont {Ferrier-Barbut}\ \emph {et~al.}(2016)\citenamefont
  {Ferrier-Barbut}, \citenamefont {Kadau}, \citenamefont {Schmitt},
  \citenamefont {Wenzel},\ and\ \citenamefont {Pfau}}]{KadauDropExp}%
  \BibitemOpen
  \bibfield  {author} {\bibinfo {author} {\bibfnamefont {I.}~\bibnamefont
  {Ferrier-Barbut}}, \bibinfo {author} {\bibfnamefont {H.}~\bibnamefont
  {Kadau}}, \bibinfo {author} {\bibfnamefont {M.}~\bibnamefont {Schmitt}},
  \bibinfo {author} {\bibfnamefont {M.}~\bibnamefont {Wenzel}}, \ and\ \bibinfo
  {author} {\bibfnamefont {T.}~\bibnamefont {Pfau}},\ }\href {\doibase
  10.1103/PhysRevLett.116.215301} {\bibfield  {journal} {\bibinfo  {journal}
  {Phys. Rev. Lett.}\ }\textbf {\bibinfo {volume} {116}},\ \bibinfo {pages}
  {215301} (\bibinfo {year} {2016})}\BibitemShut {NoStop}%
\bibitem [{\citenamefont {Böttcher}\ \emph {et~al.}(2020)\citenamefont
  {Böttcher}, \citenamefont {Schmidt}, \citenamefont {Hertkorn}, \citenamefont
  {Ng}, \citenamefont {Graham}, \citenamefont {Guo}, \citenamefont {Langen},\
  and\ \citenamefont {Pfau}}]{PfauReview}%
  \BibitemOpen
  \bibfield  {author} {\bibinfo {author} {\bibfnamefont {F.}~\bibnamefont
  {Böttcher}}, \bibinfo {author} {\bibfnamefont {J.-N.}\ \bibnamefont
  {Schmidt}}, \bibinfo {author} {\bibfnamefont {J.}~\bibnamefont {Hertkorn}},
  \bibinfo {author} {\bibfnamefont {K.~S.~H.}\ \bibnamefont {Ng}}, \bibinfo
  {author} {\bibfnamefont {S.~D.}\ \bibnamefont {Graham}}, \bibinfo {author}
  {\bibfnamefont {M.}~\bibnamefont {Guo}}, \bibinfo {author} {\bibfnamefont
  {T.}~\bibnamefont {Langen}}, \ and\ \bibinfo {author} {\bibfnamefont
  {T.}~\bibnamefont {Pfau}},\ }\href {\doibase 10.1088/1361-6633/abc9ab}
  {\bibfield  {journal} {\bibinfo  {journal} {Rep. Progr. Phys.}\ }\textbf
  {\bibinfo {volume} {84}},\ \bibinfo {pages} {012403} (\bibinfo {year}
  {2020})}\BibitemShut {NoStop}%
\bibitem [{\citenamefont {Luo}\ \emph {et~al.}(2020)\citenamefont {Luo},
  \citenamefont {Pang}, \citenamefont {Liu}, \citenamefont {Li},\ and\
  \citenamefont {Malomed}}]{MalomedLuoReview}%
  \BibitemOpen
  \bibfield  {author} {\bibinfo {author} {\bibfnamefont {Z.-H.}\ \bibnamefont
  {Luo}}, \bibinfo {author} {\bibfnamefont {W.}~\bibnamefont {Pang}}, \bibinfo
  {author} {\bibfnamefont {B.}~\bibnamefont {Liu}}, \bibinfo {author}
  {\bibfnamefont {Y.-Y.}\ \bibnamefont {Li}}, \ and\ \bibinfo {author}
  {\bibfnamefont {B.~A.}\ \bibnamefont {Malomed}},\ }\href {\doibase
  10.1007/s11467-020-1020-2} {\bibfield  {journal} {\bibinfo  {journal} {Front.
  Phys.}\ }\textbf {\bibinfo {volume} {16}},\ \bibinfo {pages} {32201}
  (\bibinfo {year} {2020})}\BibitemShut {NoStop}%
\bibitem [{\citenamefont {Malomed}(2021)}]{MalomedReview}%
  \BibitemOpen
  \bibfield  {author} {\bibinfo {author} {\bibfnamefont {B.~A.}\ \bibnamefont
  {Malomed}},\ }\href {\doibase 10.1007/s11467-020-1024-y} {\bibfield
  {journal} {\bibinfo  {journal} {Front. Phys.}\ }\textbf {\bibinfo {volume}
  {16}},\ \bibinfo {eid} {22504} (\bibinfo {year} {2021})}\BibitemShut
  {NoStop}%
\bibitem [{\citenamefont {Lee}\ \emph {et~al.}(1957)\citenamefont {Lee},
  \citenamefont {Huang},\ and\ \citenamefont {Yang}}]{LeeHuangYang1957}%
  \BibitemOpen
  \bibfield  {author} {\bibinfo {author} {\bibfnamefont {T.~D.}\ \bibnamefont
  {Lee}}, \bibinfo {author} {\bibfnamefont {K.}~\bibnamefont {Huang}}, \ and\
  \bibinfo {author} {\bibfnamefont {C.~N.}\ \bibnamefont {Yang}},\ }\href
  {\doibase 10.1103/PhysRev.106.1135} {\bibfield  {journal} {\bibinfo
  {journal} {Phys. Rev.}\ }\textbf {\bibinfo {volume} {106}},\ \bibinfo {pages}
  {1135} (\bibinfo {year} {1957})}\BibitemShut {NoStop}%
\bibitem [{\citenamefont {Cabrera}\ \emph {et~al.}(2018)\citenamefont
  {Cabrera}, \citenamefont {Tanzi}, \citenamefont {Sanz}, \citenamefont
  {Naylor}, \citenamefont {Thomas}, \citenamefont {Cheiney},\ and\
  \citenamefont {Tarruell}}]{CabreraTarruellDropExp}%
  \BibitemOpen
  \bibfield  {author} {\bibinfo {author} {\bibfnamefont {C.~R.}\ \bibnamefont
  {Cabrera}}, \bibinfo {author} {\bibfnamefont {L.}~\bibnamefont {Tanzi}},
  \bibinfo {author} {\bibfnamefont {J.}~\bibnamefont {Sanz}}, \bibinfo {author}
  {\bibfnamefont {B.}~\bibnamefont {Naylor}}, \bibinfo {author} {\bibfnamefont
  {P.}~\bibnamefont {Thomas}}, \bibinfo {author} {\bibfnamefont
  {P.}~\bibnamefont {Cheiney}}, \ and\ \bibinfo {author} {\bibfnamefont
  {L.}~\bibnamefont {Tarruell}},\ }\href {\doibase 10.1126/science.aao5686}
  {\bibfield  {journal} {\bibinfo  {journal} {Science}\ }\textbf {\bibinfo
  {volume} {359}},\ \bibinfo {pages} {301} (\bibinfo {year}
  {2018})}\BibitemShut {NoStop}%
\bibitem [{\citenamefont {Cheiney}\ \emph {et~al.}(2018)\citenamefont
  {Cheiney}, \citenamefont {Cabrera}, \citenamefont {Sanz}, \citenamefont
  {Naylor}, \citenamefont {Tanzi},\ and\ \citenamefont
  {Tarruell}}]{CheineyTarruellDropExp}%
  \BibitemOpen
  \bibfield  {author} {\bibinfo {author} {\bibfnamefont {P.}~\bibnamefont
  {Cheiney}}, \bibinfo {author} {\bibfnamefont {C.~R.}\ \bibnamefont
  {Cabrera}}, \bibinfo {author} {\bibfnamefont {J.}~\bibnamefont {Sanz}},
  \bibinfo {author} {\bibfnamefont {B.}~\bibnamefont {Naylor}}, \bibinfo
  {author} {\bibfnamefont {L.}~\bibnamefont {Tanzi}}, \ and\ \bibinfo {author}
  {\bibfnamefont {L.}~\bibnamefont {Tarruell}},\ }\href {\doibase
  10.1103/PhysRevLett.120.135301} {\bibfield  {journal} {\bibinfo  {journal}
  {Phys. Rev. Lett.}\ }\textbf {\bibinfo {volume} {120}},\ \bibinfo {pages}
  {135301} (\bibinfo {year} {2018})}\BibitemShut {NoStop}%
\bibitem [{\citenamefont {Semeghini}\ \emph {et~al.}(2018)\citenamefont
  {Semeghini}, \citenamefont {Ferioli}, \citenamefont {Masi}, \citenamefont
  {Mazzinghi}, \citenamefont {Wolswijk}, \citenamefont {Minardi}, \citenamefont
  {Modugno}, \citenamefont {Modugno}, \citenamefont {Inguscio},\ and\
  \citenamefont {Fattori}}]{SemeghiniFattoriDropExp}%
  \BibitemOpen
  \bibfield  {author} {\bibinfo {author} {\bibfnamefont {G.}~\bibnamefont
  {Semeghini}}, \bibinfo {author} {\bibfnamefont {G.}~\bibnamefont {Ferioli}},
  \bibinfo {author} {\bibfnamefont {L.}~\bibnamefont {Masi}}, \bibinfo {author}
  {\bibfnamefont {C.}~\bibnamefont {Mazzinghi}}, \bibinfo {author}
  {\bibfnamefont {L.}~\bibnamefont {Wolswijk}}, \bibinfo {author}
  {\bibfnamefont {F.}~\bibnamefont {Minardi}}, \bibinfo {author} {\bibfnamefont
  {M.}~\bibnamefont {Modugno}}, \bibinfo {author} {\bibfnamefont
  {G.}~\bibnamefont {Modugno}}, \bibinfo {author} {\bibfnamefont
  {M.}~\bibnamefont {Inguscio}}, \ and\ \bibinfo {author} {\bibfnamefont
  {M.}~\bibnamefont {Fattori}},\ }\href {\doibase
  10.1103/PhysRevLett.120.235301} {\bibfield  {journal} {\bibinfo  {journal}
  {Phys. Rev. Lett.}\ }\textbf {\bibinfo {volume} {120}},\ \bibinfo {pages}
  {235301} (\bibinfo {year} {2018})}\BibitemShut {NoStop}%
\bibitem [{\citenamefont {D'Errico}\ \emph {et~al.}(2019)\citenamefont
  {D'Errico}, \citenamefont {Burchianti}, \citenamefont {Prevedelli},
  \citenamefont {Salasnich}, \citenamefont {Ancilotto}, \citenamefont
  {Modugno}, \citenamefont {Minardi},\ and\ \citenamefont
  {Fort}}]{FortHeteroExp}%
  \BibitemOpen
  \bibfield  {author} {\bibinfo {author} {\bibfnamefont {C.}~\bibnamefont
  {D'Errico}}, \bibinfo {author} {\bibfnamefont {A.}~\bibnamefont
  {Burchianti}}, \bibinfo {author} {\bibfnamefont {M.}~\bibnamefont
  {Prevedelli}}, \bibinfo {author} {\bibfnamefont {L.}~\bibnamefont
  {Salasnich}}, \bibinfo {author} {\bibfnamefont {F.}~\bibnamefont
  {Ancilotto}}, \bibinfo {author} {\bibfnamefont {M.}~\bibnamefont {Modugno}},
  \bibinfo {author} {\bibfnamefont {F.}~\bibnamefont {Minardi}}, \ and\
  \bibinfo {author} {\bibfnamefont {C.}~\bibnamefont {Fort}},\ }\href {\doibase
  10.1103/PhysRevResearch.1.033155} {\bibfield  {journal} {\bibinfo  {journal}
  {Phys. Rev. Research}\ }\textbf {\bibinfo {volume} {1}},\ \bibinfo {pages}
  {033155} (\bibinfo {year} {2019})}\BibitemShut {NoStop}%
\bibitem [{\citenamefont {B\"ottcher}\ \emph {et~al.}(2019)\citenamefont
  {B\"ottcher}, \citenamefont {Schmidt}, \citenamefont {Wenzel}, \citenamefont
  {Hertkorn}, \citenamefont {Guo}, \citenamefont {Langen},\ and\ \citenamefont
  {Pfau}}]{Bottcher2019SupersolidDrop}%
  \BibitemOpen
  \bibfield  {author} {\bibinfo {author} {\bibfnamefont {F.}~\bibnamefont
  {B\"ottcher}}, \bibinfo {author} {\bibfnamefont {J.-N.}\ \bibnamefont
  {Schmidt}}, \bibinfo {author} {\bibfnamefont {M.}~\bibnamefont {Wenzel}},
  \bibinfo {author} {\bibfnamefont {J.}~\bibnamefont {Hertkorn}}, \bibinfo
  {author} {\bibfnamefont {M.}~\bibnamefont {Guo}}, \bibinfo {author}
  {\bibfnamefont {T.}~\bibnamefont {Langen}}, \ and\ \bibinfo {author}
  {\bibfnamefont {T.}~\bibnamefont {Pfau}},\ }\href {\doibase
  10.1103/PhysRevX.9.011051} {\bibfield  {journal} {\bibinfo  {journal} {Phys.
  Rev. X}\ }\textbf {\bibinfo {volume} {9}},\ \bibinfo {pages} {011051}
  (\bibinfo {year} {2019})}\BibitemShut {NoStop}%
\bibitem [{\citenamefont {Chomaz}\ \emph {et~al.}(2022)\citenamefont {Chomaz},
  \citenamefont {Ferrier-Barbut}, \citenamefont {Ferlaino}, \citenamefont
  {Laburthe-Tolra}, \citenamefont {Lev},\ and\ \citenamefont
  {Pfau}}]{chomaz2022dipolar}%
  \BibitemOpen
  \bibfield  {author} {\bibinfo {author} {\bibfnamefont {L.}~\bibnamefont
  {Chomaz}}, \bibinfo {author} {\bibfnamefont {I.}~\bibnamefont
  {Ferrier-Barbut}}, \bibinfo {author} {\bibfnamefont {F.}~\bibnamefont
  {Ferlaino}}, \bibinfo {author} {\bibfnamefont {B.}~\bibnamefont
  {Laburthe-Tolra}}, \bibinfo {author} {\bibfnamefont {B.~L.}\ \bibnamefont
  {Lev}}, \ and\ \bibinfo {author} {\bibfnamefont {T.}~\bibnamefont {Pfau}},\
  }\href {https://doi.org/10.48550/arXiv.2201.02672} {\bibfield  {journal}
  {\bibinfo  {journal} {arXiv:2201.02672}\ } (\bibinfo {year}
  {2022})}\BibitemShut {NoStop}%
\bibitem [{\citenamefont {Bisset}\ \emph {et~al.}(2021)\citenamefont {Bisset},
  \citenamefont {Ardila},\ and\ \citenamefont {Santos}}]{Bisset2021}%
  \BibitemOpen
  \bibfield  {author} {\bibinfo {author} {\bibfnamefont {R.~N.}\ \bibnamefont
  {Bisset}}, \bibinfo {author} {\bibfnamefont {L.~A.~P.}\ \bibnamefont
  {Ardila}}, \ and\ \bibinfo {author} {\bibfnamefont {L.}~\bibnamefont
  {Santos}},\ }\href {\doibase 10.1103/PhysRevLett.126.025301} {\bibfield
  {journal} {\bibinfo  {journal} {Phys. Rev. Lett.}\ }\textbf {\bibinfo
  {volume} {126}},\ \bibinfo {pages} {025301} (\bibinfo {year}
  {2021})}\BibitemShut {NoStop}%
\bibitem [{\citenamefont {Smith}\ \emph {et~al.}(2021)\citenamefont {Smith},
  \citenamefont {Baillie},\ and\ \citenamefont {Blakie}}]{Smith2021}%
  \BibitemOpen
  \bibfield  {author} {\bibinfo {author} {\bibfnamefont {J.~C.}\ \bibnamefont
  {Smith}}, \bibinfo {author} {\bibfnamefont {D.}~\bibnamefont {Baillie}}, \
  and\ \bibinfo {author} {\bibfnamefont {P.~B.}\ \bibnamefont {Blakie}},\
  }\href {\doibase 10.1103/PhysRevLett.126.025302} {\bibfield  {journal}
  {\bibinfo  {journal} {Phys. Rev. Lett.}\ }\textbf {\bibinfo {volume} {126}},\
  \bibinfo {pages} {025302} (\bibinfo {year} {2021})}\BibitemShut {NoStop}%
\bibitem [{\citenamefont {Ferioli}\ \emph {et~al.}(2020)\citenamefont
  {Ferioli}, \citenamefont {Semeghini}, \citenamefont {Terradas-Brians\'o},
  \citenamefont {Masi}, \citenamefont {Fattori},\ and\ \citenamefont
  {Modugno}}]{ModugnoFerioliDynamicalFormation}%
  \BibitemOpen
  \bibfield  {author} {\bibinfo {author} {\bibfnamefont {G.}~\bibnamefont
  {Ferioli}}, \bibinfo {author} {\bibfnamefont {G.}~\bibnamefont {Semeghini}},
  \bibinfo {author} {\bibfnamefont {S.}~\bibnamefont {Terradas-Brians\'o}},
  \bibinfo {author} {\bibfnamefont {L.}~\bibnamefont {Masi}}, \bibinfo {author}
  {\bibfnamefont {M.}~\bibnamefont {Fattori}}, \ and\ \bibinfo {author}
  {\bibfnamefont {M.}~\bibnamefont {Modugno}},\ }\href {\doibase
  10.1103/PhysRevResearch.2.013269} {\bibfield  {journal} {\bibinfo  {journal}
  {Phys. Rev. Research}\ }\textbf {\bibinfo {volume} {2}},\ \bibinfo {pages}
  {013269} (\bibinfo {year} {2020})}\BibitemShut {NoStop}%
\bibitem [{\citenamefont {Cappellaro}\ \emph {et~al.}(2018)\citenamefont
  {Cappellaro}, \citenamefont {Macr\`{\i}},\ and\ \citenamefont
  {Salasnich}}]{CappellaroCrossoverDynamics}%
  \BibitemOpen
  \bibfield  {author} {\bibinfo {author} {\bibfnamefont {A.}~\bibnamefont
  {Cappellaro}}, \bibinfo {author} {\bibfnamefont {T.}~\bibnamefont
  {Macr\`{\i}}}, \ and\ \bibinfo {author} {\bibfnamefont {L.}~\bibnamefont
  {Salasnich}},\ }\href {\doibase 10.1103/PhysRevA.97.053623} {\bibfield
  {journal} {\bibinfo  {journal} {Phys. Rev. A}\ }\textbf {\bibinfo {volume}
  {97}},\ \bibinfo {pages} {053623} (\bibinfo {year} {2018})}\BibitemShut
  {NoStop}%
\bibitem [{\citenamefont {Cui}\ and\ \citenamefont
  {Ma}(2021)}]{Cui2021PeriodicConfinement}%
  \BibitemOpen
  \bibfield  {author} {\bibinfo {author} {\bibfnamefont {X.}~\bibnamefont
  {Cui}}\ and\ \bibinfo {author} {\bibfnamefont {Y.}~\bibnamefont {Ma}},\
  }\href {\doibase 10.1103/PhysRevResearch.3.L012027} {\bibfield  {journal}
  {\bibinfo  {journal} {Phys. Rev. Research}\ }\textbf {\bibinfo {volume}
  {3}},\ \bibinfo {pages} {L012027} (\bibinfo {year} {2021})}\BibitemShut
  {NoStop}%
\bibitem [{\citenamefont {Ferioli}\ \emph {et~al.}(2019)\citenamefont
  {Ferioli}, \citenamefont {Semeghini}, \citenamefont {Masi}, \citenamefont
  {Giusti}, \citenamefont {Modugno}, \citenamefont {Inguscio}, \citenamefont
  {Gallem\'{\i}}, \citenamefont {Recati},\ and\ \citenamefont
  {Fattori}}]{FattoriCollisions}%
  \BibitemOpen
  \bibfield  {author} {\bibinfo {author} {\bibfnamefont {G.}~\bibnamefont
  {Ferioli}}, \bibinfo {author} {\bibfnamefont {G.}~\bibnamefont {Semeghini}},
  \bibinfo {author} {\bibfnamefont {L.}~\bibnamefont {Masi}}, \bibinfo {author}
  {\bibfnamefont {G.}~\bibnamefont {Giusti}}, \bibinfo {author} {\bibfnamefont
  {G.}~\bibnamefont {Modugno}}, \bibinfo {author} {\bibfnamefont
  {M.}~\bibnamefont {Inguscio}}, \bibinfo {author} {\bibfnamefont
  {A.}~\bibnamefont {Gallem\'{\i}}}, \bibinfo {author} {\bibfnamefont
  {A.}~\bibnamefont {Recati}}, \ and\ \bibinfo {author} {\bibfnamefont
  {M.}~\bibnamefont {Fattori}},\ }\href {\doibase
  10.1103/PhysRevLett.122.090401} {\bibfield  {journal} {\bibinfo  {journal}
  {Phys. Rev. Lett.}\ }\textbf {\bibinfo {volume} {122}},\ \bibinfo {pages}
  {090401} (\bibinfo {year} {2019})}\BibitemShut {NoStop}%
\bibitem [{\citenamefont {Cui}(2018)}]{CuiSpinOrbitBoseFermi}%
  \BibitemOpen
  \bibfield  {author} {\bibinfo {author} {\bibfnamefont {X.}~\bibnamefont
  {Cui}},\ }\href {\doibase 10.1103/PhysRevA.98.023630} {\bibfield  {journal}
  {\bibinfo  {journal} {Phys. Rev. A}\ }\textbf {\bibinfo {volume} {98}},\
  \bibinfo {pages} {023630} (\bibinfo {year} {2018})}\BibitemShut {NoStop}%
\bibitem [{\citenamefont {Rakshit}\ \emph {et~al.}(2019)\citenamefont
  {Rakshit}, \citenamefont {Karpiuk}, \citenamefont {Brewczyk},\ and\
  \citenamefont {Gajda}}]{GajdaBoseFermi}%
  \BibitemOpen
  \bibfield  {author} {\bibinfo {author} {\bibfnamefont {D.}~\bibnamefont
  {Rakshit}}, \bibinfo {author} {\bibfnamefont {T.}~\bibnamefont {Karpiuk}},
  \bibinfo {author} {\bibfnamefont {M.}~\bibnamefont {Brewczyk}}, \ and\
  \bibinfo {author} {\bibfnamefont {M.}~\bibnamefont {Gajda}},\ }\href
  {\doibase 10.21468/SciPostPhys.6.6.079} {\bibfield  {journal} {\bibinfo
  {journal} {SciPost Phys.}\ }\textbf {\bibinfo {volume} {6}},\ \bibinfo
  {pages} {79} (\bibinfo {year} {2019})}\BibitemShut {NoStop}%
\bibitem [{\citenamefont {Wang}\ \emph
  {et~al.}(2020{\natexlab{a}})\citenamefont {Wang}, \citenamefont {Pan},
  \citenamefont {Cui},\ and\ \citenamefont {Yi}}]{Wang_2020BoseFermi}%
  \BibitemOpen
  \bibfield  {author} {\bibinfo {author} {\bibfnamefont {J.-B.}\ \bibnamefont
  {Wang}}, \bibinfo {author} {\bibfnamefont {J.-S.}\ \bibnamefont {Pan}},
  \bibinfo {author} {\bibfnamefont {X.}~\bibnamefont {Cui}}, \ and\ \bibinfo
  {author} {\bibfnamefont {W.}~\bibnamefont {Yi}},\ }\href {\doibase
  10.1088/0256-307x/37/7/076701} {\bibfield  {journal} {\bibinfo  {journal}
  {Chin. Phys. Lett.}\ }\textbf {\bibinfo {volume} {37}},\ \bibinfo {pages}
  {076701} (\bibinfo {year} {2020}{\natexlab{a}})}\BibitemShut {NoStop}%
\bibitem [{\citenamefont {Sekino}\ and\ \citenamefont
  {Nishida}(2018)}]{Nishida3body}%
  \BibitemOpen
  \bibfield  {author} {\bibinfo {author} {\bibfnamefont {Y.}~\bibnamefont
  {Sekino}}\ and\ \bibinfo {author} {\bibfnamefont {Y.}~\bibnamefont
  {Nishida}},\ }\href {\doibase 10.1103/PhysRevA.97.011602} {\bibfield
  {journal} {\bibinfo  {journal} {Phys. Rev. A}\ }\textbf {\bibinfo {volume}
  {97}},\ \bibinfo {pages} {011602(R)} (\bibinfo {year} {2018})}\BibitemShut
  {NoStop}%
\bibitem [{\citenamefont {Morera}\ \emph
  {et~al.}(2021{\natexlab{a}})\citenamefont {Morera}, \citenamefont
  {Juliá-Díaz},\ and\ \citenamefont {Valiente}}]{Morera3body1D}%
  \BibitemOpen
  \bibfield  {author} {\bibinfo {author} {\bibfnamefont {I.}~\bibnamefont
  {Morera}}, \bibinfo {author} {\bibfnamefont {B.}~\bibnamefont
  {Juliá-Díaz}}, \ and\ \bibinfo {author} {\bibfnamefont {M.}~\bibnamefont
  {Valiente}},\ }\href {https://arxiv.org/abs/2103.16499} {\bibfield  {journal}
  {\bibinfo  {journal} {arXiv:2103.16499}\ } (\bibinfo {year}
  {2021}{\natexlab{a}})}\BibitemShut {NoStop}%
\bibitem [{\citenamefont {St\"urmer}\ \emph {et~al.}(2021)\citenamefont
  {St\"urmer}, \citenamefont {Tengstrand}, \citenamefont {Sachdeva},\ and\
  \citenamefont {Reimann}}]{Sturmer2021}%
  \BibitemOpen
  \bibfield  {author} {\bibinfo {author} {\bibfnamefont {P.}~\bibnamefont
  {St\"urmer}}, \bibinfo {author} {\bibfnamefont {M.~N.}\ \bibnamefont
  {Tengstrand}}, \bibinfo {author} {\bibfnamefont {R.}~\bibnamefont
  {Sachdeva}}, \ and\ \bibinfo {author} {\bibfnamefont {S.~M.}\ \bibnamefont
  {Reimann}},\ }\href {\doibase 10.1103/PhysRevA.103.053302} {\bibfield
  {journal} {\bibinfo  {journal} {Phys. Rev. A}\ }\textbf {\bibinfo {volume}
  {103}},\ \bibinfo {pages} {053302} (\bibinfo {year} {2021})}\BibitemShut
  {NoStop}%
\bibitem [{\citenamefont {Kartashov}\ \emph {et~al.}(2018)\citenamefont
  {Kartashov}, \citenamefont {Malomed}, \citenamefont {Tarruell},\ and\
  \citenamefont {Torner}}]{KartashovVortex2018}%
  \BibitemOpen
  \bibfield  {author} {\bibinfo {author} {\bibfnamefont {Y.~V.}\ \bibnamefont
  {Kartashov}}, \bibinfo {author} {\bibfnamefont {B.~A.}\ \bibnamefont
  {Malomed}}, \bibinfo {author} {\bibfnamefont {L.}~\bibnamefont {Tarruell}}, \
  and\ \bibinfo {author} {\bibfnamefont {L.}~\bibnamefont {Torner}},\ }\href
  {\doibase 10.1103/PhysRevA.98.013612} {\bibfield  {journal} {\bibinfo
  {journal} {Phys. Rev. A}\ }\textbf {\bibinfo {volume} {98}},\ \bibinfo
  {pages} {013612} (\bibinfo {year} {2018})}\BibitemShut {NoStop}%
\bibitem [{\citenamefont {Li}\ \emph {et~al.}(2018)\citenamefont {Li},
  \citenamefont {Chen}, \citenamefont {Luo}, \citenamefont {Huang},
  \citenamefont {Tan}, \citenamefont {Pang},\ and\ \citenamefont
  {Malomed}}]{Malomed2DropVortex}%
  \BibitemOpen
  \bibfield  {author} {\bibinfo {author} {\bibfnamefont {Y.}~\bibnamefont
  {Li}}, \bibinfo {author} {\bibfnamefont {Z.}~\bibnamefont {Chen}}, \bibinfo
  {author} {\bibfnamefont {Z.}~\bibnamefont {Luo}}, \bibinfo {author}
  {\bibfnamefont {C.}~\bibnamefont {Huang}}, \bibinfo {author} {\bibfnamefont
  {H.}~\bibnamefont {Tan}}, \bibinfo {author} {\bibfnamefont {W.}~\bibnamefont
  {Pang}}, \ and\ \bibinfo {author} {\bibfnamefont {B.~A.}\ \bibnamefont
  {Malomed}},\ }\href {\doibase 10.1103/PhysRevA.98.063602} {\bibfield
  {journal} {\bibinfo  {journal} {Phys. Rev. A}\ }\textbf {\bibinfo {volume}
  {98}},\ \bibinfo {pages} {063602} (\bibinfo {year} {2018})}\BibitemShut
  {NoStop}%
\bibitem [{\citenamefont {Tengstrand}\ \emph {et~al.}(2019)\citenamefont
  {Tengstrand}, \citenamefont {St\"urmer}, \citenamefont {Karabulut},\ and\
  \citenamefont {Reimann}}]{ReimannRotatingAndVortex}%
  \BibitemOpen
  \bibfield  {author} {\bibinfo {author} {\bibfnamefont {M.~N.}\ \bibnamefont
  {Tengstrand}}, \bibinfo {author} {\bibfnamefont {P.}~\bibnamefont
  {St\"urmer}}, \bibinfo {author} {\bibfnamefont {E.~O.}\ \bibnamefont
  {Karabulut}}, \ and\ \bibinfo {author} {\bibfnamefont {S.~M.}\ \bibnamefont
  {Reimann}},\ }\href {\doibase 10.1103/PhysRevLett.123.160405} {\bibfield
  {journal} {\bibinfo  {journal} {Phys. Rev. Lett.}\ }\textbf {\bibinfo
  {volume} {123}},\ \bibinfo {pages} {160405} (\bibinfo {year}
  {2019})}\BibitemShut {NoStop}%
\bibitem [{\citenamefont {Lin}\ \emph {et~al.}(2021)\citenamefont {Lin},
  \citenamefont {Xu}, \citenamefont {Chen}, \citenamefont {Yan}, \citenamefont
  {Mai},\ and\ \citenamefont {Liu}}]{ZedaLin2Dvortex}%
  \BibitemOpen
  \bibfield  {author} {\bibinfo {author} {\bibfnamefont {Z.}~\bibnamefont
  {Lin}}, \bibinfo {author} {\bibfnamefont {X.}~\bibnamefont {Xu}}, \bibinfo
  {author} {\bibfnamefont {Z.}~\bibnamefont {Chen}}, \bibinfo {author}
  {\bibfnamefont {Z.}~\bibnamefont {Yan}}, \bibinfo {author} {\bibfnamefont
  {Z.}~\bibnamefont {Mai}}, \ and\ \bibinfo {author} {\bibfnamefont
  {B.}~\bibnamefont {Liu}},\ }\href {\doibase
  https://doi.org/10.1016/j.cnsns.2020.105536} {\bibfield  {journal} {\bibinfo
  {journal} {Comm. in Non. Sci. and Num. Sim.}\ }\textbf {\bibinfo {volume}
  {93}},\ \bibinfo {pages} {105536} (\bibinfo {year} {2021})}\BibitemShut
  {NoStop}%
\bibitem [{\citenamefont {Examilioti}\ and\ \citenamefont
  {Kavoulakis}(2020)}]{Kavoulakis2DDropRotation}%
  \BibitemOpen
  \bibfield  {author} {\bibinfo {author} {\bibfnamefont {P.}~\bibnamefont
  {Examilioti}}\ and\ \bibinfo {author} {\bibfnamefont {G.~M.}\ \bibnamefont
  {Kavoulakis}},\ }\href {\doibase 10.1088/1361-6455/ab9766} {\bibfield
  {journal} {\bibinfo  {journal} {J. Phys. B: At. Mol. and Opt. Phys.}\
  }\textbf {\bibinfo {volume} {53}},\ \bibinfo {pages} {175301} (\bibinfo
  {year} {2020})}\BibitemShut {NoStop}%
\bibitem [{\citenamefont {Morera}\ \emph {et~al.}(2020)\citenamefont {Morera},
  \citenamefont {Astrakharchik}, \citenamefont {Polls},\ and\ \citenamefont
  {Juli\'a-D\'{\i}az}}]{MoreraAstrakharchikLattice}%
  \BibitemOpen
  \bibfield  {author} {\bibinfo {author} {\bibfnamefont {I.}~\bibnamefont
  {Morera}}, \bibinfo {author} {\bibfnamefont {G.~E.}\ \bibnamefont
  {Astrakharchik}}, \bibinfo {author} {\bibfnamefont {A.}~\bibnamefont
  {Polls}}, \ and\ \bibinfo {author} {\bibfnamefont {B.}~\bibnamefont
  {Juli\'a-D\'{\i}az}},\ }\href {\doibase 10.1103/PhysRevResearch.2.022008}
  {\bibfield  {journal} {\bibinfo  {journal} {Phys. Rev. Research}\ }\textbf
  {\bibinfo {volume} {2}},\ \bibinfo {pages} {022008(R)} (\bibinfo {year}
  {2020})}\BibitemShut {NoStop}%
\bibitem [{\citenamefont {Morera}\ \emph
  {et~al.}(2021{\natexlab{b}})\citenamefont {Morera}, \citenamefont
  {Astrakharchik}, \citenamefont {Polls},\ and\ \citenamefont
  {Juli\'a-D\'{\i}az}}]{MoreraAstrakharchikDimerizedDroplets}%
  \BibitemOpen
  \bibfield  {author} {\bibinfo {author} {\bibfnamefont {I.}~\bibnamefont
  {Morera}}, \bibinfo {author} {\bibfnamefont {G.~E.}\ \bibnamefont
  {Astrakharchik}}, \bibinfo {author} {\bibfnamefont {A.}~\bibnamefont
  {Polls}}, \ and\ \bibinfo {author} {\bibfnamefont {B.}~\bibnamefont
  {Juli\'a-D\'{\i}az}},\ }\href {\doibase 10.1103/PhysRevLett.126.023001}
  {\bibfield  {journal} {\bibinfo  {journal} {Phys. Rev. Lett.}\ }\textbf
  {\bibinfo {volume} {126}},\ \bibinfo {pages} {023001} (\bibinfo {year}
  {2021}{\natexlab{b}})}\BibitemShut {NoStop}%
\bibitem [{\citenamefont {Zheng}\ \emph {et~al.}(2020)\citenamefont {Zheng},
  \citenamefont {Chen}, \citenamefont {Huang}, \citenamefont {Dai},
  \citenamefont {Liu}, \citenamefont {Li},\ and\ \citenamefont
  {Wang}}]{2Dlattice}%
  \BibitemOpen
  \bibfield  {author} {\bibinfo {author} {\bibfnamefont {Y.-Y.}\ \bibnamefont
  {Zheng}}, \bibinfo {author} {\bibfnamefont {S.-T.}\ \bibnamefont {Chen}},
  \bibinfo {author} {\bibfnamefont {Z.-P.}\ \bibnamefont {Huang}}, \bibinfo
  {author} {\bibfnamefont {S.-X.}\ \bibnamefont {Dai}}, \bibinfo {author}
  {\bibfnamefont {B.}~\bibnamefont {Liu}}, \bibinfo {author} {\bibfnamefont
  {Y.-Y.}\ \bibnamefont {Li}}, \ and\ \bibinfo {author} {\bibfnamefont {S.-R.}\
  \bibnamefont {Wang}},\ }\href {\doibase 10.1007/s11467-020-1011-3} {\bibfield
   {journal} {\bibinfo  {journal} {Front. Phys.}\ }\textbf {\bibinfo {volume}
  {16}},\ \bibinfo {pages} {22501} (\bibinfo {year} {2020})}\BibitemShut
  {NoStop}%
\bibitem [{\citenamefont {Machida}\ \emph {et~al.}(2022)\citenamefont
  {Machida}, \citenamefont {Danshita}, \citenamefont {Yamamoto},\ and\
  \citenamefont {Kasamatsu}}]{KasamatsuLatticeMott}%
  \BibitemOpen
  \bibfield  {author} {\bibinfo {author} {\bibfnamefont {Y.}~\bibnamefont
  {Machida}}, \bibinfo {author} {\bibfnamefont {I.}~\bibnamefont {Danshita}},
  \bibinfo {author} {\bibfnamefont {D.}~\bibnamefont {Yamamoto}}, \ and\
  \bibinfo {author} {\bibfnamefont {K.}~\bibnamefont {Kasamatsu}},\ }\href
  {\doibase 10.1103/PhysRevA.105.L031301} {\bibfield  {journal} {\bibinfo
  {journal} {Phys. Rev. A}\ }\textbf {\bibinfo {volume} {105}},\ \bibinfo
  {pages} {L031301} (\bibinfo {year} {2022})}\BibitemShut {NoStop}%
\bibitem [{\citenamefont {Liu}\ \emph {et~al.}(2019)\citenamefont {Liu},
  \citenamefont {Zhang}, \citenamefont {Zhong}, \citenamefont {Zhang},
  \citenamefont {Qin}, \citenamefont {Huang}, \citenamefont {Li},\ and\
  \citenamefont {Malomed}}]{MalomedDualCoreTrap}%
  \BibitemOpen
  \bibfield  {author} {\bibinfo {author} {\bibfnamefont {B.}~\bibnamefont
  {Liu}}, \bibinfo {author} {\bibfnamefont {H.-F.}\ \bibnamefont {Zhang}},
  \bibinfo {author} {\bibfnamefont {R.-X.}\ \bibnamefont {Zhong}}, \bibinfo
  {author} {\bibfnamefont {X.-L.}\ \bibnamefont {Zhang}}, \bibinfo {author}
  {\bibfnamefont {X.-Z.}\ \bibnamefont {Qin}}, \bibinfo {author} {\bibfnamefont
  {C.}~\bibnamefont {Huang}}, \bibinfo {author} {\bibfnamefont {Y.-Y.}\
  \bibnamefont {Li}}, \ and\ \bibinfo {author} {\bibfnamefont {B.~A.}\
  \bibnamefont {Malomed}},\ }\href {\doibase 10.1103/PhysRevA.99.053602}
  {\bibfield  {journal} {\bibinfo  {journal} {Phys. Rev. A}\ }\textbf {\bibinfo
  {volume} {99}},\ \bibinfo {pages} {053602} (\bibinfo {year}
  {2019})}\BibitemShut {NoStop}%
\bibitem [{\citenamefont {Zhang}\ \emph {et~al.}(2019)\citenamefont {Zhang},
  \citenamefont {Xu}, \citenamefont {Zheng}, \citenamefont {Chen},
  \citenamefont {Liu}, \citenamefont {Huang}, \citenamefont {Malomed},\ and\
  \citenamefont {Li}}]{MalomedSemidiscrete}%
  \BibitemOpen
  \bibfield  {author} {\bibinfo {author} {\bibfnamefont {X.}~\bibnamefont
  {Zhang}}, \bibinfo {author} {\bibfnamefont {X.}~\bibnamefont {Xu}}, \bibinfo
  {author} {\bibfnamefont {Y.}~\bibnamefont {Zheng}}, \bibinfo {author}
  {\bibfnamefont {Z.}~\bibnamefont {Chen}}, \bibinfo {author} {\bibfnamefont
  {B.}~\bibnamefont {Liu}}, \bibinfo {author} {\bibfnamefont {C.}~\bibnamefont
  {Huang}}, \bibinfo {author} {\bibfnamefont {B.~A.}\ \bibnamefont {Malomed}},
  \ and\ \bibinfo {author} {\bibfnamefont {Y.}~\bibnamefont {Li}},\ }\href
  {\doibase 10.1103/PhysRevLett.123.133901} {\bibfield  {journal} {\bibinfo
  {journal} {Phys. Rev. Lett.}\ }\textbf {\bibinfo {volume} {123}},\ \bibinfo
  {pages} {133901} (\bibinfo {year} {2019})}\BibitemShut {NoStop}%
\bibitem [{\citenamefont {Astrakharchik}\ and\ \citenamefont
  {Malomed}(2018)}]{AstrakharchikMalomed1DDynamics}%
  \BibitemOpen
  \bibfield  {author} {\bibinfo {author} {\bibfnamefont {G.~E.}\ \bibnamefont
  {Astrakharchik}}\ and\ \bibinfo {author} {\bibfnamefont {B.~A.}\ \bibnamefont
  {Malomed}},\ }\href {\doibase 10.1103/PhysRevA.98.013631} {\bibfield
  {journal} {\bibinfo  {journal} {Phys. Rev. A}\ }\textbf {\bibinfo {volume}
  {98}},\ \bibinfo {pages} {013631} (\bibinfo {year} {2018})}\BibitemShut
  {NoStop}%
\bibitem [{\citenamefont {Tylutki}\ \emph {et~al.}(2020)\citenamefont
  {Tylutki}, \citenamefont {Astrakharchik}, \citenamefont {Malomed},\ and\
  \citenamefont {Petrov}}]{Collective1D}%
  \BibitemOpen
  \bibfield  {author} {\bibinfo {author} {\bibfnamefont {M.}~\bibnamefont
  {Tylutki}}, \bibinfo {author} {\bibfnamefont {G.~E.}\ \bibnamefont
  {Astrakharchik}}, \bibinfo {author} {\bibfnamefont {B.~A.}\ \bibnamefont
  {Malomed}}, \ and\ \bibinfo {author} {\bibfnamefont {D.~S.}\ \bibnamefont
  {Petrov}},\ }\href {\doibase 10.1103/PhysRevA.101.051601} {\bibfield
  {journal} {\bibinfo  {journal} {Phys. Rev. A}\ }\textbf {\bibinfo {volume}
  {101}},\ \bibinfo {pages} {051601(R)} (\bibinfo {year} {2020})}\BibitemShut
  {NoStop}%
\bibitem [{\citenamefont {Otajonov}\ \emph {et~al.}(2019)\citenamefont
  {Otajonov}, \citenamefont {Tsoy},\ and\ \citenamefont
  {Abdullaev}}]{AbdullaevSuperGaussian}%
  \BibitemOpen
  \bibfield  {author} {\bibinfo {author} {\bibfnamefont {S.~R.}\ \bibnamefont
  {Otajonov}}, \bibinfo {author} {\bibfnamefont {E.~N.}\ \bibnamefont {Tsoy}},
  \ and\ \bibinfo {author} {\bibfnamefont {F.~K.}\ \bibnamefont {Abdullaev}},\
  }\href {\doibase https://doi.org/10.1016/j.physleta.2019.125980} {\bibfield
  {journal} {\bibinfo  {journal} {Phys. Lett. A}\ }\textbf {\bibinfo {volume}
  {383}},\ \bibinfo {pages} {125980} (\bibinfo {year} {2019})}\BibitemShut
  {NoStop}%
\bibitem [{\citenamefont {Tengstrand}\ and\ \citenamefont
  {Reimann}(2022)}]{tengstrand2022droplet}%
  \BibitemOpen
  \bibfield  {author} {\bibinfo {author} {\bibfnamefont {M.~N.}\ \bibnamefont
  {Tengstrand}}\ and\ \bibinfo {author} {\bibfnamefont {S.~M.}\ \bibnamefont
  {Reimann}},\ }\href@noop {} {\bibfield  {journal} {\bibinfo  {journal} {Phys.
  Rev. A}\ }\textbf {\bibinfo {volume} {105}},\ \bibinfo {pages} {033319}
  (\bibinfo {year} {2022})}\BibitemShut {NoStop}%
\bibitem [{\citenamefont {Mithun}\ \emph {et~al.}(2020)\citenamefont {Mithun},
  \citenamefont {Maluckov}, \citenamefont {Kasamatsu}, \citenamefont
  {Malomed},\ and\ \citenamefont {Khare}}]{MithunMI}%
  \BibitemOpen
  \bibfield  {author} {\bibinfo {author} {\bibfnamefont {T.}~\bibnamefont
  {Mithun}}, \bibinfo {author} {\bibfnamefont {A.}~\bibnamefont {Maluckov}},
  \bibinfo {author} {\bibfnamefont {K.}~\bibnamefont {Kasamatsu}}, \bibinfo
  {author} {\bibfnamefont {B.~A.}\ \bibnamefont {Malomed}}, \ and\ \bibinfo
  {author} {\bibfnamefont {A.}~\bibnamefont {Khare}},\ }\href {\doibase
  10.3390/sym12010174} {\bibfield  {journal} {\bibinfo  {journal} {Symmetry}\
  }\textbf {\bibinfo {volume} {12}},\ \bibinfo {pages} {32201} (\bibinfo {year}
  {2020})}\BibitemShut {NoStop}%
\bibitem [{\citenamefont {Zin}\ \emph {et~al.}(2018)\citenamefont {Zin},
  \citenamefont {Pylak}, \citenamefont {Wasak}, \citenamefont {Gajda},\ and\
  \citenamefont {Idziaszek}}]{IdziaszekDimensionalCrossover}%
  \BibitemOpen
  \bibfield  {author} {\bibinfo {author} {\bibfnamefont {P.}~\bibnamefont
  {Zin}}, \bibinfo {author} {\bibfnamefont {M.}~\bibnamefont {Pylak}}, \bibinfo
  {author} {\bibfnamefont {T.}~\bibnamefont {Wasak}}, \bibinfo {author}
  {\bibfnamefont {M.}~\bibnamefont {Gajda}}, \ and\ \bibinfo {author}
  {\bibfnamefont {Z.}~\bibnamefont {Idziaszek}},\ }\href {\doibase
  10.1103/PhysRevA.98.051603} {\bibfield  {journal} {\bibinfo  {journal} {Phys.
  Rev. A}\ }\textbf {\bibinfo {volume} {98}},\ \bibinfo {pages} {051603(R)}
  (\bibinfo {year} {2018})}\BibitemShut {NoStop}%
\bibitem [{\citenamefont {Wang}\ \emph
  {et~al.}(2020{\natexlab{b}})\citenamefont {Wang}, \citenamefont {Hu},\ and\
  \citenamefont {Liu}}]{HuiThermal}%
  \BibitemOpen
  \bibfield  {author} {\bibinfo {author} {\bibfnamefont {J.}~\bibnamefont
  {Wang}}, \bibinfo {author} {\bibfnamefont {H.}~\bibnamefont {Hu}}, \ and\
  \bibinfo {author} {\bibfnamefont {X.-J.}\ \bibnamefont {Liu}},\ }\href
  {\doibase 10.1088/1367-2630/abbe55} {\bibfield  {journal} {\bibinfo
  {journal} {New J. Phys.}\ }\textbf {\bibinfo {volume} {22}},\ \bibinfo
  {pages} {103044} (\bibinfo {year} {2020}{\natexlab{b}})}\BibitemShut
  {NoStop}%
\bibitem [{\citenamefont {De~Rosi}\ \emph {et~al.}(2021)\citenamefont
  {De~Rosi}, \citenamefont {Astrakharchik},\ and\ \citenamefont
  {Massignan}}]{AstrakharchikThermal}%
  \BibitemOpen
  \bibfield  {author} {\bibinfo {author} {\bibfnamefont {G.}~\bibnamefont
  {De~Rosi}}, \bibinfo {author} {\bibfnamefont {G.~E.}\ \bibnamefont
  {Astrakharchik}}, \ and\ \bibinfo {author} {\bibfnamefont {P.}~\bibnamefont
  {Massignan}},\ }\href {\doibase 10.1103/PhysRevA.103.043316} {\bibfield
  {journal} {\bibinfo  {journal} {Phys. Rev. A}\ }\textbf {\bibinfo {volume}
  {103}},\ \bibinfo {pages} {043316} (\bibinfo {year} {2021})}\BibitemShut
  {NoStop}%
\bibitem [{\citenamefont {Guebli}\ and\ \citenamefont
  {Boudjem\^aa}(2021)}]{BoudjemaaHigherOrderThermal}%
  \BibitemOpen
  \bibfield  {author} {\bibinfo {author} {\bibfnamefont {N.}~\bibnamefont
  {Guebli}}\ and\ \bibinfo {author} {\bibfnamefont {A.}~\bibnamefont
  {Boudjem\^aa}},\ }\href {\doibase 10.1103/PhysRevA.104.023310} {\bibfield
  {journal} {\bibinfo  {journal} {Phys. Rev. A}\ }\textbf {\bibinfo {volume}
  {104}},\ \bibinfo {pages} {023310} (\bibinfo {year} {2021})}\BibitemShut
  {NoStop}%
\bibitem [{\citenamefont {Boudjemâa}(2021)}]{BoudjemaaThermal}%
  \BibitemOpen
  \bibfield  {author} {\bibinfo {author} {\bibfnamefont {A.}~\bibnamefont
  {Boudjemâa}},\ }\href {\doibase 10.1038/s41598-021-01089-6} {\bibfield
  {journal} {\bibinfo  {journal} {Sci. Rep.}\ }\textbf {\bibinfo {volume}
  {11}},\ \bibinfo {pages} {21765} (\bibinfo {year} {2021})}\BibitemShut
  {NoStop}%
\bibitem [{\citenamefont {Mithun}\ \emph {et~al.}(2021)\citenamefont {Mithun},
  \citenamefont {Mistakidis}, \citenamefont {Schmelcher},\ and\ \citenamefont
  {Kevrekidis}}]{mithun2021statistical}%
  \BibitemOpen
  \bibfield  {author} {\bibinfo {author} {\bibfnamefont {T.}~\bibnamefont
  {Mithun}}, \bibinfo {author} {\bibfnamefont {S.~I.}\ \bibnamefont
  {Mistakidis}}, \bibinfo {author} {\bibfnamefont {P.}~\bibnamefont
  {Schmelcher}}, \ and\ \bibinfo {author} {\bibfnamefont {P.~G.}\ \bibnamefont
  {Kevrekidis}},\ }\href@noop {} {\bibfield  {journal} {\bibinfo  {journal}
  {Phys. Rev. A}\ }\textbf {\bibinfo {volume} {104}},\ \bibinfo {pages}
  {033316} (\bibinfo {year} {2021})}\BibitemShut {NoStop}%
\bibitem [{\citenamefont {Astrakharchik}\ and\ \citenamefont
  {Giorgini}(2006)}]{Astrakharchik_2006}%
  \BibitemOpen
  \bibfield  {author} {\bibinfo {author} {\bibfnamefont {G.~E.}\ \bibnamefont
  {Astrakharchik}}\ and\ \bibinfo {author} {\bibfnamefont {S.}~\bibnamefont
  {Giorgini}},\ }\href {\doibase 10.1088/0953-4075/39/10/s01} {\bibfield
  {journal} {\bibinfo  {journal} {J. Phys. B: At. Mol. and Opt. Phys.}\
  }\textbf {\bibinfo {volume} {39}},\ \bibinfo {pages} {S1} (\bibinfo {year}
  {2006})}\BibitemShut {NoStop}%
\bibitem [{\citenamefont {Lavoine}\ and\ \citenamefont
  {Bourdel}(2021)}]{BourdelCrossover}%
  \BibitemOpen
  \bibfield  {author} {\bibinfo {author} {\bibfnamefont {L.}~\bibnamefont
  {Lavoine}}\ and\ \bibinfo {author} {\bibfnamefont {T.}~\bibnamefont
  {Bourdel}},\ }\href {\doibase 10.1103/PhysRevA.103.033312} {\bibfield
  {journal} {\bibinfo  {journal} {Phys. Rev. A}\ }\textbf {\bibinfo {volume}
  {103}},\ \bibinfo {pages} {033312} (\bibinfo {year} {2021})}\BibitemShut
  {NoStop}%
\bibitem [{\citenamefont {Fort}\ and\ \citenamefont
  {Modugno}(2021)}]{FortModugnoSelfEvaporation}%
  \BibitemOpen
  \bibfield  {author} {\bibinfo {author} {\bibfnamefont {C.}~\bibnamefont
  {Fort}}\ and\ \bibinfo {author} {\bibfnamefont {M.}~\bibnamefont {Modugno}},\
  }\href {\doibase 10.3390/app11020866} {\bibfield  {journal} {\bibinfo
  {journal} {Applied Sciences}\ }\textbf {\bibinfo {volume} {11(2)}},\ \bibinfo
  {pages} {866} (\bibinfo {year} {2021})}\BibitemShut {NoStop}%
\bibitem [{\citenamefont {Ancilotto}\ \emph {et~al.}(2018)\citenamefont
  {Ancilotto}, \citenamefont {Barranco}, \citenamefont {Guilleumas},\ and\
  \citenamefont {Pi}}]{AncilottoLocalDensity}%
  \BibitemOpen
  \bibfield  {author} {\bibinfo {author} {\bibfnamefont {F.}~\bibnamefont
  {Ancilotto}}, \bibinfo {author} {\bibfnamefont {M.}~\bibnamefont {Barranco}},
  \bibinfo {author} {\bibfnamefont {M.}~\bibnamefont {Guilleumas}}, \ and\
  \bibinfo {author} {\bibfnamefont {M.}~\bibnamefont {Pi}},\ }\href {\doibase
  10.1103/PhysRevA.98.053623} {\bibfield  {journal} {\bibinfo  {journal} {Phys.
  Rev. A}\ }\textbf {\bibinfo {volume} {98}},\ \bibinfo {pages} {053623}
  (\bibinfo {year} {2018})}\BibitemShut {NoStop}%
\bibitem [{\citenamefont {Mistakidis}\ \emph
  {et~al.}(2021{\natexlab{a}})\citenamefont {Mistakidis}, \citenamefont
  {Mithun}, \citenamefont {Kevrekidis}, \citenamefont {Sadeghpour},\ and\
  \citenamefont {Schmelcher}}]{Mistakidis2021}%
  \BibitemOpen
  \bibfield  {author} {\bibinfo {author} {\bibfnamefont {S.~I.}\ \bibnamefont
  {Mistakidis}}, \bibinfo {author} {\bibfnamefont {T.}~\bibnamefont {Mithun}},
  \bibinfo {author} {\bibfnamefont {P.~G.}\ \bibnamefont {Kevrekidis}},
  \bibinfo {author} {\bibfnamefont {H.~R.}\ \bibnamefont {Sadeghpour}}, \ and\
  \bibinfo {author} {\bibfnamefont {P.}~\bibnamefont {Schmelcher}},\ }\href
  {\doibase 10.1103/PhysRevResearch.3.043128} {\bibfield  {journal} {\bibinfo
  {journal} {Phys. Rev. Research}\ }\textbf {\bibinfo {volume} {3}},\ \bibinfo
  {pages} {043128} (\bibinfo {year} {2021}{\natexlab{a}})}\BibitemShut
  {NoStop}%
\bibitem [{\citenamefont {Petrov}\ and\ \citenamefont
  {Astrakharchik}(2016)}]{PetrovLowD}%
  \BibitemOpen
  \bibfield  {author} {\bibinfo {author} {\bibfnamefont {D.~S.}\ \bibnamefont
  {Petrov}}\ and\ \bibinfo {author} {\bibfnamefont {G.~E.}\ \bibnamefont
  {Astrakharchik}},\ }\href {\doibase 10.1103/PhysRevLett.117.100401}
  {\bibfield  {journal} {\bibinfo  {journal} {Phys. Rev. Lett.}\ }\textbf
  {\bibinfo {volume} {117}},\ \bibinfo {pages} {100401} (\bibinfo {year}
  {2016})}\BibitemShut {NoStop}%
\bibitem [{\citenamefont {Parisi}\ \emph {et~al.}(2019)\citenamefont {Parisi},
  \citenamefont {Astrakharchik},\ and\ \citenamefont
  {Giorgini}}]{ParisiMonteCarlo2019}%
  \BibitemOpen
  \bibfield  {author} {\bibinfo {author} {\bibfnamefont {L.}~\bibnamefont
  {Parisi}}, \bibinfo {author} {\bibfnamefont {G.~E.}\ \bibnamefont
  {Astrakharchik}}, \ and\ \bibinfo {author} {\bibfnamefont {S.}~\bibnamefont
  {Giorgini}},\ }\href {\doibase 10.1103/PhysRevLett.122.105302} {\bibfield
  {journal} {\bibinfo  {journal} {Phys. Rev. Lett.}\ }\textbf {\bibinfo
  {volume} {122}},\ \bibinfo {pages} {105302} (\bibinfo {year}
  {2019})}\BibitemShut {NoStop}%
\bibitem [{\citenamefont {Parisi}\ and\ \citenamefont
  {Giorgini}(2020)}]{ParisiGiorginiMonteCarlo}%
  \BibitemOpen
  \bibfield  {author} {\bibinfo {author} {\bibfnamefont {L.}~\bibnamefont
  {Parisi}}\ and\ \bibinfo {author} {\bibfnamefont {S.}~\bibnamefont
  {Giorgini}},\ }\href {\doibase 10.1103/PhysRevA.102.023318} {\bibfield
  {journal} {\bibinfo  {journal} {Phys. Rev. A}\ }\textbf {\bibinfo {volume}
  {102}},\ \bibinfo {pages} {023318} (\bibinfo {year} {2020})}\BibitemShut
  {NoStop}%
\bibitem [{\citenamefont {Debnath}\ \emph {et~al.}(2022)\citenamefont
  {Debnath}, \citenamefont {Khan},\ and\ \citenamefont
  {Basu}}]{DEBNATH2022MGPtrap}%
  \BibitemOpen
  \bibfield  {author} {\bibinfo {author} {\bibfnamefont {A.}~\bibnamefont
  {Debnath}}, \bibinfo {author} {\bibfnamefont {A.}~\bibnamefont {Khan}}, \
  and\ \bibinfo {author} {\bibfnamefont {S.}~\bibnamefont {Basu}},\ }\href
  {\doibase https://doi.org/10.1016/j.physleta.2022.128137} {\bibfield
  {journal} {\bibinfo  {journal} {Phys. Lett. A}\ }\textbf {\bibinfo {volume}
  {439}},\ \bibinfo {pages} {128137} (\bibinfo {year} {2022})}\BibitemShut
  {NoStop}%
\bibitem [{\citenamefont {Pathak}\ and\ \citenamefont
  {Nath}(2022)}]{PathakHarmonic2022}%
  \BibitemOpen
  \bibfield  {author} {\bibinfo {author} {\bibfnamefont {M.~R.}\ \bibnamefont
  {Pathak}}\ and\ \bibinfo {author} {\bibfnamefont {A.}~\bibnamefont {Nath}},\
  }\href {\doibase 10.1038/s41598-022-10468-6} {\bibfield  {journal} {\bibinfo
  {journal} {Sci. Rep.}\ }\textbf {\bibinfo {volume} {12}},\ \bibinfo {pages}
  {6904} (\bibinfo {year} {2022})}\BibitemShut {NoStop}%
\bibitem [{\citenamefont {Krönke}\ \emph {et~al.}(2013)\citenamefont
  {Krönke}, \citenamefont {Cao}, \citenamefont {Vendrell},\ and\ \citenamefont
  {Schmelcher}}]{Kronke_2013}%
  \BibitemOpen
  \bibfield  {author} {\bibinfo {author} {\bibfnamefont {S.}~\bibnamefont
  {Krönke}}, \bibinfo {author} {\bibfnamefont {L.}~\bibnamefont {Cao}},
  \bibinfo {author} {\bibfnamefont {O.}~\bibnamefont {Vendrell}}, \ and\
  \bibinfo {author} {\bibfnamefont {P.}~\bibnamefont {Schmelcher}},\ }\href
  {\doibase 10.1088/1367-2630/15/6/063018} {\bibfield  {journal} {\bibinfo
  {journal} {New J. Phys.}\ }\textbf {\bibinfo {volume} {15}},\ \bibinfo
  {pages} {063018} (\bibinfo {year} {2013})}\BibitemShut {NoStop}%
\bibitem [{\citenamefont {Cao}\ \emph {et~al.}(2013)\citenamefont {Cao},
  \citenamefont {Krönke}, \citenamefont {Vendrell},\ and\ \citenamefont
  {Schmelcher}}]{Cao2013}%
  \BibitemOpen
  \bibfield  {author} {\bibinfo {author} {\bibfnamefont {L.}~\bibnamefont
  {Cao}}, \bibinfo {author} {\bibfnamefont {S.}~\bibnamefont {Krönke}},
  \bibinfo {author} {\bibfnamefont {O.}~\bibnamefont {Vendrell}}, \ and\
  \bibinfo {author} {\bibfnamefont {P.}~\bibnamefont {Schmelcher}},\ }\href
  {\doibase 10.1063/1.4821350} {\bibfield  {journal} {\bibinfo  {journal} {J.
  Chem. Phys.}\ }\textbf {\bibinfo {volume} {139}},\ \bibinfo {pages} {134103}
  (\bibinfo {year} {2013})}\BibitemShut {NoStop}%
\bibitem [{\citenamefont {Cao}\ \emph {et~al.}(2017)\citenamefont {Cao},
  \citenamefont {Bolsinger}, \citenamefont {Mistakidis}, \citenamefont
  {Koutentakis}, \citenamefont {Kr{\"o}nke}, \citenamefont {Schurer},\ and\
  \citenamefont {Schmelcher}}]{cao2017unified}%
  \BibitemOpen
  \bibfield  {author} {\bibinfo {author} {\bibfnamefont {L.}~\bibnamefont
  {Cao}}, \bibinfo {author} {\bibfnamefont {V.}~\bibnamefont {Bolsinger}},
  \bibinfo {author} {\bibfnamefont {S.~I.}\ \bibnamefont {Mistakidis}},
  \bibinfo {author} {\bibfnamefont {G.~M.}\ \bibnamefont {Koutentakis}},
  \bibinfo {author} {\bibfnamefont {S.}~\bibnamefont {Kr{\"o}nke}}, \bibinfo
  {author} {\bibfnamefont {J.~M.}\ \bibnamefont {Schurer}}, \ and\ \bibinfo
  {author} {\bibfnamefont {P.}~\bibnamefont {Schmelcher}},\ }\href@noop {}
  {\bibfield  {journal} {\bibinfo  {journal} {J. Chem. Phys.}\ }\textbf
  {\bibinfo {volume} {147}},\ \bibinfo {pages} {044106} (\bibinfo {year}
  {2017})}\BibitemShut {NoStop}%
\bibitem [{\citenamefont {Pethick}\ and\ \citenamefont
  {Smith}(2008)}]{pethick2008bose}%
  \BibitemOpen
  \bibfield  {author} {\bibinfo {author} {\bibfnamefont {C.~J.}\ \bibnamefont
  {Pethick}}\ and\ \bibinfo {author} {\bibfnamefont {H.}~\bibnamefont
  {Smith}},\ }\href@noop {} {\emph {\bibinfo {title} {Bose--Einstein
  condensation in dilute gases}}}\ (\bibinfo  {publisher} {Cambridge university
  press, 2\textsuperscript{nd} edition},\ \bibinfo {year} {2008})\BibitemShut
  {NoStop}%
\bibitem [{\citenamefont {Olshanii}(1998)}]{olshanii1998atomic}%
  \BibitemOpen
  \bibfield  {author} {\bibinfo {author} {\bibfnamefont {M.}~\bibnamefont
  {Olshanii}},\ }\href@noop {} {\bibfield  {journal} {\bibinfo  {journal}
  {Phys. Rev. Lett.}\ }\textbf {\bibinfo {volume} {81}},\ \bibinfo {pages}
  {938} (\bibinfo {year} {1998})}\BibitemShut {NoStop}%
\bibitem [{\citenamefont {Chin}\ \emph {et~al.}(2010)\citenamefont {Chin},
  \citenamefont {Grimm}, \citenamefont {Julienne},\ and\ \citenamefont
  {Tiesinga}}]{chin2010feshbach}%
  \BibitemOpen
  \bibfield  {author} {\bibinfo {author} {\bibfnamefont {C.}~\bibnamefont
  {Chin}}, \bibinfo {author} {\bibfnamefont {R.}~\bibnamefont {Grimm}},
  \bibinfo {author} {\bibfnamefont {P.}~\bibnamefont {Julienne}}, \ and\
  \bibinfo {author} {\bibfnamefont {E.}~\bibnamefont {Tiesinga}},\ }\href@noop
  {} {\bibfield  {journal} {\bibinfo  {journal} {Rev. Mod. Phys.}\ }\textbf
  {\bibinfo {volume} {82}},\ \bibinfo {pages} {1225} (\bibinfo {year}
  {2010})}\BibitemShut {NoStop}%
\bibitem [{\citenamefont {K{\"o}hler}\ \emph {et~al.}(2006)\citenamefont
  {K{\"o}hler}, \citenamefont {G{\'o}ral},\ and\ \citenamefont
  {Julienne}}]{kohler2006production}%
  \BibitemOpen
  \bibfield  {author} {\bibinfo {author} {\bibfnamefont {T.}~\bibnamefont
  {K{\"o}hler}}, \bibinfo {author} {\bibfnamefont {K.}~\bibnamefont
  {G{\'o}ral}}, \ and\ \bibinfo {author} {\bibfnamefont {P.~S.}\ \bibnamefont
  {Julienne}},\ }\href@noop {} {\bibfield  {journal} {\bibinfo  {journal} {Rev.
  Mod. Phys.}\ }\textbf {\bibinfo {volume} {78}},\ \bibinfo {pages} {1311}
  (\bibinfo {year} {2006})}\BibitemShut {NoStop}%
\bibitem [{\citenamefont {G\"orlitz}\ \emph {et~al.}(2001)\citenamefont
  {G\"orlitz}, \citenamefont {Vogels}, \citenamefont {Leanhardt}, \citenamefont
  {Raman}, \citenamefont {Gustavson}, \citenamefont {Abo-Shaeer}, \citenamefont
  {Chikkatur}, \citenamefont {Gupta}, \citenamefont {Inouye}, \citenamefont
  {Rosenband},\ and\ \citenamefont {Ketterle}}]{Ketterle2001LowDexp}%
  \BibitemOpen
  \bibfield  {author} {\bibinfo {author} {\bibfnamefont {A.}~\bibnamefont
  {G\"orlitz}}, \bibinfo {author} {\bibfnamefont {J.~M.}\ \bibnamefont
  {Vogels}}, \bibinfo {author} {\bibfnamefont {A.~E.}\ \bibnamefont
  {Leanhardt}}, \bibinfo {author} {\bibfnamefont {C.}~\bibnamefont {Raman}},
  \bibinfo {author} {\bibfnamefont {T.~L.}\ \bibnamefont {Gustavson}}, \bibinfo
  {author} {\bibfnamefont {J.~R.}\ \bibnamefont {Abo-Shaeer}}, \bibinfo
  {author} {\bibfnamefont {A.~P.}\ \bibnamefont {Chikkatur}}, \bibinfo {author}
  {\bibfnamefont {S.}~\bibnamefont {Gupta}}, \bibinfo {author} {\bibfnamefont
  {S.}~\bibnamefont {Inouye}}, \bibinfo {author} {\bibfnamefont
  {T.}~\bibnamefont {Rosenband}}, \ and\ \bibinfo {author} {\bibfnamefont
  {W.}~\bibnamefont {Ketterle}},\ }\href {\doibase
  10.1103/PhysRevLett.87.130402} {\bibfield  {journal} {\bibinfo  {journal}
  {Phys. Rev. Lett.}\ }\textbf {\bibinfo {volume} {87}},\ \bibinfo {pages}
  {130402} (\bibinfo {year} {2001})}\BibitemShut {NoStop}%
\bibitem [{\citenamefont {Skov}\ \emph {et~al.}(2021)\citenamefont {Skov},
  \citenamefont {Skou}, \citenamefont {J\o{}rgensen},\ and\ \citenamefont
  {Arlt}}]{Skov2021LHYfluidexp}%
  \BibitemOpen
  \bibfield  {author} {\bibinfo {author} {\bibfnamefont {T.~G.}\ \bibnamefont
  {Skov}}, \bibinfo {author} {\bibfnamefont {M.~G.}\ \bibnamefont {Skou}},
  \bibinfo {author} {\bibfnamefont {N.~B.}\ \bibnamefont {J\o{}rgensen}}, \
  and\ \bibinfo {author} {\bibfnamefont {J.~J.}\ \bibnamefont {Arlt}},\ }\href
  {\doibase 10.1103/PhysRevLett.126.230404} {\bibfield  {journal} {\bibinfo
  {journal} {Phys. Rev. Lett.}\ }\textbf {\bibinfo {volume} {126}},\ \bibinfo
  {pages} {230404} (\bibinfo {year} {2021})}\BibitemShut {NoStop}%
\bibitem [{\citenamefont {Barashenkov}\ and\ \citenamefont
  {Panova}(1993)}]{BARASHENKOV1993Bubbles}%
  \BibitemOpen
  \bibfield  {author} {\bibinfo {author} {\bibfnamefont {I.~V.}\ \bibnamefont
  {Barashenkov}}\ and\ \bibinfo {author} {\bibfnamefont {E.~Y.}\ \bibnamefont
  {Panova}},\ }\href@noop {} {\bibfield  {journal} {\bibinfo  {journal} {Phys.
  D: Non. Phen.}\ }\textbf {\bibinfo {volume} {69}},\ \bibinfo {pages} {114}
  (\bibinfo {year} {1993})}\BibitemShut {NoStop}%
\bibitem [{\citenamefont {Triki}\ \emph {et~al.}(2017)\citenamefont {Triki},
  \citenamefont {Biswas}, \citenamefont {Moshokoa},\ and\ \citenamefont
  {Belic}}]{Optik2017Belic}%
  \BibitemOpen
  \bibfield  {author} {\bibinfo {author} {\bibfnamefont {H.}~\bibnamefont
  {Triki}}, \bibinfo {author} {\bibfnamefont {A.}~\bibnamefont {Biswas}},
  \bibinfo {author} {\bibfnamefont {S.~P.}\ \bibnamefont {Moshokoa}}, \ and\
  \bibinfo {author} {\bibfnamefont {M.}~\bibnamefont {Belic}},\ }\href@noop {}
  {\bibfield  {journal} {\bibinfo  {journal} {Optik}\ }\textbf {\bibinfo
  {volume} {128}},\ \bibinfo {pages} {63} (\bibinfo {year} {2017})}\BibitemShut
  {NoStop}%
\bibitem [{\citenamefont {Pitaevskii}\ and\ \citenamefont
  {Stringari}(2016)}]{Stringari2016BEC}%
  \BibitemOpen
  \bibfield  {author} {\bibinfo {author} {\bibfnamefont {L.}~\bibnamefont
  {Pitaevskii}}\ and\ \bibinfo {author} {\bibfnamefont {S.}~\bibnamefont
  {Stringari}},\ }\href@noop {} {\emph {\bibinfo {title} {Bose–-Einstein
  Condensation and Superfluidity}}}\ (\bibinfo  {publisher} {Oxford University
  Press},\ \bibinfo {year} {2016})\BibitemShut {NoStop}%
\bibitem [{\citenamefont {Mistakidis}\ \emph {et~al.}(2022)\citenamefont
  {Mistakidis}, \citenamefont {Volosniev}, \citenamefont {Barfknecht},
  \citenamefont {Fogarty}, \citenamefont {Busch}, \citenamefont {Foerster},
  \citenamefont {Schmelcher},\ and\ \citenamefont
  {Zinner}}]{mistakidis2022cold}%
  \BibitemOpen
  \bibfield  {author} {\bibinfo {author} {\bibfnamefont {S.~I.}\ \bibnamefont
  {Mistakidis}}, \bibinfo {author} {\bibfnamefont {A.~G.}\ \bibnamefont
  {Volosniev}}, \bibinfo {author} {\bibfnamefont {R.~E.}\ \bibnamefont
  {Barfknecht}}, \bibinfo {author} {\bibfnamefont {T.}~\bibnamefont {Fogarty}},
  \bibinfo {author} {\bibfnamefont {T.}~\bibnamefont {Busch}}, \bibinfo
  {author} {\bibfnamefont {A.}~\bibnamefont {Foerster}}, \bibinfo {author}
  {\bibfnamefont {P.}~\bibnamefont {Schmelcher}}, \ and\ \bibinfo {author}
  {\bibfnamefont {N.}~\bibnamefont {Zinner}},\ }\href@noop {} {\bibfield
  {journal} {\bibinfo  {journal} {arXiv:2202.11071}\ } (\bibinfo {year}
  {2022})}\BibitemShut {NoStop}%
\bibitem [{\citenamefont {Lode}\ \emph {et~al.}(2020)\citenamefont {Lode},
  \citenamefont {L{\'e}v{\^e}que}, \citenamefont {Madsen}, \citenamefont
  {Streltsov},\ and\ \citenamefont {Alon}}]{lode2020colloquium}%
  \BibitemOpen
  \bibfield  {author} {\bibinfo {author} {\bibfnamefont {A.~U.~J.}\
  \bibnamefont {Lode}}, \bibinfo {author} {\bibfnamefont {C.}~\bibnamefont
  {L{\'e}v{\^e}que}}, \bibinfo {author} {\bibfnamefont {L.~B.}\ \bibnamefont
  {Madsen}}, \bibinfo {author} {\bibfnamefont {A.~I.}\ \bibnamefont
  {Streltsov}}, \ and\ \bibinfo {author} {\bibfnamefont {O.~E.}\ \bibnamefont
  {Alon}},\ }\href@noop {} {\bibfield  {journal} {\bibinfo  {journal} {Rev.
  Mod. Phys.}\ }\textbf {\bibinfo {volume} {92}},\ \bibinfo {pages} {011001}
  (\bibinfo {year} {2020})}\BibitemShut {NoStop}%
\bibitem [{\citenamefont {Horodecki}\ \emph
  {et~al.}(2009{\natexlab{a}})\citenamefont {Horodecki}, \citenamefont
  {Horodecki}, \citenamefont {Horodecki},\ and\ \citenamefont
  {Horodecki}}]{horodecki2009quantum}%
  \BibitemOpen
  \bibfield  {author} {\bibinfo {author} {\bibfnamefont {R.}~\bibnamefont
  {Horodecki}}, \bibinfo {author} {\bibfnamefont {P.}~\bibnamefont
  {Horodecki}}, \bibinfo {author} {\bibfnamefont {M.}~\bibnamefont
  {Horodecki}}, \ and\ \bibinfo {author} {\bibfnamefont {K.}~\bibnamefont
  {Horodecki}},\ }\href@noop {} {\bibfield  {journal} {\bibinfo  {journal}
  {Rev. Mod. Phys.}\ }\textbf {\bibinfo {volume} {81}},\ \bibinfo {pages} {865}
  (\bibinfo {year} {2009}{\natexlab{a}})}\BibitemShut {NoStop}%
\bibitem [{\citenamefont {Frenkel}(1934)}]{frenkel1934wave}%
  \BibitemOpen
  \bibfield  {author} {\bibinfo {author} {\bibfnamefont {J.}~\bibnamefont
  {Frenkel}},\ }\href@noop {} {\enquote {\bibinfo {title} {Wave mechanics;
  elementary theory},}\ } (\bibinfo {year} {1934})\BibitemShut {NoStop}%
\bibitem [{\citenamefont {Hu}\ \emph {et~al.}(2020)\citenamefont {Hu},
  \citenamefont {Wang},\ and\ \citenamefont {Liu}}]{Hui2020LowD}%
  \BibitemOpen
  \bibfield  {author} {\bibinfo {author} {\bibfnamefont {H.}~\bibnamefont
  {Hu}}, \bibinfo {author} {\bibfnamefont {J.}~\bibnamefont {Wang}}, \ and\
  \bibinfo {author} {\bibfnamefont {X.-J.}\ \bibnamefont {Liu}},\ }\href
  {\doibase 10.1103/PhysRevA.102.043301} {\bibfield  {journal} {\bibinfo
  {journal} {Phys. Rev. A}\ }\textbf {\bibinfo {volume} {102}},\ \bibinfo
  {pages} {043301} (\bibinfo {year} {2020})}\BibitemShut {NoStop}%
\bibitem [{\citenamefont {Giamarchi}(2003)}]{giamarchi2003quantum}%
  \BibitemOpen
  \bibfield  {author} {\bibinfo {author} {\bibfnamefont {T.}~\bibnamefont
  {Giamarchi}},\ }\href@noop {} {\emph {\bibinfo {title} {Quantum physics in
  one dimension}}},\ Vol.\ \bibinfo {volume} {121}\ (\bibinfo  {publisher}
  {Clarendon press},\ \bibinfo {year} {2003})\BibitemShut {NoStop}%
\bibitem [{\citenamefont {Sakmann}\ \emph {et~al.}(2008)\citenamefont
  {Sakmann}, \citenamefont {Streltsov}, \citenamefont {Alon},\ and\
  \citenamefont {Cederbaum}}]{Sakmann2008Coher}%
  \BibitemOpen
  \bibfield  {author} {\bibinfo {author} {\bibfnamefont {K.}~\bibnamefont
  {Sakmann}}, \bibinfo {author} {\bibfnamefont {A.~I.}\ \bibnamefont
  {Streltsov}}, \bibinfo {author} {\bibfnamefont {O.~E.}\ \bibnamefont {Alon}},
  \ and\ \bibinfo {author} {\bibfnamefont {L.~S.}\ \bibnamefont {Cederbaum}},\
  }\href {\doibase 10.1103/PhysRevA.78.023615} {\bibfield  {journal} {\bibinfo
  {journal} {Phys. Rev. A}\ }\textbf {\bibinfo {volume} {78}},\ \bibinfo
  {pages} {023615} (\bibinfo {year} {2008})}\BibitemShut {NoStop}%
\bibitem [{\citenamefont {Naraschewski}\ and\ \citenamefont
  {Glauber}(1999)}]{Naraschewski1999Coher}%
  \BibitemOpen
  \bibfield  {author} {\bibinfo {author} {\bibfnamefont {M.}~\bibnamefont
  {Naraschewski}}\ and\ \bibinfo {author} {\bibfnamefont {R.~J.}\ \bibnamefont
  {Glauber}},\ }\href {\doibase 10.1103/PhysRevA.59.4595} {\bibfield  {journal}
  {\bibinfo  {journal} {Phys. Rev. A}\ }\textbf {\bibinfo {volume} {59}},\
  \bibinfo {pages} {4595} (\bibinfo {year} {1999})}\BibitemShut {NoStop}%
\bibitem [{\citenamefont {Katsimiga}\ \emph {et~al.}(2020)\citenamefont
  {Katsimiga}, \citenamefont {Mistakidis}, \citenamefont {Bersano},
  \citenamefont {Ome}, \citenamefont {Mossman}, \citenamefont {Mukherjee},
  \citenamefont {Schmelcher}, \citenamefont {Engels},\ and\ \citenamefont
  {Kevrekidis}}]{katsimiga2020observation}%
  \BibitemOpen
  \bibfield  {author} {\bibinfo {author} {\bibfnamefont {G.~C.}\ \bibnamefont
  {Katsimiga}}, \bibinfo {author} {\bibfnamefont {S.~I.}\ \bibnamefont
  {Mistakidis}}, \bibinfo {author} {\bibfnamefont {T.~M.}\ \bibnamefont
  {Bersano}}, \bibinfo {author} {\bibfnamefont {M.~K.~H.}\ \bibnamefont {Ome}},
  \bibinfo {author} {\bibfnamefont {S.~M.}\ \bibnamefont {Mossman}}, \bibinfo
  {author} {\bibfnamefont {K.}~\bibnamefont {Mukherjee}}, \bibinfo {author}
  {\bibfnamefont {P.}~\bibnamefont {Schmelcher}}, \bibinfo {author}
  {\bibfnamefont {P.}~\bibnamefont {Engels}}, \ and\ \bibinfo {author}
  {\bibfnamefont {P.~G.}\ \bibnamefont {Kevrekidis}},\ }\href@noop {}
  {\bibfield  {journal} {\bibinfo  {journal} {Phys. Rev. A}\ }\textbf {\bibinfo
  {volume} {102}},\ \bibinfo {pages} {023301} (\bibinfo {year}
  {2020})}\BibitemShut {NoStop}%
\bibitem [{\citenamefont {Bersano}\ \emph {et~al.}(2018)\citenamefont
  {Bersano}, \citenamefont {Gokhroo}, \citenamefont {Khamehchi}, \citenamefont
  {D’Ambroise}, \citenamefont {Frantzeskakis}, \citenamefont {Engels},\ and\
  \citenamefont {Kevrekidis}}]{bersano2018three}%
  \BibitemOpen
  \bibfield  {author} {\bibinfo {author} {\bibfnamefont {T.~M.}\ \bibnamefont
  {Bersano}}, \bibinfo {author} {\bibfnamefont {V.}~\bibnamefont {Gokhroo}},
  \bibinfo {author} {\bibfnamefont {M.~A.}\ \bibnamefont {Khamehchi}}, \bibinfo
  {author} {\bibfnamefont {J.}~\bibnamefont {D’Ambroise}}, \bibinfo {author}
  {\bibfnamefont {D.~J.}\ \bibnamefont {Frantzeskakis}}, \bibinfo {author}
  {\bibfnamefont {P.}~\bibnamefont {Engels}}, \ and\ \bibinfo {author}
  {\bibfnamefont {P.~G.}\ \bibnamefont {Kevrekidis}},\ }\href@noop {}
  {\bibfield  {journal} {\bibinfo  {journal} {Phys. Rev. Lett.}\ }\textbf
  {\bibinfo {volume} {120}},\ \bibinfo {pages} {063202} (\bibinfo {year}
  {2018})}\BibitemShut {NoStop}%
\bibitem [{\citenamefont {Tolra}\ \emph {et~al.}(2004)\citenamefont {Tolra},
  \citenamefont {O'Hara}, \citenamefont {Huckans}, \citenamefont {Phillips},
  \citenamefont {Rolston},\ and\ \citenamefont {Porto}}]{Tolra20043BR}%
  \BibitemOpen
  \bibfield  {author} {\bibinfo {author} {\bibfnamefont {B.~L.}\ \bibnamefont
  {Tolra}}, \bibinfo {author} {\bibfnamefont {K.~M.}\ \bibnamefont {O'Hara}},
  \bibinfo {author} {\bibfnamefont {J.~H.}\ \bibnamefont {Huckans}}, \bibinfo
  {author} {\bibfnamefont {W.~D.}\ \bibnamefont {Phillips}}, \bibinfo {author}
  {\bibfnamefont {S.~L.}\ \bibnamefont {Rolston}}, \ and\ \bibinfo {author}
  {\bibfnamefont {J.~V.}\ \bibnamefont {Porto}},\ }\href {\doibase
  10.1103/PhysRevLett.92.190401} {\bibfield  {journal} {\bibinfo  {journal}
  {Phys. Rev. Lett.}\ }\textbf {\bibinfo {volume} {92}},\ \bibinfo {pages}
  {190401} (\bibinfo {year} {2004})}\BibitemShut {NoStop}%
\bibitem [{\citenamefont {Horodecki}\ \emph
  {et~al.}(2009{\natexlab{b}})\citenamefont {Horodecki}, \citenamefont
  {Horodecki}, \citenamefont {Horodecki},\ and\ \citenamefont
  {Horodecki}}]{Horodecki2009}%
  \BibitemOpen
  \bibfield  {author} {\bibinfo {author} {\bibfnamefont {R.}~\bibnamefont
  {Horodecki}}, \bibinfo {author} {\bibfnamefont {P.}~\bibnamefont
  {Horodecki}}, \bibinfo {author} {\bibfnamefont {M.}~\bibnamefont
  {Horodecki}}, \ and\ \bibinfo {author} {\bibfnamefont {K.}~\bibnamefont
  {Horodecki}},\ }\href {\doibase 10.1103/RevModPhys.81.865} {\bibfield
  {journal} {\bibinfo  {journal} {Rev. Mod. Phys.}\ }\textbf {\bibinfo {volume}
  {81}},\ \bibinfo {pages} {865} (\bibinfo {year}
  {2009}{\natexlab{b}})}\BibitemShut {NoStop}%
\bibitem [{\citenamefont {Hu}\ and\ \citenamefont {Liu}(2020)}]{Hui20203D}%
  \BibitemOpen
  \bibfield  {author} {\bibinfo {author} {\bibfnamefont {H.}~\bibnamefont
  {Hu}}\ and\ \bibinfo {author} {\bibfnamefont {X.-J.}\ \bibnamefont {Liu}},\
  }\href {\doibase 10.1103/PhysRevLett.125.195302} {\bibfield  {journal}
  {\bibinfo  {journal} {Phys. Rev. Lett.}\ }\textbf {\bibinfo {volume} {125}},\
  \bibinfo {pages} {195302} (\bibinfo {year} {2020})}\BibitemShut {NoStop}%
\bibitem [{\citenamefont {Ronzheimer}\ \emph {et~al.}(2013)\citenamefont
  {Ronzheimer}, \citenamefont {Schreiber}, \citenamefont {Braun}, \citenamefont
  {Hodgman}, \citenamefont {Langer}, \citenamefont {McCulloch}, \citenamefont
  {Heidrich-Meisner}, \citenamefont {Bloch},\ and\ \citenamefont
  {Schneider}}]{ronzheimer2013expansion}%
  \BibitemOpen
  \bibfield  {author} {\bibinfo {author} {\bibfnamefont {J.~P.}\ \bibnamefont
  {Ronzheimer}}, \bibinfo {author} {\bibfnamefont {M.}~\bibnamefont
  {Schreiber}}, \bibinfo {author} {\bibfnamefont {S.}~\bibnamefont {Braun}},
  \bibinfo {author} {\bibfnamefont {S.~S.}\ \bibnamefont {Hodgman}}, \bibinfo
  {author} {\bibfnamefont {S.}~\bibnamefont {Langer}}, \bibinfo {author}
  {\bibfnamefont {I.~P.}\ \bibnamefont {McCulloch}}, \bibinfo {author}
  {\bibfnamefont {F.}~\bibnamefont {Heidrich-Meisner}}, \bibinfo {author}
  {\bibfnamefont {I.}~\bibnamefont {Bloch}}, \ and\ \bibinfo {author}
  {\bibfnamefont {U.}~\bibnamefont {Schneider}},\ }\href@noop {} {\bibfield
  {journal} {\bibinfo  {journal} {Phys. Rev. Lett.}\ }\textbf {\bibinfo
  {volume} {110}},\ \bibinfo {pages} {205301} (\bibinfo {year}
  {2013})}\BibitemShut {NoStop}%
\bibitem [{\citenamefont {Fukuhara}\ \emph {et~al.}(2013)\citenamefont
  {Fukuhara}, \citenamefont {Kantian}, \citenamefont {Endres}, \citenamefont
  {Cheneau}, \citenamefont {Schau{\ss}}, \citenamefont {Hild}, \citenamefont
  {Bellem}, \citenamefont {Schollw{\"o}ck}, \citenamefont {Giamarchi},
  \citenamefont {Gross}, ,\ and\ \citenamefont {Bloch}}]{fukuhara2013quantum}%
  \BibitemOpen
  \bibfield  {author} {\bibinfo {author} {\bibfnamefont {T.}~\bibnamefont
  {Fukuhara}}, \bibinfo {author} {\bibfnamefont {A.}~\bibnamefont {Kantian}},
  \bibinfo {author} {\bibfnamefont {M.}~\bibnamefont {Endres}}, \bibinfo
  {author} {\bibfnamefont {M.}~\bibnamefont {Cheneau}}, \bibinfo {author}
  {\bibfnamefont {P.}~\bibnamefont {Schau{\ss}}}, \bibinfo {author}
  {\bibfnamefont {S.}~\bibnamefont {Hild}}, \bibinfo {author} {\bibfnamefont
  {D.}~\bibnamefont {Bellem}}, \bibinfo {author} {\bibfnamefont
  {U.}~\bibnamefont {Schollw{\"o}ck}}, \bibinfo {author} {\bibfnamefont
  {T.}~\bibnamefont {Giamarchi}}, \bibinfo {author} {\bibfnamefont
  {C.}~\bibnamefont {Gross}}, , \ and\ \bibinfo {author} {\bibfnamefont
  {I.}~\bibnamefont {Bloch}},\ }\href@noop {} {\bibfield  {journal} {\bibinfo
  {journal} {Nature Phys.}\ }\textbf {\bibinfo {volume} {9}},\ \bibinfo {pages}
  {235} (\bibinfo {year} {2013})}\BibitemShut {NoStop}%
\bibitem [{\citenamefont {Salasnich}(2000)}]{Salasnich2000}%
  \BibitemOpen
  \bibfield  {author} {\bibinfo {author} {\bibfnamefont {L.}~\bibnamefont
  {Salasnich}},\ }\href {\doibase 10.1142/S0217979200000029} {\bibfield
  {journal} {\bibinfo  {journal} {Int. J. Mod. Phys. B}\ }\textbf {\bibinfo
  {volume} {14}},\ \bibinfo {pages} {1} (\bibinfo {year} {2000})}\BibitemShut
  {NoStop}%
\bibitem [{\citenamefont {Mistakidis}\ \emph
  {et~al.}(2021{\natexlab{b}})\citenamefont {Mistakidis}, \citenamefont
  {Koutentakis}, \citenamefont {Grusdt}, \citenamefont {Sadeghpour},\ and\
  \citenamefont {Schmelcher}}]{mistakidis2021radiofrequency}%
  \BibitemOpen
  \bibfield  {author} {\bibinfo {author} {\bibfnamefont {S.~I.}\ \bibnamefont
  {Mistakidis}}, \bibinfo {author} {\bibfnamefont {G.~M.}\ \bibnamefont
  {Koutentakis}}, \bibinfo {author} {\bibfnamefont {F.}~\bibnamefont {Grusdt}},
  \bibinfo {author} {\bibfnamefont {H.~R.}\ \bibnamefont {Sadeghpour}}, \ and\
  \bibinfo {author} {\bibfnamefont {P.}~\bibnamefont {Schmelcher}},\
  }\href@noop {} {\bibfield  {journal} {\bibinfo  {journal} {New J. Phys.}\
  }\textbf {\bibinfo {volume} {23}},\ \bibinfo {pages} {043051} (\bibinfo
  {year} {2021}{\natexlab{b}})}\BibitemShut {NoStop}%
\bibitem [{\citenamefont {Mistakidis}\ \emph {et~al.}(2020)\citenamefont
  {Mistakidis}, \citenamefont {Katsimiga}, \citenamefont {Koutentakis},
  \citenamefont {Busch},\ and\ \citenamefont
  {Schmelcher}}]{mistakidis2020pump}%
  \BibitemOpen
  \bibfield  {author} {\bibinfo {author} {\bibfnamefont {S.~I.}\ \bibnamefont
  {Mistakidis}}, \bibinfo {author} {\bibfnamefont {G.~C.}\ \bibnamefont
  {Katsimiga}}, \bibinfo {author} {\bibfnamefont {G.~M.}\ \bibnamefont
  {Koutentakis}}, \bibinfo {author} {\bibfnamefont {T.}~\bibnamefont {Busch}},
  \ and\ \bibinfo {author} {\bibfnamefont {P.}~\bibnamefont {Schmelcher}},\
  }\href@noop {} {\bibfield  {journal} {\bibinfo  {journal} {Phys. Rev.
  Research}\ }\textbf {\bibinfo {volume} {2}},\ \bibinfo {pages} {033380}
  (\bibinfo {year} {2020})}\BibitemShut {NoStop}%
\bibitem [{\citenamefont {Ma}\ \emph {et~al.}(2021)\citenamefont {Ma},
  \citenamefont {Peng},\ and\ \citenamefont {Cui}}]{ma2021borromean}%
  \BibitemOpen
  \bibfield  {author} {\bibinfo {author} {\bibfnamefont {Y.}~\bibnamefont
  {Ma}}, \bibinfo {author} {\bibfnamefont {C.}~\bibnamefont {Peng}}, \ and\
  \bibinfo {author} {\bibfnamefont {X.}~\bibnamefont {Cui}},\ }\href@noop {}
  {\bibfield  {journal} {\bibinfo  {journal} {Phys. Rev. Lett.}\ }\textbf
  {\bibinfo {volume} {127}},\ \bibinfo {pages} {043002} (\bibinfo {year}
  {2021})}\BibitemShut {NoStop}%
\bibitem [{\citenamefont {Keiler}\ \emph {et~al.}(2021)\citenamefont {Keiler},
  \citenamefont {Mistakidis},\ and\ \citenamefont
  {Schmelcher}}]{keiler2021polarons}%
  \BibitemOpen
  \bibfield  {author} {\bibinfo {author} {\bibfnamefont {K.}~\bibnamefont
  {Keiler}}, \bibinfo {author} {\bibfnamefont {S.~I.}\ \bibnamefont
  {Mistakidis}}, \ and\ \bibinfo {author} {\bibfnamefont {P.}~\bibnamefont
  {Schmelcher}},\ }\href@noop {} {\bibfield  {journal} {\bibinfo  {journal}
  {Phys. Rev. A}\ }\textbf {\bibinfo {volume} {104}},\ \bibinfo {pages}
  {L031301} (\bibinfo {year} {2021})}\BibitemShut {NoStop}%
\end{thebibliography}%

\end{document}